\documentclass[final,notitlepage,11pt, a4paper]{article}

\usepackage{xcolor}
\usepackage{tcolorbox}
\usepackage{amsfonts}
\usepackage{amssymb}
\usepackage{graphicx}
\usepackage{amsmath}
\usepackage{latexsym}
\usepackage{rotating}
\usepackage{lscape}
\usepackage{color}
\usepackage{enumerate}
\usepackage{epstopdf}
\usepackage{tabularx}
\usepackage{array}
\usepackage[center]{caption}
\usepackage[authoryear]{natbib}
\usepackage{setspace}
\usepackage{geometry}
\usepackage{enumitem}
\usepackage{subfig}
\usepackage[unicode=true,pdfusetitle,
 bookmarks=true,bookmarksnumbered=false,bookmarksopen=false,
 breaklinks=false,pdfborder={0 0 1},backref=false,colorlinks=false,hidelinks]{hyperref}
\usepackage{amsthm}

\newcolumntype{C}[1]{>{\centering\arraybackslash}p{#1}}
\newcolumntype{L}[1]{>{\raggedright\let\newline\\\arraybackslash\hspace{0pt}}m{#1}}
\theoremstyle{definition}
\newtheorem{theorem}{Theorem}

\newtheorem{assumption}{Assumption}

\newtheorem{example}{Example}

\newtheorem{lemma}{Lemma}

\newtheorem{remark}{Remark}

\numberwithin{equation}{section}
\makeatletter
\renewcommand{\@cite}[1]{#1}
\makeatother
\makeatletter
\def\@biblabel#1{\hspace*{-\labelsep}}
\makeatletter
\renewcommand\@biblabel[1]{}

\makeatother
\begin{document}

\title{$\ell_2$-Relaxation: With Applications to \\
Forecast Combination and Portfolio Analysis}
\author{Zhentao Shi, Liangjun Su and Tian Xie}
\date{}

\maketitle

\begin{abstract}
This paper tackles forecast combination with many forecasts
or minimum variance portfolio selection with many assets. A novel convex
problem called $\ell_{2}$-\emph{relaxation} is proposed. In contrast to
standard formulations, $\ell _{2}$-relaxation minimizes the squared
Euclidean norm of the weight vector subject to a set of relaxed linear
inequality constraints. The magnitude of relaxation, controlled by a tuning
parameter, balances the bias and variance. When the variance-covariance (VC)
matrix of the individual forecast errors or financial assets exhibits latent
group structures --- a block equicorrelation matrix plus a VC for
idiosyncratic noises, the solution to $\ell_{2}$-relaxation delivers
roughly equal within-group weights. Optimality of the new method is
established under the asymptotic framework when the number of the
cross-sectional units $N$ potentially grows much faster than the time
dimension $T$. Excellent finite sample performance of our method is
demonstrated in Monte Carlo simulations. Its wide applicability is
highlighted in three real data examples concerning empirical applications of
microeconomics, macroeconomics, and finance.

{\footnotesize \medskip{} }


\end{abstract}

\singlespacing

\singlespacing

%\begin{acknowledgement*}
{\footnotesize We thank the editor and two referees for their valuable
suggestions. We are also grateful to Bruce Hansen, Yingying Li, Esfandiar
Maasoumi, Jack Porter, Yuying Sun, Aman Ullah, Xia Wang, Xinyu Zhang, and
Xinghua Zheng for their helpful comments. Shi acknowledges financial support
from Hong Kong Research Grants Council (RGC) No.14500118. Su gratefully
acknowledges the support from Natural Science Foundation of China
(No.72133002). Xie's research is supported by the Natural Science Foundation
of China (No.72173075), the Shanghai Research Center for Data Science and
Decision Technology, and the Fundamental Research Funds for the Central
Universities. Address correspondence: Zhentao Shi: \texttt{%
zhentao.shi@gatech.edu}, School of Economics, Georgia Institute of
Technology, 205 Old C.E. Building, 221 Bobby Dodd Way, Atlanta, GA 30332,
U.S.A., and Department of Economics, 928 Esther Lee Building, the Chinese
University of Hong Kong, Shatin, New Territories, Hong Kong SAR, China.
Liangjun Su: \texttt{sulj@sem.tsinghua.edu.cn}, School of Economics and
Management, Tsinghua University, Beijing, China. Tian Xie: \texttt{%
xietian@shufe.edu.cn}, College of Business, Shanghai University of Finance
and Economics, Shanghai, China.} %\end{acknowledgement*}

\newpage

\singlespacing

\section{Introduction}

Forecast combination assigns weights to individual experts to reduce
forecast errors, and portfolio management assigns weights to financial
assets to reduce risk exposure. The classical approach to both problems, to
be formally laid out in Section \ref{subsec:classical}, solves a quadratic
optimization 
\begin{equation}
\min_{\mathbf{w}\in \mathbb{R}^{N}}\,\frac{1}{2}\mathbf{w}^{\prime }\widehat{%
\boldsymbol{\Sigma }}\mathbf{w}\ \ \text{subject to \ }\mathbf{w}^{\prime }%
\boldsymbol{1}_{N}=1,  \label{eq:bates-granger}
\end{equation}%
where $N$ is the number of forecasts or assets, $\mathbf{1}_{N}$ is a column
of $N$ ones, and $\widehat{\boldsymbol{\Sigma }}$ is a variance-covariance
(VC) matrix estimate computed from $T$ time series observations. 
% of an $N$-vector $(s_{1t},s_{2t},\ldots,s_{Nt})'$. 
% Here $s_{it}$ represents a forecast error in 
% forecast combination \citep{bates1969combination}, and an asset's excess return in portfolio \citep{markowitz1952portfolio}.
When $\widehat{\boldsymbol{\Sigma }}$ is invertible, the explicit solution
to \eqref{eq:bates-granger} is 
\begin{equation}
\widehat{\mathbf{w}}^{\mathrm{\mathrm{C}}}=(\boldsymbol{1}_{N}^{\prime }%
\widehat{\boldsymbol{\Sigma }}^{-1}\boldsymbol{1}_{N})^{-1}\widehat{%
\boldsymbol{\Sigma }}^{-1}\boldsymbol{1}_{N},
\label{eq:bates-granger-explicit}
\end{equation}%
where \textquotedblleft C\textquotedblright\ in the superscript refers to
the \emph{classical} approach.

Consider, for simplicity, the case when $\widehat{\boldsymbol{\Sigma }}$ is
estimated as the plain sample VC matrix computed from $T$ observations of $%
N\times 1$ vectors, where each entry represents a forecast error in forecast
combination \citep{bates1969combination}, or an asset's excess return in a
portfolio \citep{markowitz1952portfolio}. When $T\gg N$, the classical
solution \eqref{eq:bates-granger-explicit} is valid as $\widehat{\boldsymbol{%
\Sigma }}$ is generally invertible. However, $\widehat{\boldsymbol{\Sigma }}$
may suffer ill-posedness when $N$ is comparable to $T$, and it is singular
when $N$ is larger than $T$, which invalidates %
\eqref{eq:bates-granger-explicit}. Such defects are well recognized in the
empirical literature. When $N$ is large, $\widehat{\mathbf{w}}^{\mathrm{C}}$
often performs poorly because of the difficulty in estimating the large
population VC matrix with precision. Rather, the simple average (namely, 
\textit{equal-weight}) routinely outperforms. In forecast combination, this
empirical fact is known as the \textit{forecast combination puzzle} %
\citep{Clemen1989, Stock_Watson2004}. Parallelly, in portfolio management
the so-called \emph{naive $1/N$ diversification strategy} is found to
achieve robust out-of-sample gains compared to many sophisticated
alternatives \citep{demiguel2009optimal, demiguel2009generalized}.

In this paper, we propose a new estimation technique on the weights, to be
presented in Section \ref{subsec:relaxation} below. This method minimizes
the squared Euclidean norm ($\ell _{2}$-norm) of the weight vector subject
to a relaxed version of the first order conditions from the minimization
problem in (\ref{eq:bates-granger}), yielding the $\ell _{2}$-\emph{%
relaxation} problem. The strategy is similar in spirit to the $\ell _{1}$%
-relaxation in Dantzig selector \emph{a la} \citet{candes2007dantzig}.
Interestingly, \ $\ell _{2}$-\emph{relaxation} incorporates as special cases
the simple average (equal-weight) strategy by setting the tuning parameter
to be sufficiently large, and the classical optimal weighting scheme when
the tuning parameter is zero. The tuning parameter helps to balance the bias
and variance and to deliver roughly equal groupwise weights when the VC
matrix exhibits a certain latent group structure. This is consistent with
the intuition that when the VC matrix displays an exactly block
equicorrelation structure, one should assign the same weight to all
individual units within the same group whereas potentially distinct weights
to the units in different groups. When the VC matrix is contaminated with a
noise component, we show that the resultant $\ell _{2}$-relaxed weights are
close to the infeasible groupwise equal weights.

The approximate latent group structure is not a man-made artifact. It is
inherent in many models and applications. For example, it emerges when a
factor structure dwells in the forecasting errors and the factor loadings
are directly governed by certain latent group structures or are approximable
by a few values (see Examples \ref{exa:one_factor} and \ref{exa:q_factor} in
Section 3.1). It emerges when forecast combinations are based on a large
number of forecast models with a fixed number of predictive regressors (see
Example \ref{exa:regression} in Section 3.1). It also emerges when
industrial classification serves as a proxy of the clustering pattern in the
VC matrix of returns of stocks (see Example \ref{exa:portfolio} in Section
3.1). % In the latter case, we argue that the $p
% $ regressors serve as a part of the\ \textquotedblleft
% latent\textquotedblright\ factors.

Given the latent group structure, we establish two main theoretical results
under a high dimensional asymptotic framework in which the number of
individual units can be much larger than the time series dimension. (i) The
estimated weights of $\ell _{2}$-relaxation converges to the within-group
equal-weight solution (see Theorem \ref{thm:w_converg} in Section 3.2), and
(ii) The empirical risk based on $\ell _{2}$-relaxation approaches the risk
given by the oracle group information (see Theorem \ref{thm:oracle} in
Section 3.2). We assess the finite sample behavior of $\ell _{2}$-relaxation
in Monte Carlo simulations. Compared with the oracle estimator and some
popular off-the-shelf machine learning estimators, $\ell _{2}$-relaxation
performs well under various data generating processes (DGPs). We further
evaluate its empirical accuracy in three real data examples covering box
office prediction, inflation forecast by surveyed professionals, and
financial market portfolios. These examples showcase the wide applicability
of $\ell _{2}$-relaxation.

\textbf{Literature Review.} This paper stands on several strands of vast
literature. Forecast combination is reviewed by \citet{Clemen1989} and %
\citet{elliott2016} up to the points of their writing. Averaging forecasts
appear to be a more robust procedure than the so-called optimal combination %
\citep{bates1969combination}, and a reasonable explanation suggests that the
errors on the estimation of the weights can be large and thus dominate the
gains from the use of optimal combination (see, e.g., %
\citeauthor{Smith_Wallis2009}, \citeyear{Smith_Wallis2009}, %
\citeauthor{CMVW2016}, \citeyear{CMVW2016}).

A lesson learned from this literature is that it is unwise to include all
possible variables; limiting the number of unknown parameters can help
reduce estimation errors. This stylized fact has catalyzes the adoption of
various shrinkage, regularization, and machine learning techniques. See,
e.g., \citet{hansen2007least}, %\citet{Li_Chen2014}, 
\citet{Conflitti2015optimal}, %
%\citet{Konzen_Ziegelmann2016}, 
% \citet{Stasinakis2016forecasting}, %
\citet{Bayer2018combining}, \citet{Wilms2018}, %
\citet{Kotchoni2019macroeconomic}, \citet{Coulombe2020machine}, %
\citet{diebold2021aggregation}, and \citet{Roccazzella2020optimal}. In
particular, \citet{elliott2013complete} propose a complete subset regression
(CSR) approach to forecast combinations by using equal weights to combine
forecasts based on the same number of predictive variables. 
%They show that in many cases subset regression
%combinations amount to a form of shrinkage that is more general than the
%conventional variable-by-variable shrinkage implied by ridge regression. 
\citet{Diebold_Shin2019} bring forth the partially egalitarian Lasso
(peLASSO) procedures that discard some forecasts and then shrink the
remaining forecasts toward the equal weights. % In essence, both %
% These two procedures assign two distinct
% weights to two subsets of models: zero weight to a large subset of the
% models and equal or roughly equal weights to the remaining subset of the
% models.
Our $\ell _{2}$-relaxation adds a new way to regularize the weight
estimation and includes the strategy of \citet{Diebold_Shin2019} as a
special case.

Our paper is related to the burgeoning literature on latent group structures
in panel data analysis; see, e.g., \citet{bonhomme2015grouped}, %
\citet{su2016identifying}, \citet{Su_Ju2018}, \citet{Su_Wang_Jin2019}, %
\citet{vogt2017classification}, \citet{vogt2020multiscale}, and %
\citet{bonhomme2022discretizing}. While most of these previous studies focus
on the recovery of the latent group structures in the conditional mean
model, in our paper the group pattern is a latent structure in the VC matrix
that encourages parameter parsimony and facilitates estimation accuracy. We
do not attempt to recover the group identities.

Lastly, there is statistical and financial econometric literature on the
estimation of large VC matrix. See \citet{Disatnik_Katz2012}, %
\citet{fanli2012vast}, \citet{Fan_Liao_Mincheva2013}, %
\citet{fan2016incorporating}, \citet{Ledoit_Wolf2017}, and %
\citet{ao2019approaching}, among many others. In particular, %
\citet{Ledoit_Wolf2004} use Bayesian methods for shrinking the sample
correlation matrix to an equicorrelated target and show that this helps
select portfolios with low volatility compared to those based on the sample
correlation; \citet{Ledoit_Wolf2017} promote a nonlinear shrinkage estimator
that is more flexible than the previous linear shrinkage estimators. 
% and has just the right number of free parameters.
Instead of regularizing the VC matrix, $\ell _{2}$-relaxation shrinks the
weights and it can be used in conjunction with a high dimensional VC
estimator.

Our paper complements the literature from the following aspects. Firstly, in
terms of combination techniques, we corroborate in theory and in numerical
experiments that $\ell_2$-relaxation is a competitive and easy-to-implement
procedure. Secondly, unlike most panel data group structure papers, we focus
on improvement of the out-of-sample performance and do not attempt to
recover the membership for each individual, and a latent community or group
structure is assumed for statistical optimality. Finally, while the
dominating method in the large scale portfolio analysis shrinks the entries
of the VC matrix, we take a viable alternative to directly discipline the
weights. In summary, within the unified framework of forecast combination
and portfolio optimization, $\ell _{2}$-relaxation is an innovative method
with asymptotic guarantee under latent group structures.

\textbf{Organization.} The rest of the paper is organized as follows.
Section \ref{sec:-Relaxation} motivates and introduces the $\ell _{2}$%
-relaxation problem. Section \ref{sec:theory} studies the statistical
properties of the estimator and establishes its asymptotic optimality under
the latent group structures. Section \ref{sec:simulations} reports Monte
Carlo simulation results. The new method is applied to three datasets in
Section \ref{sec:empirical_app}. All theoretical results are proved in
Appendix A, and additional numerical results are contained in Appendix B.

\medskip

\textbf{Notation.} Let \textquotedblleft $:=$\textquotedblright\ signify a
definition, \textquotedblleft $\otimes $\textquotedblright\ be the Kronecker
product, and $a\wedge b=\min \left\{ a,b\right\} $. We write $a\asymp b$
when both $a/b$ and $b/a$ are stochastically bounded. For a random variable $%
x$, we write its population mean as $E\left[ x\right] $; for a sample $%
\left( x_{1},\ldots ,x_{T}\right) $, we write its sample mean as $\mathbb{E}%
_{T}\left[ x_{t}\right] :=T^{-1}\sum_{t=1}^{T}x_{t}$. A plain $b$ denotes a
scalar, a boldface lowercase $\mathbf{b}$ denotes a vector, and a boldface
uppercase $\mathbf{B}$ denotes a matrix. The $\ell _{1}$-norm and $\ell _{2}$%
-norm of $\mathbf{b}=(b_{1},...,b_{n})^{\prime }$ are written as $\left\Vert 
\mathbf{b}\right\Vert _{1}:=\sum_{i=1}^{n}\left\vert b_{i}\right\vert $ and $%
\left\Vert \mathbf{b}\right\Vert _{2}:=\left( \sum_{i=1}^{n}b_{i}^{2}\right)
^{1/2},$ respectively. For a generic index set $\mathcal{G}\subset \left[ N%
\right] :=\left\{ 1,\ldots ,N\right\} $, we denote $\left\vert \mathcal{G}%
\right\vert $ as the cardinality of $\mathcal{G}$, and $\mathbf{b}_{\mathcal{%
G}}=\left( b_{i}\right) _{i\in \mathcal{G}_{k}}$ as the $\left\vert \mathcal{%
G}\right\vert $-dimensional subvector. $\phi _{\max }\left( \mathbf{\cdot }%
\right) $ and $\phi _{\min }\left( \cdot \right) $ represent the maximum and
minimum eigenvalues of a real symmetric matrix, respectively. For a generic $%
n\times m$ matrix $\mathbf{B}$, define the sup-norm as $\left\Vert \mathbf{B}%
\right\Vert _{\infty }:=\max_{i\leq n,j\leq m}\left\vert b_{ij}\right\vert $%
, and the maximum column $\ell _{2}$ matrix norm as $\left\Vert \mathbf{B}%
\right\Vert _{c2}:=\max\nolimits_{j\leq m}\Vert \mathbf{B}_{\cdot j}\Vert
_{2}$, where $\mathbf{B}_{\cdot j}$ is the $j$-th column. $\boldsymbol{0}%
_{n} $ and $\boldsymbol{1}_{n}$ are $n\times 1$ vectors of zeros and ones,
respectively, and $\mathbf{\mathbf{I}}_{n}$ is the $n\times n$ identity
matrix.

\section{ Formulation}

\label{sec:-Relaxation}

\subsection{Classical Approaches}

\label{subsec:classical}

In this section, we fix the ideas by characterizing the similarities between
the classical forecast combination problem and portfolio optimization. We
start with the former. Suppose that $y_{t+1}$ is an outcome variable, and
there are $N$ forecasts for $y_{t+1}$, stacked as $\mathbf{f}%
_{t}=(f_{1t},...,f_{Nt})^{\prime }$, available at time $t$. The time
dimension is indexed by $t\in \left[ T\right] :=\{1,2,...,T\}$, and the
cross-sectional units are indexed by $i\in \left[ N\right] $. An $N\times 1$
weight vector $\mathbf{w}=(w_{i})_{i\in \lbrack N]}$ will linearly combine
the forecasts into $\mathbf{w}^{\prime }\mathbf{f}_{t}$. We are interested
in finding the weight $\mathbf{w}$ to minimize the mean squared forecast
error (MSFE) of the combined forecast error $(y_{t+1}-\mathbf{w}^{\prime }%
\mathbf{f}_{t})$. % One way to estimate $\mathbf{w}$ is to
% run the restricted least squares (RLS): 
% \begin{equation}
% \min_{\mathbf{w}\in \mathbb{R}^{N}}\,\frac{1}{2T}\sum_{t=1}^{T}\left(
% y_{t+1}-\mathbf{w}^{\prime }\mathbf{f}_{t}\right) ^{2}\ \ \text{\ \ subject
% to \ }\mathbf{w}^{\prime }\boldsymbol{1}_{N}=1,  \label{eq:BG_regression}
% \end{equation}%
% where $\boldsymbol{1}_{N}$ is an $N\times 1$ vector of ones. Alternatively,
% The two formulations in (\ref{eq:bates-granger}) and (\ref{eq:BG_regression}%
% ) are numerically equivalent, as demonstrated by \citet{granger1984improved}. 

% \citet{bates1969combination} proposes the following procedure.
We collect the individual forecast error $e_{it}:=y_{t+1}-f_{it}$, and
denote $\mathbf{e}_{t}:=(e_{it})_{i\in \lbrack N]}$ and $\bar{\mathbf{e}}%
:=T^{-1}\sum_{t=1}^{T}\mathbf{e}_{t}$.\footnote{\citet{bates1969combination}
assume unbiased forecasts and thus no demeaning is necessary in the
construction of $\widehat{\boldsymbol{\Sigma }}$ in (\ref{eq:vanilla_vc})
below. Here we accommodate potential biases of individual forecasts by the
centered sample variance in order to present a unified framework for both
forecast combination and portfolio optimization (see \eqref{eq:markowitz}).
Section \ref{subsec:bg_w} in the Online Appendix shows that $\widehat{%
\boldsymbol{\Sigma }}$ in (\ref{eq:vanilla_vc}) copes with biased forecasts.}
We compute its plain sample VC matrix\footnote{%
We use the plain sample VC here to simplify the presentation. Alternative VC
estimators tailored for high dimensional contexts can also be employed. In
Section \ref{sec:simulations}, we report simulation results from the
shrinkage VC estimators \citep{Ledoit_Wolf2004, Ledoit_Wolf2020} along with
those from the plain sample VC estimator.} 
\begin{equation}
\widehat{\boldsymbol{\Sigma }}:=T^{-1}\sum_{t=1}^{T}(\mathbf{e}_{t}-\bar{%
\mathbf{e}})(\mathbf{e}_{t}-\bar{\mathbf{e}})^{\prime }.
\label{eq:vanilla_vc}
\end{equation}%
Traditionally, the weights are determined by solving \eqref{eq:bates-granger}%
.

Forecast combination is intrinsically related to the mean-variance analysis
of portfolio selection \citep{markowitz1952portfolio}. Given $N$ financial
assets of excess (relative to a risk-free asset) return $\mathbf{r}%
_{t}=(r_{it})_{i\in \lbrack N]}$, write the sample average return $\bar{%
\mathbf{r}}:=T^{-1}\sum_{t=1}^{T}\mathbf{r}_{t}$ and the plain sample VC
matrix 
\begin{equation}
\widehat{\boldsymbol{\Sigma }}:=T^{-1}\sum_{t=1}^{T}(\mathbf{r}_{t}-\bar{%
\mathbf{r}})(\mathbf{r}_{t}-\bar{\mathbf{r}})^{\prime }.
\label{eq:vanilla_vc2}
\end{equation}%
The weight vector $\mathbf{w}$ can be solved from 
\begin{equation}
\min_{\mathbf{w}\in \mathbb{R}^{N}}\,\frac{1}{2}\mathbf{w}^{\prime }\widehat{%
\boldsymbol{\Sigma }}\mathbf{w}\ \ \text{subject to \ }\mathbf{w}^{\prime }%
\boldsymbol{1}_{N}=1\mbox{ and }\bar{\mathbf{r}}^{\prime }\mathbf{w}\geq
r^{\ast },  \label{eq:markowitz}
\end{equation}%
where $r^{\ast }$ is a user-specified target return. It is recognized that
when there are many assets, precise estimation of the mean returns is a
challenging task \citep{merton1980estimating}, and the recent literature
shifts to the minimum variance portfolio (MVP). MVP drops the linear
restriction on returns (i.e., $\bar{\mathbf{r}}^{\prime }\mathbf{w}\geq
r^{\ast }$), which leads to an optimization problem identical to %
\eqref{eq:bates-granger}.\footnote{%
See \citet[Chapter 7.1]{linton2019financial} for a textbook treatment, %
\citet{demiguel2009generalized} and \citet{demiguel2009generalized} for
extensive empirical comparisons, and \citet{fan2012vast}, \citet{cai2020high}
and \citet{ding2021high} for latest advancements of MVP.}

As forecast combination and MVP share the same form, we focus on the
optimization problem in \eqref{eq:bates-granger}. It can be rewritten as an
unconstrained Lagrangian problem $\mathbf{w}^{\prime }\widehat{\boldsymbol{%
\Sigma }}\mathbf{w}/2 +\gamma \left( \mathbf{w}^{\prime }\boldsymbol{1}%
_{N}-1\right) ,$ where $\gamma $ is the Lagrangian multiplier. The
corresponding Kuhn-Karush-Tucker (KKT) conditions are: 
\begin{equation}
\begin{array}{rcl}
\widehat{\boldsymbol{\Sigma }}\mathbf{w}+\gamma \boldsymbol{1}_{N} & = & 
\boldsymbol{0}_{N}, \\ 
\mathbf{w}^{\prime }\boldsymbol{1}_{N}-1 & = & 0.%
\end{array}
\label{eq:KKT}
\end{equation}

%In both the forecast combination problem and the MVP problem, 
The invertibility of the estimated VC matrix is not innocuous in high
dimensional settings.\footnote{%
The term \textquotedblleft high dimensional\textquotedblright\ means that
the number of unknown parameters (in our context, $N$) is comparable to or
larger than the sample size $T$. We will allow for $N/T\rightarrow c\in
(0,\infty )$ as $\left( N,T\right) \rightarrow \infty $.} For example, when $%
\widehat{\boldsymbol{\Sigma }}$ is the plain sample VC matrix, it must be
singular when $N>T.$ 
%Potential problems arise when $N$ is large relatively to the time dimension $T$. 
Consider the case where $N$ is of similar magnitude to $T$ but $N<T$. 
%, say $N=80$ and $T=100$. 
Even if $\widehat{\boldsymbol{\Sigma }}$ is non-singular, a few sample
eigenvalues of $\widehat{\boldsymbol{\Sigma }}$ are likely to be close to
zero, leading to a numerically unstable solution when taking the matrix
inverse in (\ref{eq:bates-granger-explicit}).

\subsection{Relaxation}

\label{subsec:relaxation}

To stabilize the numerical solution, we are inspired by the Dantzig selector %
\citep{candes2007dantzig} and the relaxed empirical likelihood %
\citep{shi2016econometric} to consider relaxing the sup-norm of the KKT
condition as follows: 
\begin{equation}  \label{eq:relax}
\min_{\left( \mathbf{w},\gamma \right) \in \mathbb{R}^{N+1}}\ \frac{1}{2}%
\left\Vert \mathbf{w}\right\Vert _{2}^{2}\text{ \ \ subject to \ }\mathbf{w}%
^{\prime }\boldsymbol{1}_{N}=1\ \text{and }\Vert \widehat{\boldsymbol{\Sigma 
}}\mathbf{w}+\gamma \boldsymbol{1}_{N}\Vert _{\infty }\leq \tau ,
\end{equation}%
where $\tau $ is a tuning parameter to be specified by the user. We call the
programming in (\ref{eq:relax}) the $\ell _{2}$\textit{-relaxation problem},
and denote its solution as $\widehat{\mathbf{w}}=\widehat{\mathbf{w}}_{\tau
},$ where the dependence of $\widehat{\mathbf{w}}$ on $\tau $ is often
suppressed for notational conciseness.

When $\tau =0$, the solution $\widehat{\mathbf{w}}$ is characterized by the
KKT conditions in (\ref{eq:KKT}), and $\widehat{\mathbf{w}}^{\mathrm{C}}$ in
(\ref{eq:bates-granger-explicit}) is the unique solution when $\widehat{%
\boldsymbol{\Sigma }}$ is invertible. Thus $\ell _{2}$-relaxation keeps the
classical approach as a special case. Constraints in (\ref{eq:relax}) are
feasible for any $\tau \geq 0$. The solution to (\ref{eq:relax}) is always
unique because the objective is a strictly convex function and the feasible
set is a closed convex set. The tuning parameter $\tau $ plays a crucial
role in balancing the bias and variance: the bias is small when $\tau $ is
small, whereas the variance is small when $\tau $ is large.\footnote{%
A numerical illustration is provided in Appendix \ref{subsec:num_demo}.} If $%
\tau $ is sufficiently large, say $\tau \geq \max_{i\in \left[ N\right] }|%
\widehat{\boldsymbol{\Sigma }}_{i\cdot }\boldsymbol{1}_{N}|/N$, the second
constraint in (\ref{eq:relax}) is slack and thus irrelevant to the
minimization. As a result, the simple average weight $N^{-1}\boldsymbol{1}%
_{N}$ solves (\ref{eq:relax}). In addition, relaxing $\tau $ from 0 reduces
the sensitivity of the weights to the noise in the estimated VC matrix to
prevent in-sample over-fitting.

On the other hand, $\ell _{2}$-relaxation can be motivated from the
information theory as in \citet{diebold2021aggregation}. The choice of $\ell
_{2}$-norm can be viewed as a special case of R{\'{e}}nyi's cross-entropy %
\citep[Eq.(1.21) with $\alpha = 2$]{renyi1961measures}. This particular
choice is for convenience because: (i) the dual of the Euclidean norm with
respect to the inner product is the Euclidean norm itself and (ii) it
accommodates $w_{i}<0$, which should not be ruled out in applications of
forecast combination and portfolio analysis. By choosing a positive value of 
$\tau ,$ the constraints in (\ref{eq:relax}) yield a feasible set that is
larger than that associated with $\tau =0.$ Let $\overline{w}:=\frac{1}{N}%
\sum_{i=1}^{N}w_{i}.$ Then 
\begin{equation*}
\left\Vert \mathbf{w}\right\Vert _{2}^{2}
=\sum_{i=1}^{N}w_{i}^{2}=\sum_{i=1}^{N}\left( w_{i}-\bar{w}\right) ^{2}+N%
\bar{w}^{2} =\sum_{i=1}^{N}\left( w_{i}-\frac{1}{N}\right) ^{2}+\frac{1}{N}, 
\end{equation*}
where the last equality holds under the constraint $\mathbf{w}^{\prime }%
\boldsymbol{1}_{N}=1.$ Clearly, the $\ell _{2}$-relaxation aims to minimize
the sample variance of the weights over the feasible set and it effectively
eliminates unnecessary variations across the individual weights. As we shall
see, in the presence of a latent group structure in the dominant component
of the VC matrix, the $\ell _{2}$-relaxation shrinks the individual weights
to the group mean. This allows our estimator to include the widely used
simple average (SA) estimator as a special case.\footnote{%
There are other possibilities for new estimators that combine a particular
entropy and a feasible set defined by a geometric structure tailored for a
high-dimensional economic or financial problem of interest. We will need
further exploration to see whether they can include some popular estimators
as special cases.}

\section{Theoretical Analysis}

\label{sec:theory}

\subsection{Latent Group Structures}

\label{subsec:latent}

In this section, we impose latent group structures on $\widehat{\boldsymbol{%
\Sigma}}$ or its population expectation $E[\widehat{\boldsymbol{\Sigma}}]$
and then study the implications on the $\ell_{2}$-relaxed estimates of the
weights.

Statistical analysis of high dimensional problems typically postulates
certain structures on the data generating process for dimension reduction.
For example, variable selection methods such as Lasso %
\citep{tibshirani1996regression} and SCAD \citep{fan2001variable} are
motivated from regressions with sparsity, meaning most of the regression
coefficients are either exactly zero or approximately zero. Similarly, in
large VC estimation, various structures have been considered in the
literature. \citet{bickel2008regularized} impose many off-diagonal elements
to be zero, \citet{engle2012dynamic} assume a block equicorrelation
structure, and \citet{Ledoit_Wolf2004} use Bayesian methods for shrinking
the sample correlation matrix to an equicorrelated target, to name just a
few.

Imposing latent group structures is an alternative way to reduce dimensions,
which now has grown into a burgeoning literature. To analyze \eqref{eq:relax}
in depth in the high dimensional framework, we assume $\widehat{\boldsymbol{%
\Sigma }}=\{\widehat{\Sigma }_{ij}\}_{i,j\in \left[ N\right] }$ can be
approximated by a block equicorrelation matrix: 
\begin{equation}
\widehat{\boldsymbol{\Sigma }}=\widehat{\boldsymbol{\Sigma }}^{\ast }+%
\widehat{\boldsymbol{\Sigma }}^{e},  \label{eq:decomposition}
\end{equation}%
where $\widehat{\boldsymbol{\Sigma }}^{\ast }=\{\widehat{\Sigma }_{ij}^{\ast
}\}_{i,j\in \left[ N\right] }$ is a block equicorrelation matrix and $%
\widehat{\boldsymbol{\Sigma }}^{e}=\{\widehat{\Sigma }_{ij}^{e}\}_{i,j\in %
\left[ N\right] }$ denotes the deviation of $\widehat{\boldsymbol{\Sigma }}$
from the block equicorrelation matrix. 
% We will treat $\widehat{\boldsymbol{%
% \Sigma }}^{\ast }$ as the oracle object in Section \ref%
% {subsec:Oracle-Problem}, instead of its population counterpart.
% , so that we
% can study the finite-sample numerical properties in Section \ref%
% {subsec:Numerical-properties} without resorting to asymptotics. 
We write 
\begin{equation}
\underset{\left( N\times N\right) }{\widehat{\boldsymbol{\Sigma }}^{\ast }}=%
\underset{\left( N\times K\right) }{\mathbf{Z}}\underset{\left( K\times
K\right) }{\widehat{\mathbf{\Sigma }}^{\mathrm{co}}}\mathbf{Z}^{\prime }
\label{eq:core}
\end{equation}%
where $\mathbf{Z}=\{Z_{ik}\}$ denotes an $N\times K$ binary matrix providing
the cluster membership of each individual forecast, i.e., $Z_{ik}=1$ if
forecast $i$ belongs to group $\mathcal{G}_{k}\subset \left[ N\right] $ and $%
Z_{ik}=0$ otherwise, and $\widehat{\boldsymbol{\Sigma }}^{\mathrm{co}}=\{%
\widehat{\Sigma }_{kl}^{\mathrm{co}}\}_{k,l\in \left[ K\right] }$ is a $%
K\times K$ symmetric positive definite matrix. Here, the superscript
\textquotedblleft co\textquotedblright\ stands for \textquotedblleft
core\textquotedblright . Note that $\widehat{\Sigma }_{ij}^{\ast }=\widehat{%
\Sigma }_{kl}^{\mathrm{co}}$ if $i\in \mathcal{G}_{k}\text{ and }j\in 
\mathcal{G}_{l}.$

One can observe $\widehat{\boldsymbol{\Sigma }}$ from the data but not $%
\widehat{\boldsymbol{\Sigma }}^{\ast }$. We will be precise about the
definition of \textquotedblleft approximation\textquotedblright\ for $%
\widehat{\boldsymbol{\Sigma }}^{e}$ in Assumption \ref{assu:idiosyn} later.
Let $N_{k}:=\left\vert \mathcal{G}_{k}\right\vert $ be the number of
individuals in the $k$th group, and thus $N = \sum_{k=1}^{K}N_{k}$. For ease
of notation and after necessary re-ordering the $N$ forecast units, we write 
\begin{equation}
\widehat{\boldsymbol{\Sigma}}^{\ast}=(\widehat{\Sigma}_{kl}^{\mathrm{co}%
}\cdot\boldsymbol{1}_{N_{k}}\boldsymbol{1}_{N_{l}}^{\prime})_{k,l\in\lbrack
K],}  \label{eq:sigma_star}
\end{equation}
in which the units in the same group cluster together in a block. The
re-ordering is for the convenience of notation only. The theory to be
developed is irrelevant to the ordering of individuals, and does not require
the knowledge about the membership matrix $\mathbf{Z}$.

We now motivate the decomposition (\ref{eq:decomposition}) using five
examples.

\begin{example}
\label{exa:one_factor} \citet{Chan_Pauwels2018} assume the existence of a
\textquotedblleft best\textquotedblright\ unbiased forecast $f_{0t}$ of
variable $y_{t+1}$ with an associated forecast error $e_{0t}$, and the
forecast error $e_{it}$ of model $i$ can be decomposed as 
\begin{equation*}
e_{it}=e_{0t}+u_{it},
\end{equation*}%
where $e_{0t}$ represents the forecast error from the best forecasting
model, and $u_{it}$ is the deviation of $e_{it}$ from the best forecasting
model. Assuming $E\left[ u_{it}\right] =0 $ and $E\left[ e_{0t}u_{it}\right]
=0$ for each $i,$ the VC of $\mathbf{e}_{t}$ can be written as $\mathbf{%
\Sigma }_0 =E\left[ \mathbf{e}_{t}\mathbf{e}_{t}^{\prime }\right] =E\left[
e_{0t}^{2}\right] \mathbf{1}_{N} \mathbf{1}_{N}^{\prime }+E[\mathbf{u}_{t}%
\mathbf{u}_{t}^{\prime }],$ where $\mathbf{u}_{t}=(u_{1t},...,u_{Nt})^{%
\prime }$. At the sample level, we have $\widehat{\boldsymbol{\Sigma }}=%
\widehat{\boldsymbol{\Sigma }}^{\ast }+\widehat{\boldsymbol{\Sigma }}^{e},$
where 
\begin{equation*}
\widehat{\boldsymbol{\Sigma }}=\mathbb{E}_{T}\left[ \mathbf{e}_{t}\mathbf{e}%
_{t}^{\prime }\right] ,\text{ }\widehat{\boldsymbol{\Sigma }}^{\ast }=%
\mathbb{E}_{T}\left[ e_{0t}^{2}\right] \mathbf{1}_{N}\mathbf{1}_{N}^{\prime
},\text{ and }\widehat{\boldsymbol{\Sigma }}^{e}=\mathbb{E}_{T}[\mathbf{u}%
_{t}\mathbf{u}_{t}^{\prime }]+\mathbf{1}_{N}\mathbb{E}_{T}[e_{0t}\mathbf{u}%
_{t}^{\prime }]+\mathbb{E}_{T}[e_{0t}\mathbf{u}_{t}]\mathbf{1}_{N}^{\prime }.
\end{equation*}%
In this case, all the $N$ forecast units belong to the same group $\mathcal{G%
}_{1}$ as rank$(\widehat{\boldsymbol{\Sigma }}^{\ast })=1$.
\end{example}

\begin{example}
\label{exa:q_factor} Consider that each individual forecast $f_{it}$ is
generated from a factor model 
\begin{equation}
f_{it}=\boldsymbol{\lambda }_{g_{i}}^{\prime }\boldsymbol{\eta }_{t}+u_{it},
\label{eq:r_factor}
\end{equation}%
where $\boldsymbol{\lambda }_{g_{i}}$ is a $q\times 1$ vector of factor
loadings, $\boldsymbol{\eta }_{t}$ is a $q\times 1$ vector of latent
factors, and $u_{it}$ is an idiosyncratic shock. Here $g_{i}$ denotes
individual $i$'s membership, i.e., it takes value $k$ if individual $i$
belongs to group $\mathcal{G}_{k}$ for $k\in \lbrack K]$ and $i\in \lbrack
N] $. Similarly, assume $y_{t+1}=\boldsymbol{\lambda }_{y}^{\prime }%
\boldsymbol{\eta }_{t}+u_{y,t+1}$, with $E\left[ u_{it}|\boldsymbol{\eta }%
_{t}\right] =0$ and $E\left[ u_{y,t+1}|\boldsymbol{\eta }_{t}\right] =0$ and 
$E\left[ u_{it}u_{y,t+1}|\boldsymbol{\eta }_{t}\right] =0.$\footnote{%
Other than those $q$ factors in $\boldsymbol{\eta }_{t}$, the additional
latent factor $u_{y,t+1}$ in $y_{t+1}$ is unforeseeable at time $t$. In
other words, given the information set $\mathcal{I}_{t}$ that contains $%
\left( \{f_{it}\}_{i\in \left[ N\right] },\boldsymbol{\eta }_{t}\right) $
and $\mathbf{u}_{t}$ at time $t$, the error $u_{y,t+1}=y_{t+1}-\boldsymbol{%
\lambda }_{y}^{\prime }\boldsymbol{\eta }_{t}=y_{t+1}-E\left( y_{t+1}|%
\mathcal{I}_{t}\right) $ must be orthogonal to $\mathcal{I}_{t}.$ Then $E%
\left[ u_{it}u_{y,t+1}|\mathcal{I}_{t}\right] =0$ implies $E\left[
u_{it}u_{y,t+1}|\boldsymbol{\eta }_{t}\right] =0$ by the law of iterated
expectations.} For simplicity, we also assume conditional homoskedasticity
var$\left( \mathbf{u}_{t}|\boldsymbol{\eta }_{t}\right) =\mathbf{\mathbf{%
\Omega }}_{u}$ and the factor loadings are nonstochastic. Then individual $i$%
's forecast error is 
\begin{equation*}
e_{it}=y_{t+1}-f_{it}=[\left( \boldsymbol{\lambda }_{y}-\boldsymbol{\lambda }%
_{g_{i}}\right) ^{\prime }\boldsymbol{\eta }_{t}+u_{y,t+1}]-u_{it}=\lambda
_{g_{i}}^{\dag \prime }\boldsymbol{\eta }_{t}^{\dag }-u_{it},
\end{equation*}%
where $\boldsymbol{\eta }_{t}^{\dag }:=(\boldsymbol{\eta }_{t}^{\prime
},u_{y,t+1})^{\prime }$ and $\boldsymbol{\lambda }_{g_{i}}^{\dag }:=((%
\boldsymbol{\lambda }_{y}-\boldsymbol{\lambda }_{g_{i}})^{\prime
},1)^{\prime }$, or equivalently $\mathbf{e}_{t}=\mathbf{\Lambda }^{\dag }%
\boldsymbol{\eta }_{t}^{\dag }-\mathbf{u}_{t}$ in a vector form, where $%
\mathbf{\Lambda }^{\dag }\mathbf{:}=\left( \lambda _{g_{1}}^{\dag },\ldots
,\lambda _{g_{N}}^{\dag }\right) ^{\prime }$. The population VC of $\mathbf{e%
}_{t}$ is given by $\mathbf{\Sigma }_{0}=E\left[ \mathbf{e}_{t}\mathbf{e}%
_{t}^{\prime }\right] =\mathbf{\Lambda }^{\dag }E[\boldsymbol{\eta }%
_{t}^{\dag }\boldsymbol{\eta }_{t}^{\dag \prime }]\mathbf{\Lambda }^{\dag
\prime }+\mathbf{\Omega }_{u}.$ Decompose the sample VC as $\widehat{%
\boldsymbol{\Sigma }}=\widehat{\boldsymbol{\Sigma }}^{\ast }+\widehat{%
\boldsymbol{\Sigma }}^{e},$ where\footnote{%
The conclusion here also holds for the centered version of the VC matrix: $%
\widehat{\boldsymbol{\Sigma }}=\mathbb{E}_{T}\left[ \left( \mathbf{e}_{t}-%
\mathbf{\bar{e}}\right) (\mathbf{e}_{t}-\mathbf{\bar{e}})^{\prime }\right] $
with more complicated notation.} 
\begin{equation*}
\widehat{\boldsymbol{\Sigma }}=\mathbb{E}_{T}\left[ \mathbf{e}_{t}\mathbf{e}%
_{t}^{\prime }\right] ,\text{ }\widehat{\boldsymbol{\Sigma }}^{\ast }=%
\mathbf{\Lambda }^{\dag }\mathbb{E}_{T}[\boldsymbol{\eta }_{t}^{\dag }%
\boldsymbol{\eta }_{t}^{\dag \prime }]\mathbf{\Lambda }^{\dag \prime },\text{
and }\widehat{\boldsymbol{\Sigma }}^{e}=\mathbb{E}_{T}[\mathbf{u}_{t}\mathbf{%
u}_{t}^{\prime }]-\mathbf{\Lambda }^{\dag }\mathbb{E}_{T}[\boldsymbol{\eta }%
_{t}^{\dag }\mathbf{u}_{t}^{\prime }]-\mathbb{E}_{T}[\mathbf{u}_{t}%
\boldsymbol{\eta }_{t}^{\dag \prime }]\mathbf{\Lambda }^{\dag \prime }.
\end{equation*}%
By construction, the core matrix has element $\widehat{\Sigma }_{kl}^{%
\mathrm{co}}=\lambda _{k}^{\dag \prime }\mathbb{E}_{T}[\boldsymbol{\eta }%
_{t}^{\dag }\boldsymbol{\eta }_{t}^{\dag \prime }]\lambda _{l}^{\dag }$ for $%
k,l\in \left[ K\right] $, the equicorrelation matrix has element $\widehat{%
\Sigma }_{ij}^{\ast }=\widehat{\Sigma }_{kl}^{\mathrm{co}}$ if $i\in 
\mathcal{G}_{k}$ and $j\in \mathcal{G}_{l}$, and rank$(\widehat{\boldsymbol{%
\Sigma }}^{\ast })\leq \left( q+1\right) \wedge K$.
\end{example}

\begin{remark}
We emphasize that our theory below does not require the knowledge about the
group membership for individual forecasts. Alternatively, one can estimate
the multi-factor structure in (\ref{eq:r_factor}) by the principal component
analysis (PCA) and then apply either the $K$-means algorithm or the
sequential binary segmentation algorithm \citep{Wang_Su2021} to the
estimated factor loadings to identity the true group membership. Then one
can impose the recovered group structure before computing classical weights.
This method is computationally involved and is subject to the usual
classification error issue: the presence of classification error in finite
samples is carried upon and thus adversely affects the subsequent estimation
of the weights. In contrast, the advantage of $\ell _{2}$-relaxation is that
it is computationally simple as it directly works with the sample moments
and hence bypasses the factor structure and the group membership.\footnote{%
It is worth mentioning that \citet{Hsiao_Wan2014} assume that the forecast
errors exhibit a multi-factor structure, but they do not assume the presence
of $K$ latent groups in the $N$ factor loadings and write $\boldsymbol{%
\lambda }_{i}$ in place of $\boldsymbol{\lambda }_{g_{i}}$. In the absence
of the latent group structures among the factor loadings $\{\boldsymbol{%
\lambda }_{i}\}_{i\in \lbrack N]}$, the dominant component in $\widehat{%
\boldsymbol{\Sigma }}$ will have a low-rank structure but not a latent group
structure. Analyses of this case will be different from the current paper,
which we leave for future research.}
\end{remark}

Latent groups may be present not only in approximate factor models, as in
the above two motivating examples, but also in some forecast problems in
which multi-factor structures are implicit. Here follows such an example.

\begin{example}
\label{exa:regression} Suppose that the outcome variable $y_{t+1}$ is
generated via the process 
\begin{equation}
y_{t+1}=\mathbf{x}_{t}^{\prime }\boldsymbol{\theta }^{0}+u_{t+1}\,\,\,%
\mbox{
for }t=-T_{0},...,-1,0,1,...  \label{Reg1}
\end{equation}%
where $\mathbf{x}_{t}=(x_{j,t})_{j=1}^{p}$ is a $p\times 1$ vector of
potential predictive variables, $\boldsymbol{\theta }^{0}=(\theta
_{j}^{0})_{j=1}^{p}$ is a $p\times 1$ vector of regression coefficients, and 
$u_{t+1}$ is the error term such that $E\left[ u_{t+1}|\mathbf{x}_{t}\right]
=0$ and $E\left[ u_{t+1}^{2}|\mathbf{x}_{t}\right] =\sigma _{u}^{2}$. 
Due to costly data collection or ignorance, the forecaster $i$ utilizes only
a subset $\mathbf{x}_{S_{i},t}$ of $\mathbf{x}_{t}$, where $S_{i}\subset
\lbrack p]$, to exercise prediction with the OLS estimate. Let $\widehat{%
\boldsymbol{\theta }}_{S_{i},t}=(\sum_{l=-T_{0}+1}^{t}\mathbf{x}_{S_{i},l-1}%
\mathbf{x}_{S_{i},l-1}^{\prime })^{-1}\sum_{l=-T_{0}+1}^{t}\mathbf{x}%
_{S_{i},l-1}\mathbf{y}_{l},$ and $\widehat{\boldsymbol{\theta }}_{i,t}$ be
the sparse $p\times 1$ vector that embeds the corresponding $\widehat{%
\boldsymbol{\theta }}_{S_{i},t}$ so that $(\widehat{\boldsymbol{\theta }}%
_{i,t})_{S_{i}}=\widehat{\boldsymbol{\theta }}_{S_{i},t}$ and $(\widehat{%
\boldsymbol{\theta }}_{i,t})_{[p]\backslash S_{i}}=\mathbf{0}$. We consider
two forecasting schemes: the fixed window and the rolling window.

(i) In the case of a fixed estimation window, the $i$th forecast of $y_{t+1}$
is given by $f_{it}:=\mathbf{x}_{S_{i},t}^{\prime }\widehat{\boldsymbol{%
\theta }}_{S_{i},0}$ for $t\geq 1$. The associated forecast error\ is%
\begin{equation*}
e_{it}=y_{t+1}-f_{it}=y_{t+1}-\mathbf{x}_{t}^{\prime }\widehat{\boldsymbol{%
\theta }}_{i,0}=u_{t+1}+\mathbf{x}_{t}^{\prime }(\boldsymbol{\theta }^{0}-%
\boldsymbol{\theta }_{i}^{0})+\epsilon _{i,t},
\end{equation*}%
where $\boldsymbol{\theta }_{i}^{0}:=$ plim$_{T_{0}}\widehat{\boldsymbol{%
\theta }}_{i,0}$ and $\epsilon _{i,t}:=\mathbf{x}_{t}^{\prime }(\widehat{%
\boldsymbol{\theta }}_{i}-\boldsymbol{\theta }_{i}^{0}).$ This is a $\left(
p+1\right) $-factor model with factors $(\mathbf{x}_{t}^{\prime
},u_{t+1})^{\prime }$ and factor loadings $((\boldsymbol{\theta }^{0}-%
\boldsymbol{\theta }_{i}^{0})^{\prime },1)^{\prime }.$

(ii) In the case of a rolling window, the forecast error is 
\begin{equation*}
e_{it}=y_{t+1}-\mathbf{x}_{S_{i},t}^{\prime }\widehat{\boldsymbol{\theta }}%
_{S_{i},t}=u_{t+1}+\mathbf{x}_{t}^{\prime }(\boldsymbol{\theta }^{0}-%
\widehat{\boldsymbol{\theta }}_{i,t})=u_{t+1}+\mathbf{x}_{t}^{\prime }(%
\boldsymbol{\theta }^{0}-\boldsymbol{\theta }_{i}^{0})+\epsilon _{i,t},
\end{equation*}%
where $\widehat{\boldsymbol{\theta }}_{i,t}\overset{p}{\rightarrow }%
\boldsymbol{\theta }_{i}^{0}$ as $T_{0}\rightarrow \infty $ is assumed to
hold uniformly in $\left( i,t\right) $ under some regularity conditions that
include the covariance stationarity, and $\epsilon _{i,t}:=\mathbf{x}%
_{t}^{\prime }(\widehat{\boldsymbol{\theta }}_{i,t}-\boldsymbol{\theta }%
_{i}^{0})$. Therefore, we have an approximate $\left( p+1\right) $-factor
model with factors $(\mathbf{x}_{t}^{\prime },u_{t+1})^{\prime }$ and factor
loadings $((\boldsymbol{\theta }^{0}-\boldsymbol{\theta }_{i}^{0})^{\prime
},1)^{\prime }$. Similar analysis applies to the rolling window of fixed
length $L$, in which the forecaster $i$ estimates the coefficient by $%
\widehat{\boldsymbol{\theta }}_{S_{i},t}^{L}=(\sum_{l=t-L+1}^{t}\mathbf{x}%
_{S_{i},t-1}\mathbf{x}_{S_{i},t-1}^{\prime })^{-1}\sum_{l=t-L+1}^{t}\mathbf{x%
}_{S_{i},t-1}\mathbf{y}_{t}$.

In either case, $e_{it}$ exhibits a factor structure where the factor
loadings are given by $((\boldsymbol{\theta }^{0}-\boldsymbol{\theta }%
_{i}^{0})^{\prime },1)^{\prime }.$ When $\boldsymbol{\theta }_{i}^{0}$
exhibits a latent group structure (see the next example), say, $\boldsymbol{%
\theta }_{i}^{0}=\boldsymbol{\theta }_{g_{i}}^{0}$ with $g_{i}$ being as
defined in the last example, the forecast error reduces to that in Example %
\ref{exa:q_factor} with 
\begin{equation*}
e_{it}=\lambda _{g_{i}}^{\dag \prime }\boldsymbol{\eta }_{t}^{\dag }-u_{i,t},
\end{equation*}
where $\lambda _{g_{i}}^{\dag }:=((\boldsymbol{\theta }^{0}-\boldsymbol{%
\theta }_{g_{i}}^{0})^{\prime },1)^{\prime },$ $\boldsymbol{\eta }_{t}^{\dag
}=(\mathbf{x}_{t}^{\prime },u_{t+1})^{\prime },$ and $u_{i,t}=-\epsilon
_{i,t}.$ Then $\widehat{\boldsymbol{\Sigma }},$ $\widehat{\boldsymbol{\Sigma 
}}^{\ast },$ and $\widehat{\boldsymbol{\Sigma }}^{e}$ can be defined as in
Example \ref{exa:q_factor}.
\end{example}

The next example is a simple linear regression that yields a two-group
structure in the VC matrix, and it can be easily extended to the multiple
group structure by allowing for multiple regressors to have predictive power.

\begin{example}
\label{exa:regression2} We reuse the notation in Example \ref{exa:regression}
while focus on a special case where only one regressor inside $\mathbf{x}%
_{t},$ say, $x_{1,t},$ has predictive power and we employ the fixed window
scheme to forecast. Then $\theta _{1}^{0}\neq 0$ and $\theta _{j}^{0}=0$ for
all $j=2,...,p,$ where we allow $p$ to diverge to infinity slowly. We can
divide the $N$ forecasts into two groups according to whether $1\in S_{i}$,
i.e., whether the only predictive regressor $x_{1,t}$ is included in the $i$%
th forecasting model. Without loss of generality, we assume $E\left[ \mathbf{%
x}_{t}\right] =\mathbf{0}$ and $E\left[ \mathbf{x}_{t}\mathbf{x}_{t}^{\prime
}\right] =\mathbf{I}_{p}$. Furthermore, we assume that $x_{1,t}$ is included
in forecasting model $i$ as the first element in $\mathbf{x}_{S_{i},t}$ for\ 
$i\in \mathcal{G}_{1}=[N_{1}]$ while it is excluded for $i\in \mathcal{G}%
_{2}=\{N_{1}+1,....,N\}$. Intuitively, the first $N_{1}$ forecasting models
are correctly specified for the conditional mean of $y_{t+1}$ while the
other $N_{2}:=N-N_{1}$ models are misspecified. Note that 
\begin{equation*}
e_{it}=y_{t+1}-\mathbf{x}_{S_{i},t}^{\prime }\widehat{\boldsymbol{\theta }}%
_{S_{i},0}=\left[ u_{t+1}+(x_{1,t}\theta _{1}^{0}-\mathbf{x}%
_{S_{i},t}^{\prime }\boldsymbol{\theta }_{S_{i},0}^{0})\right] +\mathbf{x}%
_{S_{i},t}^{\prime }(\boldsymbol{\theta }_{S_{i},0}^{0}-\widehat{\boldsymbol{%
\theta }}_{S_{i},0})=v_{it}+s_{it},
\end{equation*}%
where $v_{it}:=u_{t+1}+(x_{1,t}\theta _{1}^{0}-\mathbf{x}_{S_{i},t}^{\prime }%
\boldsymbol{\theta }_{S_{i},0}^{0}),\ s_{it}:=\mathbf{x}_{S_{i},t}^{\prime }(%
\boldsymbol{\theta }_{S_{i},0}^{0}-\widehat{\boldsymbol{\theta }}_{S_{i},0})$
and $\boldsymbol{\theta }_{S_{i},0}^{0}$ is the probability limit of $%
\widehat{\boldsymbol{\theta }}_{S_{i},0}.$ Under some regularity conditions,
the effect of the parameter estimation error $s_{it}$ can be made as small
as possible for a sufficiently large $T_{0}$ as $\Vert \boldsymbol{\theta }%
_{S_{i},0}^{0}-\widehat{\boldsymbol{\theta }}_{S_{i},0}\Vert
_{2}=O_{p}((T_{0}/p_{i})^{-1/2})$, where $p_{i}$ is the number of regressors
in the $i$th model that can be divergent to infinity too. The orthonormal
regressors imply $\boldsymbol{\theta }_{S_{i},0}^{0}=\left( \theta _{1}^{0},%
\mathbf{0}_{p_{i}-1}^{\prime }\right) ^{\prime }$ for $i\in \mathcal{G}_{1}$
and $\boldsymbol{\theta }_{S_{i},0}^{0}=\mathbf{0}_{p_{i}}$ for $i\in 
\mathcal{G}_{2}$. Let $\mathbf{v}_{t}:=\left( v_{1t},...,v_{Nt}\right)
^{\prime }$, $\mathbf{\bar{v}}:=\mathbb{E}_{T}(\mathbf{v}_{t})$, and $%
\widehat{\mathbf{V}}:=\mathbb{E}_{T}\left[ \left( \mathbf{v}_{t}-\mathbf{%
\bar{v}}\right) (\mathbf{v}_{t}-\mathbf{\bar{v}})^{\prime }\right] $. Define 
$\mathbf{s}_{t}$ and $\mathbf{\bar{s}}$ analogously. Then $\widehat{%
\boldsymbol{\Sigma }}=\mathbb{E}_{T}\left[ \left( \mathbf{e}_{t}-\mathbf{%
\bar{e}}\right) (\mathbf{e}_{t}-\mathbf{\bar{e}})^{\prime }\right] =\widehat{%
\boldsymbol{\Sigma }}^{\ast }+\widehat{\boldsymbol{\Sigma }}^{e},$ where 
\begin{eqnarray*}
\widehat{\boldsymbol{\Sigma }}^{\ast }&:= &\text{plim}_{T\rightarrow \infty }%
\widehat{\mathbf{V}}=\left( 
\begin{array}{cc}
\sigma _{u}^{2}\boldsymbol{1}_{N_{1}}\boldsymbol{1}_{N_{1}}^{\prime } & 
\sigma _{u}^{2}\boldsymbol{1}_{N_{1}}\boldsymbol{1}_{N_{2}}^{\prime } \\ 
\sigma _{u}^{2}\boldsymbol{1}_{N_{2}}\boldsymbol{1}_{N_{1}}^{\prime } & 
\left[ \sigma _{u}^{2}+\left( \theta _{1}^{0}\right) ^{2}\right] \boldsymbol{%
1}_{N_{2}}\boldsymbol{1}_{N_{2}}^{\prime }%
\end{array}%
\right) , \\
\widehat{\boldsymbol{\Sigma }}^{e}&:= &(\widehat{\mathbf{V}}-\text{plim}%
_{T\rightarrow \infty }\widehat{\mathbf{V}})+\mathbb{E}_{T}\left[ \left( 
\mathbf{s}_{t}-\mathbf{\bar{s}}\right) (\mathbf{s}_{t}-\mathbf{\bar{s}}%
)^{\prime }+\left( \mathbf{v}_{t}-\mathbf{\bar{v}}\right) (\mathbf{s}_{t}-%
\mathbf{\bar{s}})^{\prime }+\left( \mathbf{s}_{t}-\mathbf{\bar{s}}\right) (%
\mathbf{v}_{t}-\mathbf{\bar{v}})^{\prime }\right] .
\end{eqnarray*}%
can be easily verified.

% In this example it is easy to verify 
% \begin{equation*}
% \widehat{\boldsymbol{\Sigma }}^{\ast }=\left( 
% \begin{array}{cc}
% \sigma _{u}^{2}\boldsymbol{1}_{N_{1}}\boldsymbol{1}_{N_{1}}^{\prime } & 
% \sigma _{u}^{2}\boldsymbol{1}_{N_{1}}\boldsymbol{1}_{N_{2}}^{\prime } \\ 
% \sigma _{u}^{2}\boldsymbol{1}_{N_{2}}\boldsymbol{1}_{N_{1}}^{\prime } & 
% \left[ \sigma _{u}^{2}+\left( \theta _{1}^{0}\right) ^{2}\right] \boldsymbol{%
% 1}_{N_{2}}\boldsymbol{1}_{N_{2}}^{\prime }%
% \end{array}%
% \right).
% \end{equation*}%
\end{example}

\begin{remark}
Example \ref{exa:regression2} offers a setting in which the strategy of %
\citet{Diebold_Shin2019} is optimal. Intuitively, in the presence of two
groups of forecasts (say $\mathcal{G}_{1}$ and $\mathcal{G}_{2}$) with the
same forecast variance among each group, if the covariance between the good
(those in $\mathcal{G}_{1},$ say) and bad (those in $\mathcal{G}_{2},$ say)
forecasts is the same as the variance of the good forecasts, an optimal
forecast combination should assign zero weight to the group of bad forecasts
and equal nonzero weight to the group of good forecasts. Lemma \ref{lem:w0}
below suggests that if $\widehat{\boldsymbol{\Sigma }}^{\ast }$ was observed
and used, the optimal strategy would assign $1/N_{1}$ weight to each of the
first $N_{1}$ forecasts and 0 weight to each of the last $N_{2}$ forecasts.
When $\widehat{\boldsymbol{\Sigma }}^{\ast }$ is replaced by the feasible
version $\widehat{\boldsymbol{\Sigma }},$ our theory below ensures that the $%
\ell _{2}$-relaxation assigns approximately $1/N_{1}$ weight to each of the
first $N_{1}$ forecasts and approximately 0 weight to each of the last $N_{2}
$ forecasts.
\end{remark}

Lastly, we give an example that illustrates the use of group structure in
portfolio analysis.

\begin{example}
\label{exa:portfolio} Volatility matrix is a fundamental component for
portfolio analysis. To reduce the complexity in estimating a vast VC matrix, %
\citet{engle2012dynamic} employ the Standard Industrial Classification (SIC)
to assign the equicorrelated blocks. Using MVP, \citet{clements2015benefits}
find evidence in favor of equicorrelation across portfolio sizes. Each of
these papers explicitly specifies a criterion, either SIC or portfolio size,
to allocate an individual's group identity. In contrast, no knowledge about
the membership is required to implement $\ell _{2}$-relaxation; block
equicorrelation is taken as a latent structure.
\end{example}

Next, we specify an asymptotic target for the $\ell _{2}$-relaxation
estimator $\widehat{\mathbf{w}}$. Consider the oracle problem of $\ell _{2}$%
-relaxation with an infeasible $\widehat{\boldsymbol{\Sigma }}^{\ast }$: 
\begin{equation}
\min_{\left( \mathbf{w},\gamma \right) \in \mathbb{R}^{N+1}}\frac{1}{2}%
\left\Vert \mathbf{w}\right\Vert _{2}^{2}\text{\ \ subject to \ }\mathbf{w}%
^{\prime }\boldsymbol{1}_{N}=1\text{ and }\widehat{\boldsymbol{\Sigma }}%
^{\ast }\mathbf{w}+\gamma \mathbf{1}_{N}=0.  \label{eq:oracle_primal_0}
\end{equation}%
Denote the solution to the above problem as $\mathbf{w}^{\ast }$. Lemma \ref%
{lem:w0} below shows that the squared $\ell _{2}$-norm objective function
produces the \textit{within-group equally weighted solution $\mathbf{w}%
^{\ast }$}. 
% , which in general does not accommodate an explicit form due to
% the presence of sup-norm in the inequality constraint. 
% However, in the
% special case of $\tau =0$, the inequality constraint can be equivalently
% written as $K$ equality constraints so that we can solve for $\mathbf{w}%
% _{0}^{\ast }:=\mathbf{w}_{\tau =0}^{\ast }$ in closed-form. 
The problem (\ref{eq:relax}) with $\left( N+1\right) $ free parameters is
effectively reduced to merely $\left( K+1\right) $ free parameters in the
oracle problem (\ref{eq:oracle_primal_0}).

\begin{lemma}
\label{lem:w0} The solution to (\ref{eq:oracle_primal_0}) takes within-group
equal values in the form 
\begin{equation*}
\mathbf{w}^{\ast }=\left( N^{-1}b_{01}^{\ast }\boldsymbol{1}_{N_{1}}^{\prime
},\cdots ,N^{-1}b_{0K}^{\ast }\boldsymbol{1}_{N_{K}}^{\prime }\right)
^{\prime },
\end{equation*}%
where the expression of $\left( b_{0k}^{\ast }\right) _{k\in \lbrack K]}$ is
given in Equation (\ref{eq:b0star}) in the Online Appendix.
\end{lemma}

The use of squared $\ell _{2}$-norm in (\ref{eq:oracle_primal_0}) yields the
same weight across units in each group for the oracle problem. When $%
\widehat{\boldsymbol{\Sigma }}^{\ast }$ is replaced by its feasible version $%
\widehat{\boldsymbol{\Sigma }},$ we will show that $\ell _{2}$-relaxation
guarantees that the weights are approximately equal within each group so
that $\widehat{\mathbf{w}}$ and $\mathbf{w}^{\ast }$ are sufficiently close.

\subsection{Asymptotic Theory}

We study the asymptotic properties of the $\ell _{2}$-relaxed estimator in
this section. We consider a triangular array of models indexed by $T$ and $N$%
, both passing to infinity. Let $\phi _{NT}:=\sqrt{ (\log N) / (T\wedge N)}%
\rightarrow 0$. Note that we allow both $N\gg T $ (as in standard high
dimensional problems) and $T\gg N$ or $T\asymp N$ in view of $\phi _{NT}$.
But we rule out the traditional case of \textquotedblleft fixed $N$ and
large $T$\textquotedblright , which has been covered by the classical
approach \eqref{eq:bates-granger} and \eqref{eq:bates-granger-explicit}.

In \eqref{eq:decomposition} $\widehat{\mathbf{\Sigma }}$ is decomposed into $%
\widehat{\mathbf{\Sigma }}^{\ast }$ and $\widehat{\mathbf{\Sigma }}^{e}$,
and in \eqref{eq:sigma_star} $\widehat{\mathbf{\Sigma }}^{\ast }$ is
characterized by $\widehat{\mathbf{\Sigma }}^{\mathrm{co}}$. Let $%
\boldsymbol{\Sigma }_{0}^{e}:=E[\widehat{\boldsymbol{\Sigma }}^{e}]$, $%
\boldsymbol{\Delta }^{e}:=\widehat{\boldsymbol{\Sigma }}^{e}-\boldsymbol{%
\Sigma }_{0}^{e}$, $\mathbf{\Sigma }_{0}^{\ast }=E[\widehat{\mathbf{\Sigma }}%
^{\ast }],$ $\boldsymbol{\Sigma }_{0}^{\mathrm{co}}:=E[\widehat{\boldsymbol{%
\Sigma }}^{\mathrm{co}}]$, and $\boldsymbol{\Delta }^{\mathrm{co}}:=\widehat{%
\boldsymbol{\Sigma }}^{\mathrm{co}}-\boldsymbol{\Sigma }_{0}^{\mathrm{co}}$.
We impose regularity conditions on the population matrices and the sampling
errors.

\renewcommand\theenumi{(\alph{enumi})} \renewcommand\labelenumi{\theenumi}

\begin{assumption}
\label{assu:idiosyn} There are positive finite constants $C_{e0}$, $%
\underline{c}$, and $\overline{c}$ such that:

\begin{enumerate}
\item $\phi_{\max}\left(\boldsymbol{\Sigma}_{0}^{e}\right)=O(\sqrt{N}%
\phi_{NT})$, $\Vert\boldsymbol{\Sigma}_{0}^{e}\Vert_{c2}\leq C_{e0} \cdot
\phi_{\max}\left(\boldsymbol{\Sigma}_{0}^{e}\right)$, and $\left\Vert 
\boldsymbol{\Delta}^{e}\right\Vert _{\infty}=O_{p}((T/\log N)^{-1/2})$;

\item $\underline{c}\leq\phi_{\min}\left(\boldsymbol{\Sigma}_{0}^{\mathrm{co}%
}\right)\leq\phi_{\max}\left(\boldsymbol{\Sigma}_{0}^{\mathrm{co}}\right)\leq%
\overline{c}$, and $\left\Vert \boldsymbol{\Delta}^{\mathrm{co}}\right\Vert
_{\infty}=O_{p}((T/\log N)^{-1/2})$.
\end{enumerate}

\noindent
\end{assumption}

The first condition in Assumption \ref{assu:idiosyn}(a) allows the maximum
eigenvalue of the $N\times N$ matrix $\boldsymbol{\Sigma }_{0}^{e}$ to
diverge to infinity, but at a limited rate $\sqrt{N}\phi _{NT}$. The second
condition in (a) is similar to but weaker than the absolute row-sum
condition that is frequently used to model weak cross-sectional dependence;
see, e.g., \citet{Fan_Liao_Mincheva2013}. The third condition in (a)
requires that the sampling error of $\Delta _{ij}^{e}$ be controlled by $%
(T/\log N)^{-1/2}$ uniformly over $i$ and $j$ so that each element of $%
\widehat{\boldsymbol{\Sigma }}^{e}$ should not deviate too much from its
population mean $\boldsymbol{\Sigma }_{0}^{e}$. This condition can be
established under some low-level assumptions; see, e.g., Chapter 6 in %
\citet{Wainwright2019high}. Assumption \ref{assu:idiosyn}(b) bounds all
eigenvalues of the population core away from 0 and infinity, and impose
similar stochastic order on $\boldsymbol{\boldsymbol{\Delta }}^{\mathrm{co}}$
as that on $\boldsymbol{\Delta }^{e}$ in Assumption \ref{assu:idiosyn}(a).
Because ${\widehat{\mathbf{\Sigma }}}^{\mathrm{co}}$ is a low-rank matrix,
the restriction on $\mathbf{\Delta }^{\mathrm{co}}$ is very mild and the
sample error of the feasible VC $\widehat{\mathbf{\Sigma }}-E[\widehat{%
\mathbf{\Sigma }}]=\mathbf{\Delta }^{e}+\mathbf{\Delta }^{\mathrm{co}}$ is
primarily determined by $\mathbf{\Delta }^{e}$.

\bigskip {}

\begin{example}
(Example \ref{exa:q_factor}, cont.) Following the notation of Example \ref%
{exa:q_factor}, we can decompose the population variance-covariance matrix $%
\mathbf{\Sigma }:=E\left[ \mathbf{e}_{t}\mathbf{e}_{t}^{\prime }\right] $ as 
$\mathbf{\Sigma =\Sigma }_{0}^{\ast }+\mathbf{\Sigma }_{0}^{e},$ where $%
\mathbf{\Sigma }_{0}^{\ast }=\mathbf{\Lambda }^{\dag }E[\boldsymbol{\eta }%
_{t}^{\dag }\boldsymbol{\eta }_{t}^{\dag \prime }]\mathbf{\Lambda }^{\dag
\prime },$ and $\mathbf{\Sigma }_{0}^{e}=\mathbf{\Omega }_{x}=\left\{ \Omega
_{x,ij}\right\} .$ The corresponding sampling error is 
\begin{equation*}
\Delta _{ij}^{e}=\left\{ \mathbb{E}_{T}\left[ \epsilon _{i,t}\epsilon _{j,t}%
\right] -\Omega _{x,ij}\right\} -\sum_{l\in \{i,j\}}\left\{ (\boldsymbol{%
\lambda }_{y}-\boldsymbol{\lambda }_{g_{l}})^{\prime }\mathbb{E}_{T}\left[ 
\boldsymbol{\eta }_{t}(u_{y,t+1}-u_{l,t})\right] +\mathbb{E}_{T}\left[
u_{y,t+1}u_{l,t}\right] \right\} .
\end{equation*}%
Then the first part of Assumption \ref{assu:idiosyn}(a) is satisfied as long
as $\phi _{\max }\left( \boldsymbol{\Omega }_{x}\right) _{\text{{}}}=O(\sqrt{%
N}\phi _{NT})$. For the sampling error matrix, if 
\begin{equation*}
\max_{i,j\in \lbrack N]}\left\{ \left\vert \mathbb{E}_{T}\left[ \boldsymbol{%
\eta }_{t}(u_{y,t+1}-u_{i,t})\right] \right\vert +\left\vert \mathbb{E}_{T}%
\left[ u_{y,t}u_{i,t}\right] \right\vert +\left\vert \mathbb{E}_{T}\left[
u_{i,t}u_{i,j}\right] -\Omega _{ij,x}\right\vert \right\} =O_{p}((T/\log
N)^{-1/2}),
\end{equation*}%
then $\left\Vert \boldsymbol{\boldsymbol{\Delta }}^{e}\right\Vert _{\infty
}=O_{p}((T/\log N)^{-1/2})$ is satisfied as well.
\end{example}

The extent of relaxation in (\ref{eq:relax}) is controlled by the tuning
parameter $\tau $, which is to be chosen by cross validations (CV) in
simulations and applications. We spell out admissible range of $\tau $ in
Assumption \ref{assu:rate}(a) below. Assumption \ref{assu:rate}(b) restricts 
$\underline{r} := \min_{ k\in [K] } N_k / N$ relative to $K$.

\begin{assumption}
\label{assu:rate}As $\left(N,T\right)\rightarrow\infty,$

\begin{enumerate}
\item $\sqrt{K}\phi_{NT}/\tau+K^{5/2}\tau\to0$;

\item $\underline{r}\asymp K^{-1}$.
\end{enumerate}
\end{assumption}

In order to meet the condition $\sqrt{K}\phi _{NT}/\tau \rightarrow 0$ in
Assumption \ref{assu:rate}(a), it suffices to specify 
\begin{equation*}
\tau =D_{\tau }\sqrt{K}\phi _{NT}
\end{equation*}%
for some slowly diverging sequence $D_{\tau }$ as $\left( N,T\right)
\rightarrow \infty $, for example, $\log \log \left( N\wedge T\right)$. If $%
K $ is finite, this specification implies that $\tau $ should shrink to zero
at a rate slightly slower than $\phi _{NT}$. We allow $K\rightarrow \infty $%
, provided $K^{5/2}\tau \rightarrow 0$ so that the sampling error in $%
\widehat{\boldsymbol{\Sigma }}^{e}$ would not offset the dominant grouping
effect of the $\ell _{2}$-relaxation in the presence of latent groups in $%
\widehat{\boldsymbol{\Sigma }}^{\mathrm{\ast }}$. The particular rate $%
K^{5/2}\tau $ will appear as the order of convergence in Theorem \ref%
{thm:oracle} below. Though the exact number of groups $K$ is usually unknown
in reality, if the researcher believes that $K$ is asymptotically dominated
by some explicit rate function $\bar{K}_{N,T}$ of $N$ and $T$ in that $%
\limsup K/\bar{K}_{N,T}<1$, say $\bar{K}_{N,T}=\left( N\wedge T\right)
^{1/7} $, then all the following theoretical results still hold if $K$ is
replaced by $\bar{K}_{N,T}$ and $\tau $ is replaced by $\tau _{\bar{K}%
}=C_{\tau }\bar{K}_{N,T}^{1/2}\phi _{NT}$ for some positive constant $%
C_{\tau }$, and Assumption \ref{assu:rate}(a) is replaced by $\bar{K}%
_{N,T}^{1/2}\phi _{NT}/\tau +\bar{K}_{N,T}^{5/2}\tau \rightarrow 0$
accordingly.

Assumption \ref{assu:rate}(b) requires that the smallest relative group size 
$\underline{r}$ be proportional to the reciprocal of $K$. If a group
included too few members, the weight of the group would be too small to
matter and thus the associated coefficients too difficult to estimate.
Assumption \ref{assu:rate} (b) is, indeed, a simplifying condition for
notational conciseness. If we drop it, $\underline{r}$ will appear in the
rates of convergence in all the following results, which complicates the
expressions but adds no new insight.

Recall that the oracle weight vector $\mathbf{w}^{*}$ shares equal weights
within each group. Theorem \ref{thm:w_converg} establishes meaningful
convergence for $\widehat{\mathbf{w}}$ to $\mathbf{w}^{*}$.

\begin{theorem}
\label{thm:w_converg} Under Assumptions \ref{assu:idiosyn} and \ref%
{assu:rate}, we have 
\begin{equation*}
\left\Vert \widehat{\mathbf{w}}-\mathbf{w}^{*}\right\Vert
_{2}=O_{p}\left(N^{-1/2}K^{2}\tau\right)=o_{p}\left(N^{-1/2}\right)\text{
and }\left\Vert \widehat{\mathbf{w}}-\mathbf{w}^{*}\right\Vert
_{1}=O_{p}\left(K^{2}\tau\right)=o_{p}\left(1\right).
\end{equation*}
\end{theorem}

\begin{remark}
When we work with weight vectors of growing dimension, we must be cautious
about the rate of convergence. As a trivial example, consider the simple
average weight $\widehat{\mathbf{w}}^{\mathrm{SA}}=\boldsymbol{1}_{N}/N$ and
an \emph{ad hoc} oracle weight of two groups $\mathbf{w}^{\ast }=(0.5\cdot 
\mathbf{1}_{0.5N}^{\prime },1.5\cdot \mathbf{1}_{0.5N}^{\prime })^{\prime
}/N $.\footnote{%
Without loss of generality, we assume $N$ is an even number here.} In this
case, the $\ell _{2}$-distance 
\begin{equation*}
\left\Vert \widehat{\mathbf{w}}^{\mathrm{SA}}-\mathbf{w}^{\ast }\right\Vert
_{2}=\Vert 0.5\cdot \boldsymbol{1}_{N}/N\Vert _{2}=0.5/\sqrt{N}\rightarrow 0
\end{equation*}%
while $\left\Vert \widehat{\mathbf{w}}^{\mathrm{SA}}-\mathbf{w}^{\ast
}\right\Vert _{1}=0.5$. It is thus only non-trivial if we manage to show $%
\left\Vert \widehat{\mathbf{w}}-\mathbf{w}^{\ast }\right\Vert
_{1}=o_{p}\left( 1\right) $ and $\left\Vert \widehat{\mathbf{w}}-\mathbf{w}%
_{0}^{\ast }\right\Vert _{2}=o_{p}\left( N^{-1/2}\right) $, which is
achieved by Theorem \ref{thm:w_converg}.
\end{remark}

The convergence further implies desirable oracle (in)equalities in Theorem %
\ref{thm:oracle} below. It shows that the empirical risk under $\widehat{%
\mathbf{w}}$ would be asymptotically as small as if we knew the oracle
object $\widehat{\boldsymbol{\Sigma }}^{\ast }$.

\begin{theorem}[Oracle (in)equalities]
\label{thm:oracle} Under the assumptions in Theorem \ref{thm:w_converg}, we
have

\begin{enumerate}
\item $\widehat{\mathbf{w}}^{\prime }\widehat{\boldsymbol{\Sigma }}\widehat{%
\mathbf{w}}=\mathbf{w}^{\ast \prime }\widehat{\boldsymbol{\Sigma }}^{\ast }%
\mathbf{w}^{\ast }+O_{p}\left( \tau K^{5/2}\right) .$
\end{enumerate}

\noindent Furthermore, let $\widehat{\boldsymbol{\Sigma }}^{\mathrm{new}}$
and $\widehat{\boldsymbol{\Sigma }}^{\ast \mathrm{new}}$ be the counterparts
of $\widehat{\boldsymbol{\Sigma }}$ and $\widehat{\boldsymbol{\Sigma }}%
^{\ast }$ from a new testing sample of $T^{\mathrm{new}}$ observations where 
$T^{\mathrm{new}}\asymp T$. The testing sample can be either dependent or
independent of the training dataset used to estimate $\widehat{\mathbf{w}}$
and $\mathbf{w}^{\ast }$. If the testing dataset is generated by the same
DGP as that of the training dataset, then

\begin{enumerate}
\setcounter{enumi}{1}

\item $\widehat{\mathbf{w}}^{\prime }\widehat{\boldsymbol{\Sigma }}^{\mathrm{%
new}}\widehat{\mathbf{w}}=\mathbf{w}^{\ast \prime }\widehat{\boldsymbol{%
\Sigma }}^{\ast \mathrm{new}}\mathbf{w}^{\ast }+O_{p}\left( \tau
K^{5/2}\right) $;

\item $\widehat{\mathbf{w}}^{\prime }\widehat{\boldsymbol{\Sigma }}\widehat{%
\mathbf{w}}\leq \widehat{\mathbf{w}}^{\prime }\widehat{\boldsymbol{\Sigma }}%
^{\mathrm{new}}\widehat{\mathbf{w}}\leq Q(\boldsymbol{\Sigma }%
_{0})+O_{p}(\tau K^{5/2}),$ where $\boldsymbol{\Sigma }_{0}=\boldsymbol{%
\Sigma }_{0}^{\ast }+\boldsymbol{\Sigma }_{0}^{e}$ and $Q(\boldsymbol{\Sigma 
}_{0}):=\min_{\mathbf{w}^{\prime }\boldsymbol{1}_{N}=1}\,\mathbf{w}^{\prime }%
\boldsymbol{\Sigma }_{0}\mathbf{w} $.
\end{enumerate}
\end{theorem}

Theorem \ref{thm:oracle}(a) is an in-sample oracle equality, and (b) is an
out-of-sample oracle equality. %The proof is a
% term-by-term analysis of the difference between $\widehat{\mathbf{w}}%
% ^{\prime }\widehat{\boldsymbol{\Sigma }}\widehat{\mathbf{w}}$ and $\mathbf{w}%
% _{0}^{\ast \prime }\widehat{\boldsymbol{\Sigma }}^{\ast }\mathbf{w}%
% _{0}^{\ast }$ caused by the idiosyncratic shock. 
Because the magnitude of the idiosyncratic shock is controlled by Assumption %
\ref{assu:idiosyn}, the convergence of the weight estimator in Theorem \ref%
{thm:w_converg} allows the sample risk $\widehat{\mathbf{w}}^{\prime }%
\widehat{\boldsymbol{\Sigma }}\widehat{\mathbf{w}}$ to approximate the
oracle risk $\mathbf{w}^{\ast \prime }\widehat{\boldsymbol{\Sigma }}^{\ast }%
\mathbf{w}^{\ast }$. The approximation is nontrivial by noting that $\mathbf{%
w}^{\ast \prime }\widehat{\boldsymbol{\Sigma }}^{\ast }\mathbf{w}^{\ast }$
and $\mathbf{w}^{\ast \prime }\widehat{\boldsymbol{\Sigma }}^{\ast \mathrm{%
new}}\mathbf{w}^{\ast }$ are bounded away from 0 given the low rank
structure of $\widehat{\boldsymbol{\Sigma }}^{\ast \mathrm{new}}$ and that $%
\tau K^{5/2}\rightarrow 0$ under Assumption \ref{assu:rate}(a). In other
words, the risk of our sample estimator would be as low as if we were
informed of the infeasible oracle group membership, up to an asymptotically
negligible term $O_{p}\left( \tau K^{5/2}\right)$.

While our $\ell _{2}$-relaxation regularizes the combination weights, there
is another line of literature of regularizing the high dimensional VC
estimation or its inverse (the precision matrix); see %
\citet{bickel2008regularized}, \citet{Fan_Liao_Mincheva2013}, and the
overview by \citet{fan2016overview}. Theorem \ref{thm:oracle}(c) implies
that the in-sample and out-of-sample risks coming out of $\ell _{2}$%
-relaxation are comparable with the resultant risk from estimating the high
dimensional VC matrix. Given that the population VC $\boldsymbol{\Sigma }%
_{0} $ is the target of high dimensional VC estimation, in forecast
combination $Q(\boldsymbol{\Sigma }_{0})$ is \citet{bates1969combination}'s
optimal risk, and in portfolio analysis $Q(\boldsymbol{\Sigma }_{0})$ is the
global minimum risk. VC $\boldsymbol{\Sigma }_{0}$ takes into account both
the low rank component $\boldsymbol{\Sigma }_{0}^{\ast }$ and the high rank
component $\boldsymbol{\Sigma }_{0}^{e}$. Even if $\boldsymbol{\Sigma }_{0}$
can be estimated so well that the estimation error is completely eliminated,
our out-of-sample risk $\widehat{\mathbf{w}}^{\prime }\widehat{\boldsymbol{%
\Sigma }}^{\mathrm{new}}\widehat{\mathbf{w}}$ is within an $O_{p}\left( \tau
K^{5/2}\right) $ tolerance level of $Q(\boldsymbol{\Sigma }_{0})$.\medskip {}

\begin{remark}
We establish original asymptotic results to support this new $\ell _{2}$%
-relaxation method, although they are relegated to the Online Appendix due
to their technical nature and the limitations of space. Here we give a
roadmap of the theoretical construction. The duality between the sup-norm
constraint and the $\ell _{1}$-norm leads to the dual problem (\ref%
{eq:sim_dual}), which is a linearly constrained quadratic optimization. This
dual resembles Lasso \citep{tibshirani1996regression} in view of its $\ell
_{1}$-penalty. Rather than directly working with the primal problem (\ref%
{eq:relax}), we first develop the asymptotic convergence in the dual.
Studies of high dimensional regressions have offered a few inequalities for
Lasso to handle sparse regressions. We sharpen these techniques in our
context to cope with groupwise sparsity in an innovative way. Once the
convergence of the high dimensional parameters in the dual problem is
established (See Theorem \ref{thm:alpha_converg}), the convergence of the
combination weights follows in Theorem \ref{thm:w_converg}, and then the
asymptotic optimality in Theorem \ref{thm:oracle} proceeds.
\end{remark}

\section{Monte Carlo Simulations}

\label{sec:simulations}

In this section, we illustrate the performance of the proposed $\ell _{2}$%
-relaxation method via Monte Carlo simulations. We consider two different
simulation settings corresponding to the forecasting combination and
portfolio optimization in Section \ref{sec:empirical_app}.

This paper's numerical works are implemented in MATLAB, with the VC
estimates described in Box 1. With modern convex optimization modeling
languages and open-source convex solvers, the quadratic optimization with
constraints such as (\ref{eq:relax}) can be handled with ease even when $N$
is in hundreds or thousands. Proprietary convex solvers can also be called
upon for further speed gain in numerical operations; see %
\citet{gaoimplementing}.

\begin{center}
{\footnotesize 
\begin{tcolorbox}[width = 15cm,colback=white,,colframe=blue!65!,title=\textbf{Box 1. $\ell_{2}$-relaxation}
]
\begin{enumerate}[label=\arabic*]
\item Compute $\widehat{\boldsymbol{\Sigma}}$ via one of the three options: 
\begin{itemize}
\item $\widehat{\boldsymbol{\Sigma}}_s$: the plain sample variance-covariance as in \eqref{eq:vanilla_vc};
\item $\widehat{\boldsymbol{\Sigma}}_1$: \citet{Ledoit_Wolf2004}'s shrinkage VC;
\item $\widehat{\boldsymbol{\Sigma}}_2$: \citet{Ledoit_Wolf2020}'s nonlinear shrinkage VC.\footnote{ MATLAB codes for $\widehat{\boldsymbol{\Sigma}}_1$ and $\widehat{\boldsymbol{\Sigma}}_2$ are available in \\
\url{https://www.econ.uzh.ch/en/people/faculty/wolf/publications.html\#Programming\_Code}}
\end{itemize}
\item Given an estimated VC matrix $\widehat{\mathbf{\Sigma}}$ and a predetermined $\tau$, we solve the convex minimization problem in \eqref{eq:relax} and obtain $\widehat{\mathbf{w}}$.\footnote{
Given strict convexity, the generic MATLAB function {\tt fmincon} works in our experiment. 
For speed gain, we call the convex solver {\tt MOSEK} via {\tt CVX} \citep{cvx};
check \url{http://cvxr.com/cvx/} for details. 
}  

\end{enumerate}
\end{tcolorbox}}
\end{center}

\subsection{Forecast Combination}

We assume the simulated data follow a group pattern with the same number of
members in each group, i.e., $N_{k}=N/K$ for each $k\in \left[ K\right] $.
Let $\mathbf{\Psi }^{\text{co}}$ be a $K\times K$ symmetric positive
definite matrix, and $\mathbf{\Psi }=\mathbf{\Psi }^{\text{co}}\otimes (%
\mathbf{1}_{N_{1}}\mathbf{1}_{N_{1}}^{\prime })$ be its $N\times N$ block
equicorrelation matrix. We consider three DGPs, in which we start with a
baseline model of independent factors, and then allow dynamic factors and
approximate factors.

\begin{description}
\item \textbf{DGP 1.} The baseline model generates $N$ forecasters from 
\begin{equation}
\mathbf{f}_{t}=\mathbf{\Psi }^{1/2}\boldsymbol{\eta }_{t}+\mathbf{u}_{t},
\label{sim_ft}
\end{equation}%
where $\mathbf{\Psi }^{1/2}=N_{1}^{-1/2}(\mathbf{\Psi }^{\text{co}%
})^{1/2}\otimes (\mathbf{1}_{N_{1}}\mathbf{1}_{N_{1}}^{\prime })$, $%
\boldsymbol{\eta }_{t}\sim N(\boldsymbol{0},\mathbf{I}_{N})$ is independent
of the idiosyncratic noise $\mathbf{u}_{t}\sim N(\boldsymbol{0},\mathbf{%
\Omega } _{u})$, and the latter is independent across $t$.

\item \textbf{DGP 2.} We extend the baseline model by allowing for temporal
serial dependence in $\left\{ \boldsymbol{\eta }_{t}\right\} $.
Specifically, for each $i$, we generate $\eta _{it}$ from an AR(1) model 
\begin{equation*}
\eta _{it}=\rho _{i}\eta _{i,t-1}+\epsilon _{it}^{\eta },
\end{equation*}
where $\rho _{i}\sim \mathrm{Uniform}(0,0.9)$ is a random autoregressive
coefficient, the noise $\epsilon _{it}^{\eta }\sim \mathrm{i.i.d.\;}%
N(0,1-\rho _{i}^{2})$, and the initial values $\eta_{i0}\sim \mathrm{i.i.d.\;%
}N(0,1)$.

\item \textbf{DGP 3.} The equal factor loadings within a group can be an
approximation of more general factor loading configurations. This DGP
experiments with another extension to the baseline model by defining $\tilde{%
\mathbf{\Psi }}^{1/2}:={\mathbf{\Psi }}^{1/2}+\mathrm{i.i.d.\;}%
N(0,N_{1}^{-1/2})$ as a perturbed factor loading matrix to replace ${\mathbf{%
\Psi }}^{1/2}$ in \eqref{sim_ft}.
\end{description}

The target variable is generated as $y_{t+1}={\mathbf{w}_{\psi }^{\ast }}%
^{\prime }\mathbf{\Psi }^{1/2}\boldsymbol{\eta }_{t}+u_{y,t+1},$ where $%
u_{y,t+1}\sim N(0,\sigma _{y}^{2})$ is independent of $\boldsymbol{\eta }%
_{t} $ and $\mathbf{u}_{t},$ and $\mathbf{w}_{\psi }^{\ast }=[(\mathbf{\Psi }%
^{\text{co}})^{-1}\mathbf{1}_{K}]\otimes \mathbf{1}_{N_{1}}/[N_{1}\mathbf{1}%
_{K}^{\prime }(\mathbf{\Psi }^{\text{co}})^{-1}\mathbf{1}_{K}].$ The
forecast error vector is 
\begin{equation*}
\mathbf{e}_{t}=y_{t+1}\mathbf{1}_{N}-\mathbf{f}_{t} =[(\mathbf{1}_{N}{%
\mathbf{w}_{\psi }^{\ast }}^{\prime }-\mathbf{I}_{N})\mathbf{\Psi }^{1/2}%
\boldsymbol{\eta }_{t}+u_{y,t+1}\mathbf{1}_{N}]-\mathbf{u}_{t},
\end{equation*}%
and its population VC can be written as 
\begin{equation*}
E\left[ \mathbf{e}_{t}\mathbf{e}_{t}^{\prime }\right] =\underbrace{\left( 
\mathbf{I}_{N}-\mathbf{1}_{N}\mathbf{w}_{\psi }^{\ast \prime }\right) 
\mathbf{\Psi }\left( \mathbf{I}_{N}-\mathbf{w}_{\psi }^{\ast }\mathbf{1}%
_{N}^{\prime }\right) +\sigma _{y}^{2}\mathbf{1}_{N}\mathbf{1}_{N}^{\prime }}%
_{\mathbf{\Sigma }_{0}}+\underbrace{\mathbf{\Omega }_{u}}_{\mathbf{\Omega }%
_{0}}.
\end{equation*}%
By construction, $\mathbf{w}_{\psi }^{\ast } = \arg \underset{\mathbf{w}%
^{\prime }\mathbf{1}_{N}=1}{\min }\,\mathbf{w}^{\prime }\mathbf{\Sigma }_{0}%
\mathbf{w}$.

We compare the following estimators of $\mathbf{w}$, all subject to the
restriction $\mathbf{w}^{\prime }\mathbf{1}_{N}=1$: (i) the oracle estimator
with known group membership; (ii) simple averaging (SA); (iii) the $\ell
_{2} $-relaxation estimator with $\tau = 0$ ($\ell_2$-relax$^0$); (iv)
Lasso; (v) Ridge; (vi) the principle component (PC) grouping estimator; and
(vii) $\ell _{2}$-relaxation with three different $\widehat{\boldsymbol{%
\Sigma }}$ estimators in Box 1.

\begin{remark}
We elaborate the rivalries. The oracle estimator takes advantage of the true
group membership in the DGP. Given information about the group membership,
we reduce the $N$ forecasters to $K$ forecasters $f_{(g_{k}),t}=N_{k}^{-1}%
\sum_{i\in \mathcal{G}_{k}}f_{it}$ for $k\in \lbrack K]$, and use the
low-dimensional (\ref{eq:bates-granger-explicit}) to find the optimal
weights. For Lasso and Ridge, we recenter the weights toward the SA weights
for a fair comparison: 
\begin{eqnarray*}
\widehat{\mathbf{w}}_{\text{Lasso}}\ &=&\arg \underset{\mathbf{w}^{\prime }%
\mathbf{1}_{N}=1}{\min }\,\frac{1}{2}\mathbf{w}^{\prime }\widehat{%
\boldsymbol{\Sigma }}\mathbf{w}+\tau \Vert \mathbf{w}-\mathbf{1}_{N}/N\Vert
_{1} \\
\widehat{\mathbf{w}}_{\text{Ridge}}\ &=&\arg \underset{\mathbf{w}^{\prime }%
\mathbf{1}_{N}=1}{\min }\,\frac{1}{2}\mathbf{w}^{\prime }\widehat{%
\boldsymbol{\Sigma }}\mathbf{w}+\tau \Vert \mathbf{w}-\mathbf{1}_{N}/N\Vert
_{2}^{2}
\end{eqnarray*}%
where $\tau $ is the tuning parameter. Furthermore, we estimate the group
membership in PC as follows. We compute the $T\times N$ in-sample
forecasters' error matrix $\widehat{\mathbf{E}}=(\widehat{\mathbf{e}}%
_{1},...,\widehat{\mathbf{e}}_{T})^{\prime }$, save the associated $N\times N
$ factor loading matrix $\widehat{\mathbf{\Gamma }}$ of the singular
decomposition $\widehat{\mathbf{E}}=\widehat{\mathbf{U}}\widehat{\mathbf{D}}%
\widehat{\mathbf{\Gamma }}^{\prime }$, where $\widehat{\mathbf{D}}$ is the
\textquotedblleft diagonal\textquotedblright\ matrix of the singular values
in descending order. We extract the the first $q$ columns of $\widehat{%
\mathbf{\Gamma }}$, and perform the standard $K$-means clustering algorithm
to partition the factor loading vectors into $K$ estimated groups $\widehat{%
\mathcal{G}}_{k}$, $k\in \lbrack K]$. We use the true $K$ and try $q=5,10,20$
to avoid tuning on these hyperparameters in this PC grouping procedure.
\end{remark}

We estimate the weights $\widehat{\mathbf{w}}$ with the training sample $%
\{(y_{t+1},\mathbf{f}_{t}),$ $t\in \lbrack T])$, and then cast the
one-step-ahead prediction $\widehat{\mathbf{w}}^{\prime }\mathbf{f}_{T+1}$
for $y_{T+2}$. The above exercise is repeated to evaluate the MSFE $E\left[
(y_{T+2}-\widehat{\mathbf{w}}^{\prime }\mathbf{f}_{T+1})^{2}\right] -\sigma
_{y}^{2}$ of each estimator, where the unpredictable components $\sigma
_{y}^{2}$ in the MSFE is subtracted and the mathematical expectations are
approximated by empirical averages of 1000 simulation replications.\footnote{%
We also report the mean absolute forecast error (MAFE), which is not covered
by our theory; see Online Appendix \ref{mafe} for reference.}

We experiment with three training sample sizes $T=50,100,200$ with the
corresponding $K=2,4,6$ and $N=100,200,300$, respectively. We specify 
\begin{equation*}
\mathbf{\Psi }^{\text{co}}\ =\left[ 
\begin{array}{ccccc}
1 & 0.1 & 0 & \cdots & 0 \\ 
0.1 & \frac{3}{2} & 0.1 & \cdots & 0 \\ 
0 & 0.1 & 2 & \cdots & 0 \\ 
\vdots & \vdots & \vdots &  & \vdots \\ 
0 & 0 & 0 & \cdots & \frac{K+1}{2}%
\end{array}%
\right] ,\ 
\end{equation*}%
and $\mathbf{\Omega }_{u}=\sigma _{u}^{2}\mathbf{I}_{N}$ with $\sigma _{u}=5$%
. To highlight the effect of the signal-to-noise ratio (SNR) on the forecast
accuracy, we specify $\sigma _{y}=1$ as the low-signal design (with SNR
around $3:7$) and $\sigma _{y}=0.1$ as the high-signal design (with SNR
around $7:3$). Online Appendix \ref{signal} details the formula of the SNR
for our setting.

To implement $\ell _{2}$-relaxation, one needs to choose the tuning
parameter $\tau .$ Even though it is beyond the scope of the current paper
to provide a formal theoretical analysis on the choice of $\tau $, according
to our experience gained from extensive experiments, the commonly used
cross-validation (CV) method or its time-series-adjusted version works
fairly well in simulations and applications. Here, the tuning parameters for
DGPs 1 and 3 are obtained by the conventional 5-fold CV through a grid
search, detailed in Box 2; we have also tried the 3-fold CV and the 10-fold
CV, and the results are qualitatively intact. This 5-fold CV that randomly
permutes the data accounts for neither the chronological order nor the
serial correlation of the time series data, however. Practitioners usually
resort to the out-of-sample (OOS) evaluation instead.\footnote{%
See \cite{bergmeir2012} and \cite{mirakyan2017}, among others. See also \cite%
{arlot2010} for a survey of cross-validation procedures for model selection.}
Algorithm for the OOS evaluation, applied to DGP 2, is provided in Box 3.

\begin{center}
{\footnotesize 
\begin{tcolorbox}[width = 15cm,colback=white,,colframe=blue!65!,title=\textbf{Box 2. 
$5$-fold Cross Validation by MSFE}]
\begin{enumerate}[label=\arabic*]
\item The $T$ observations are {\it randomly} divided  into $5$ equal-sized  (up to rounding to integers) folds. 
\item For each fold:
\begin{itemize}
\item Take this fold as the test data and the other 4 folds as the training data.
\item For each value of the tuning parameter from 0.1 to 1 with increment 0.1, fit the method on the training data and evaluate it on the test data.
\end{itemize}
\item Summarize the forecast errors for all folds and compute the MSFE for each value of the tuning parameter. 

\item Choose the value of the  tuning parameter that yields the smallest MSFE. 
\end{enumerate}
\end{tcolorbox}}

{\footnotesize 
\begin{tcolorbox}[width = 15cm,colback=white,,colframe=blue!65!,title=\textbf{Box 3.
Out-of-sample Evaluation by MSFE}]
\begin{enumerate}[label=\arabic*]
\item The full dataset,   order {\it chronologically}, is divided into 5 blocks of equal-sized (up to rounding to integers) folds, indexed as $\tt b = 1,\ldots, 5$.
\item When each block $\tt b \in \{2,\ldots,5\}$ serves as the test dataset, respectively,
\begin{itemize}
\item Take all earlier folds as the training data.
\item For each value of the tuning parameter within the grid search range from 0.1 to 1 with increment 0.1, fit the method on the training data and evaluate it on the test data.
\end{itemize}
\item Summarize the forecast errors for blocks $\tt b \in \{ 2,\ldots, 5\}$ and compute the MSFE for each value of the tuning parameter. 

\item Choose the value of the  tuning parameter that yields the smallest MSFE. 
\end{enumerate}
\end{tcolorbox}}
\end{center}

\begin{figure}[ht!]
\centering
\includegraphics[scale = 0.85]{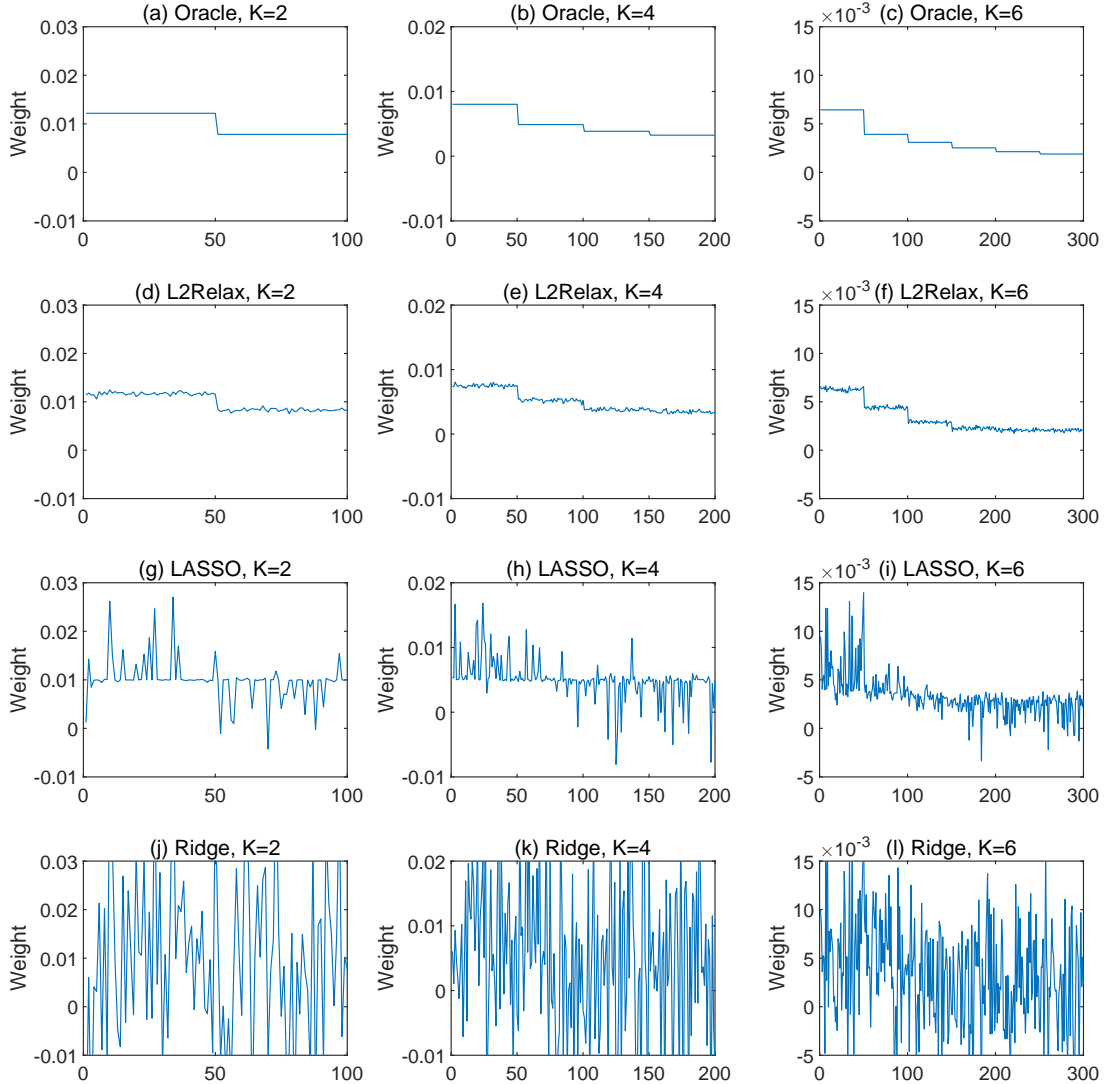}
\caption{Illustration of the Estimated Weights in DGP 1}
\label{sim1}
\end{figure}

Figure \ref{sim1} illustrates the estimated weights of a typical replication
under DGP 1. The four rows of sub-figures correspond to the oracle, the $%
\ell _{2}$-relaxation with $\widehat{\boldsymbol{\Sigma }}_{2}$, Lasso, and
ridge, respectively; the three columns represent the results under $K=2,4,6$%
, respectively. We unify the scale of axes for the four subplots in each
column to facilitate comparison. For each sub-figure, the estimated weights
are plotted against $\left[ N\right] $. Although individuals are not
explicitly classified into groups, $\ell _{2}$-relaxation estimates exhibits
grouping patterns that mimic the oracle weights. Such patterns are observed
in neither Lasso nor Ridge.

The six panels in Table \ref{table1} report the out-of-sample prediction
accuracy by MSFE for all three DGPs with low and high SNRs, respectively.%
\footnote{%
Besides $\ell_2$-relaxation, Ridge and Lasso
also require tuning parameters, which are selected in the same fashion. For
DGPs 1 and 3, we use the 5-fold cross-validation (Box 2); and for DGP 2, we
use the out-of-sample evaluation approach (Box 3). For the best empirical
performance, we use the nonlinear shrinkage VC estimator $\widehat{%
\boldsymbol{\Sigma }}_{2}$ for $\widehat{\boldsymbol{\Sigma }}$. In
addition, for the Ridge estimation, it is easy to verify that centering the
weights around $1/N$ or any other constant yields the same estimator $%
\widehat{\mathbf{w}}_{\text{Ridge}}$ when the constraint $\mathbf{w}^{\prime
}\mathbf{1}_{N}=1$ is imposed.} The first three columns show the settings of 
$T$, $N$ and $K$, and the following columns show the MSFEs of the labeled
estimators. All estimators have stronger performance under a high SNR than
that under a low SNR. The rankings of relative performance among the six
estimators are similar across different DGPs despite that the additional
factor loading noises enlarge the MSFEs of all estimators from DGP 3
relative to those from DGP 1.

\begin{table}[h!]
\caption{Results of Prediction Accuracy by MSFE}
\label{table1}\centering 
\scriptsize 

\begin{tabular}{cccccccccccccccccc} 
                        \hline 
& $T$ & $N$ & $K$ & Oracle & SA & $\ell_2$-relax$^0$  & Lasso & Ridge  & \multicolumn{3}{c}{PC}  & \multicolumn{3}{c}{$\ell_2$-relax} \\
% \cline{9-11}  \cline{13-15}
&&&&&&&&& $q = 5$ & $q=10$ & $q=20$ & $\widehat{\boldsymbol{\Sigma}}_s$ & $\widehat{\boldsymbol{\Sigma}}_1$ & $\widehat{\boldsymbol{\Sigma}}_2$   \\
\hline 
\multicolumn{10}{l}{\emph{Panel A: DGP 1 with Low SNR}}\\
&   50 &  100 &    2 &    0.312 &    0.891 &  1.536 &  0.402 &    1.253 &    0.679 &    0.662 &    0.667 &    0.393 &    0.366 &    0.342   \\ 
&  100 &  200 &    4 &    0.175 &    3.707 &  0.922 &   0.289 &    1.054 &    1.209 &    1.454 &    1.438 &    0.268 &    0.274 &    0.256   \\ 
&  200 &  300 &    6&    0.077 &    4.469 &   0.295 &  0.139 &    0.386 &    1.233 &    1.358 &    1.415 &    0.133 &    0.122 &    0.102  \\ 

\\
\multicolumn{10}{l}{\emph{Panel B: DGP 1 with  High SNR}}\\
& 50 & 100 & 2 & 0.259 & 0.938  & 0.574 & 0.311 & 0.502 & 0.702 & 0.730 & 0.710 &  0.274 & 0.271 & 0.267 \\ 
& 100 & 200 & 4 & 0.132 & 2.915 & 0.254	& 0.166 & 0.258 & 1.017 & 1.206 & 1.266 &  0.147 & 0.146 & 0.141 \\ 
& 200 & 300 & 6 & 0.101 & 3.975 & 0.213	& 0.123 & 0.136 & 1.274 & 1.285 & 1.284 &  0.120 & 0.122 & 0.114 \\

\\
\multicolumn{10}{l}{\emph{Panel C: DGP 2 with  Low SNR}}\\
&   50 &  100 &    2 &    0.292 &    1.032 & 1.251 &    0.401 &    1.235 &    0.787 &    0.800 &    0.763 &    0.383 &    0.390 &    0.366 \\ 
&  100 &  200 &    4 &    0.133 &    3.052 & 0.763 &    0.236 &    1.010 &    1.134 &    1.275 &    1.531 &    0.219 &    0.258 &    0.274 \\ 
&  200 &  300 &    6 &    0.066 &    3.699 & 0.411 &   0.131 &    0.412 &    1.109 &    1.050 &    1.173 &    0.173 &    0.144 &    0.124 \\ 

\\
 \multicolumn{10}{l}{\emph{Panel D: DGP 2 with  High SNR}}\\

& 50 & 100 & 2 & 0.262 & 0.993 & 0.401 & 0.323 & 0.488 & 0.702 & 0.747 & 0.751 &  0.280 & 0.279 & 0.271 \\ 
& 100 & 200 & 4 & 0.146 & 3.210 & 0.255 & 0.185 & 0.274 & 1.030 & 1.257 & 1.428 & 0.160 & 0.167 & 0.154 \\ 
& 200 & 300 & 6 & 0.106 & 4.524 & 0.135 & 0.126 & 0.139 & 1.343 & 1.601 & 1.639 & 0.126 & 0.122 & 0.114 \\ 

\\
\multicolumn{10}{l}{\emph{Panel E: DGP 3 with  Low SNR}}\\
&   50 &  100 &    2 &    0.430 &    1.017 &  1.581 &   1.055 &    1.518 &    0.837 &    0.952 &    0.901 &    0.541 &    0.486 &    0.443 \\ 
&  100 &  200 &    4 &    0.287 &    3.718 &  1.116 &  0.710 &    1.394 &    1.821 &    1.775 &    2.002 &    0.384 &    0.415 &    0.357 \\ 
&  200 &  300 &    6 &    0.234 &    4.702 &  0.935 &   0.484 &    0.869 &    1.856 &    2.175 &    2.400 &    0.256 &    0.281 &    0.262  \\ 
\\
\multicolumn{10}{l}{\emph{Panel F: DGP 3 with  High SNR}}\\
& 50 & 100 & 2 & 0.378 & 1.148  & 0.761 & 0.757 & 0.765 & 0.941 & 0.969 & 0.993 &  0.412 & 0.410 & 0.392 \\ 
& 100 & 200 & 4 & 0.279 & 3.191 & 0.482 & 0.393 & 0.482 & 1.518 & 1.665 & 1.812 &  0.301 & 0.285 & 0.285 \\ 
& 200 & 300 & 6 & 0.263 & 4.536 & 0.458 & 0.317 & 0.347 & 2.055 & 2.096 & 2.180 &  0.297 & 0.313 & 0.278 \\ 

\hline                  
        \end{tabular}
\end{table}

The infeasible grouping information helps the oracle estimator to prevail in
all cases. Regardless which $\widehat{\boldsymbol{\Sigma }}$ estimator is
employed, $\ell _{2}$-relaxation outperforms feasible competitors and its
MSFE approaches that of the oracle estimator. The $\ell _{2}$-relaxation
with $\widehat{\boldsymbol{\Sigma }}_{2}$ generally achieves the best
performance among all feasible estimators. Lasso and Ridge are in general
better than the PC estimator. Given the group structures in the DGP, SA in
general lags far behind the other feasible estimators that learn the
combination weights from the data. Notice that the MSFEs by Oracle, $\ell_2$%
-relax$^0$, Lasso, Ridge, and the $\ell _{2}$-relaxation decrease as $T$
grows along with $N$. However, the results by SA and PC under all values of $%
q$ may diverge as $(T,N)$ increases.

Our results are not sensitive to different evaluation methods. \cite%
{bergmeir2018} argue that the standard 5-fold CV is valid in purely
autoregressive models with uncorrelated errors. Simulation results for DGP 2
by the conventional 5-fold CV are reported in Appendix \ref{con5fold}. In
summary, we observe robust performance of $\ell _{2}$-relaxation superior to
the other feasible estimators across the DGP designs, signal strength, and
CV methods.

\subsection{Portfolio Analysis}

\label{subsec:sim_port}

We extend \citet{fan2012vast}'s design to a simulated Fama-French
five-factor model. Besides the market factor, \citet{fama2015} identify four
additional factors capturing the size, value, profitability, and investment
patterns in average stock returns. Let $R_{i}$ be the excessive return of
the $i$th stock. The five-factor model is similar to \eqref{eq:r_factor}: 
\begin{equation}
r_{it}=\boldsymbol{\lambda }_{i}^{\prime }\boldsymbol{\eta }_{t}+u_{it},
\label{return}
\end{equation}%
where $\boldsymbol{\lambda }_{i}=\{\lambda _{ij}\}_{j=1}^{5}$ is the vector
of 5 factor loadings.

\begin{table}[h]
\caption{Parameters for the Portfolio Simulation}
\label{cali}\centering 
{\footnotesize 
\begin{tabular}{crrrrrrrrrrrr}
\hline
\multicolumn{6}{c}{Parameters for Factor Returns} &  & \multicolumn{6}{c}{
Parameters for Factor Loadings (Size-BM)} \\ \cline{1-6}\cline{8-13}
$\mathbf{\mu}_f$ & \multicolumn{5}{c}{$\mathbf{cov}_f$} &  & $\mathbf{\mu}%
_{\lambda}$ & \multicolumn{5}{c}{$\mathbf{cov}_{\lambda}$} \\ \hline
0.644 & 20.388 & 4.175 & 1.324 & -4.530 & -1.351 &  & 1.009 & 0.013 & 0.000
& 0.006 & 0.002 & -0.005 \\ 
0.280 & 4.175 & 7.129 & 2.111 & -1.646 & 0.378 &  & 0.617 & 0.001 & 0.165 & 
-0.029 & -0.028 & -0.006 \\ 
-0.091 & 1.324 & 2.111 & 8.346 & 0.990 & 2.869 &  & 0.175 & 0.007 & -0.030 & 
0.143 & 0.028 & 0.002 \\ 
0.306 & -4.530 & -1.646 & 0.990 & 4.855 & 0.751 &  & -0.040 & 0.002 & -0.028
& 0.028 & 0.061 & 0.015 \\ 
0.107 & -1.351 & 0.378 & 2.869 & 0.752 & 3.431 &  & 0.005 & -0.005 & -0.006
& 0.002 & 0.015 & 0.058 \\ 
&  &  &  &  &  &  &  &  &  &  &  &  \\ 
\multicolumn{6}{c}{Parameters for Factor Loadings (Size-INV)} &  & 
\multicolumn{6}{c}{Parameters for Factor Loadings (Size-OP)} \\ 
\cline{1-6}\cline{8-13}
$\mathbf{\mu}_{\lambda}$ & \multicolumn{5}{c}{$\mathbf{cov}_{\lambda}$} &  & 
$\mathbf{\mu}_{\lambda}$ & \multicolumn{5}{c}{$\mathbf{cov}_{\lambda}$} \\ 
\hline
1.053 & 0.015 & 0.002 & -0.008 & -0.011 & -0.000 &  & 1.060 & 0.017 & 0.002
& -0.008 & -0.004 & -0.009 \\ 
0.621 & 0.002 & 0.156 & 0.012 & -0.033 & -0.026 &  & 0.630 & 0.002 & 0.175 & 
0.011 & 0.041 & 0.000 \\ 
0.128 & -0.007 & 0.012 & 0.026 & 0.031 & 0.010 &  & 0.169 & -0.008 & 0.011 & 
0.055 & 0.054 & -0.008 \\ 
-0.079 & -0.010 & -0.032 & 0.031 & 0.086 & 0.020 &  & -0.033 & -0.004 & 0.041
& 0.054 & 0.233 & 0.021 \\ 
-0.055 & -0.000 & -0.025 & 0.010 & 0.020 & 0.159 &  & -0.139 & -0.009 & 0.000
& -0.009 & 0.021 & 0.041 \\ \hline
\end{tabular}
}
\end{table}

We simulate the returns for $N=100$ assets and $T=240$ months. The factors
and factor loadings are generated from the multivariate normal distributions 
$N(\mathbf{\mu }_{f},\mathbf{cov}_{f})$ and $N(\mathbf{\mu }_{\lambda },%
\mathbf{cov}_{\lambda })$, respectively. The values of the parameters $(%
\mathbf{\mu }_{b},\mathbf{cov}_{b},\mathbf{\mu }_{f},\mathbf{cov}_{f})$ are
displayed in Table \ref{cali}, which are calibrated to the 2001--2020 real
market data of Fama and French 100 portfolios on the \emph{size and
book-to-market} (Size-BM), \emph{size and investment} (Size-INV), and \emph{%
size and operating profitability} (Size-OP).\footnote{%
The factor and portfolio data are available at %
\url{https://mba.tuck.dartmouth.edu/pages/faculty/ken.french/data%
\_library.html}.} The idiosyncratic noises are generated from $N(\mathbf{0},%
\mathbf{cov}_{u})$, where $\mathbf{cov}_{u}$ is the sample VC matrix of the
residuals from the OLS estimation of \eqref{return}.

Instead of MFSE, we use the Sharpe ratio as the criterion for MVP. The $\ell
_{2} $-relaxation estimator allows negative weights, which correspond to
short positions of financial assets. For each repetition, the following
rolling window estimation is considered with window length $L$. We avoid
recursively training models for each month. Following a similar strategy in %
\citet{xiu2020}, we train and roll forward once every year, as elaborated in
Box 4.

\begin{center}
{\footnotesize 
\begin{tcolorbox}[width = 15cm,colback=white,,colframe=blue!65!,title=\textbf{
Box 4. Algorithm: Portfolio Optimization}]

\begin{enumerate}[label=\arabic*]
\item Given $T$ monthly observations in total, fix $L$ as the length of the rolling windows.
 Start with the first $L$ observations as training data. 
\item Estimate the weight $\widehat{\mathbf{w}}$ using $L$ training observations. Apply $\widehat{\mathbf{w}}$ to forecast the next 12 months, the validation data.
Among a grid system from 0.1 to 1 with increment 0.1,
choose $\tau$ that yields the highest Sharpe ratio in the validation data.
\item Roll both the training data and the validation data one year forward. 

\item Repeat Steps 2--3 until the end of the sample.

\end{enumerate}
\end{tcolorbox}}
\end{center}

\begin{table}[ht!] 
\centering 
\caption{Sharpe Ratios of Estimated Portfolios} \label{tps}
% \footnotesize 
\begin{tabular}{cccccccccccc} 
\hline 
$L$ & $N$ & SA && \multicolumn{2}{c}{GEC} && \multicolumn{3}{c}{$\ell_2$-relax}\\ 
 
&&&& $c = 1$ & $c = 2$ &&  $\widehat{\boldsymbol{\Sigma}}_s$ & $\widehat{\boldsymbol{\Sigma}}_1$ & $\widehat{\boldsymbol{\Sigma}}_2$ \\
\hline 
\multicolumn{5}{l}{\emph{Panel A: Size-BM}}\\
60 &	100 &    0.131 &&    0.171 &    0.234 &&    0.378 &    0.382 &    0.363  \\ 
120&	100 &    0.128 &&    0.175 &    0.255 &&    0.463 &    0.478 &    0.470  \\ 
\\
\multicolumn{5}{l}{\emph{Panel B: Size-INV}}\\
60 &	100 &    0.154 &&    0.176 &    0.206 &&    0.283 &    0.284 &    0.271  \\ 
120&	100 &    0.158 &&    0.180 &    0.216 &&    0.339 &    0.353 &    0.344  \\ 
\\
\multicolumn{5}{l}{\emph{Panel C: Size-OP}}\\
60 &	100 &    0.170 &&    0.213 &    0.270 &&    0.404 &    0.409 &    0.389  \\ 
120&	100 &    0.168 &&    0.214 &    0.283 &&    0.515 &    0.528 &    0.516  \\ 
\hline 
\end{tabular} 
\end{table} 

We consider training data of lengths $L=60$ (5 years) and 120 (10 years).
The $\ell _{2}$-relaxation estimator is compared with SA and the \textit{%
gross exposure constraints} (GEC) methods \citep{fan2012vast}: 
\begin{equation*}
\min_{\mathbf{w}\in \mathbb{R}^{N}}\,\frac{1}{2}\mathbf{w}^{\prime }\widehat{%
\boldsymbol{\Sigma }}\mathbf{w}\ \ \text{subject to \ }\mathbf{w}^{\prime }%
\boldsymbol{1}_{N}=1\mbox{ and }\Vert \mathbf{w}\Vert _{1}\leq c,
\end{equation*}%
where the exposure constraint is set as $c=1$ (no short exposure) or $c=2$
(allowing 50\% short exposure).\footnote{%
Similar to the numerical implementation of Lasso and Ridge, throughout this
paper GEC is estimated with the VC matrix $\widehat{\boldsymbol{\Sigma }}%
_{2} $ for a fair comparison with the best $\ell _{2}$-relaxation outcomes
in most cases.} Table \ref{tps} reports the Sharpe ratios averaged over 1000
replications. The three panels correspond to portfolios sorted by Size-BM,
Size-INV, and Size-OP, respectively. SA performs poorly in terms of yielding
the lowest Sharpe ratios in all cases. GEC with a short exposure of 50\% is
better than those without. For the case of $\ell _{2}$-relaxation, the three
choices of $\widehat{\boldsymbol{\Sigma }}$ lead to similar Sharpe ratios,
which all outperform GEC and SA.

\section{Empirical Applications}

\label{sec:empirical_app}

In this section, we explore three empirical examples. In the first two
applications, we assess the MSFE\footnote{%
The MAFE results are available in Appendix \ref{mafe2}.} of a microeconomic
study of forecasting box office and a macroeconomic exercise for the survey
of professional forecasters (SPF). The last one is a financial application
of MVP evaluated by the Sharpe ratio.

\subsection{Box Office}

The motion picture industry devotes enormous resources to marketing in order
to influence consumer sentiment toward their products. These resources are
intended to reduce the supply-demand friction on the market. On the supply
side, movie making is an expensive business; on the demand side, however,
the audience's taste is notoriously fickle. Accurate prediction of box
office is financially crucial for motion picture investors.

Based on the data of Hollywood movies released in North America between
October 1, 2010 and June 30, 2012, \cite{lehrer2017} demonstrate the sound
out-of-sample performance of the \textit{prediction model averaging} (PMA).
We revisit their dataset of 94 cross-sectional observations (movies), 28
non-constant explanatory variables and 95 candidate forecasters according to
a multitude of model specifications. Guided by the intuition that the input
variables capturing similar characteristics are \textquotedblleft
closer\textquotedblright\ to one another, \cite{lehrer2017} cluster input
variables into six groups in their Appendix D.1: 
\begin{align*}
\text{\textbf{Key variables}}:& \mathtt{\
Constant,Animation,Family,Weeks,Screens,VOL:T-1/-3} \\
\text{\textbf{Twitter Volume}}:& \mathtt{\ T-21/-27,T-14/-20,T-7/-13,T-4/-6}
\\
\text{\textbf{Twitter Sentiment}}:& \mathtt{\
T-21/-27,T-14/-20,T-7/-13,T-4/-6,T-1/-3} \\
\text{\textbf{Rating Related}}:& \mathtt{\ PG,PG13,R,Budget} \\
\text{\textbf{Male Genre}}:& \mathtt{\
Action,Adventure,Crime,Fantasy,Sci-Fi,Thriller} \\
\text{\textbf{Female Genre}}:& \mathtt{\ Comedy,Drama,Mystery,Romance}
\end{align*}%
Since the 95 forecasters are generated based on these input variables, the
potential group patterns may help $\ell _{2}$-relaxation achieve more
accurate forecasts than other off-the-shelf machine learning shrinkage
methods in this setting.

Following \cite{lehrer2017}, we randomly rearrange the full sample with $%
n=94 $ movies into a training set of the size $n_{\mathrm{tr}}$ and an
evaluation set of the size $n_{\mathrm{ev}}=n-n_{\mathrm{tr}}$, which we
experiment with $n_{\mathrm{ev}}=10,$ $20,$ $30$, and 40. We repeat this
procedure for 1,000 times and evaluate the MSFE of PMA, $\ell_2$-relax$^0$,
the CSR \citep{elliott2013complete}, the peLASSO \citep{Diebold_Shin2019},
Lasso, Ridge, and $\ell _{2}$-relaxation. We choose the number of subset
regressors to be 10 and 15 for the CSR approach, denoted as CSR$_{10}$ and
CSR$_{15}$, respectively. Since the total number of candidate models is too
large to handle, we follow \citet{genre2013} and randomly pick 10000
candidate models instead. For peLASSO, we follow \citet{Diebold_Shin2019}
and conduct a two-step estimation (first Lasso, then Ridge). Movies are
viewed as independent observations and thus the tuning parameters are chosen
by the conventional 5-fold CV. We conduct a grid search from 0 to 5 with
increment 0.1.

\begin{table}[tph]
\caption{Relative MSFE of Movie Forecasting }
\label{out}\centering
\begin{tabular}{ccccccccccc}
\hline
$n_{\mathrm{ev}}$ & PMA & $\ell_2$-relax$^0$ & CSR$_{10}$ & CSR$_{15}$ & 
peLasso & Lasso & Ridge & \multicolumn{3}{c}{$\ell_2$-relax} \\ 
&  &  &  &  &  &  &  & $\widehat{\boldsymbol{\Sigma}}_s$ & $\widehat{%
\boldsymbol{\Sigma}}_1$ & $\widehat{\boldsymbol{\Sigma}}_2$ \\ \hline
10 & 1.000 & 5.075 & 2.564 & 3.475 & 4.518 & 1.085 & 1.073 & 0.933 & 0.903 & 
0.863 \\ 
20 & 1.000 & 3.088 & 2.461 & 3.697 & 3.837 & 1.167 & 1.167 & 0.924 & 0.864 & 
0.836 \\ 
30 & 1.000 & 3.122 & 1.660 & 2.689 & 3.004 & 1.161 & 1.236 & 0.914 & 0.825 & 
0.722 \\ 
40 & 1.000 & 3.112 & 1.107 & 1.891 & 1.736 & 1.028 & 1.169 & 0.904 & 0.851 & 
0.692 \\ \hline
\multicolumn{5}{l}{\scriptsize Note: The MSFE of PMA is normalized as 1.} & 
&  &  &  &  & 
\end{tabular}%
\end{table}

Since the magnitude of MSFEs varies on the evaluation sizes $n_{\mathrm{ev}}$%
, we report in Table \ref{out} the mean risk relative to that of PMA for
convenience of comparison. Entries smaller than 1 indicate better
performance relative to that of PMA. While PMA is known to outperform Lasso
and Ridge in \cite{lehrer2017}, in this exercise it also outperforms $\ell_2$%
-relax$^0$, CSR, and peLasso. Shrinkage toward the global equal weight $1/N$
or toward 0 is not favored in this experiment. $\ell _{2}$-relaxation, on
the other hand, yields lower risk than PMA under any $\widehat{\boldsymbol{%
\Sigma }}$, and the edge generally increases with the value of $n_{\mathrm{ev%
}}$.

\begin{figure}[ht!]
\centering
\includegraphics[scale = 0.8]{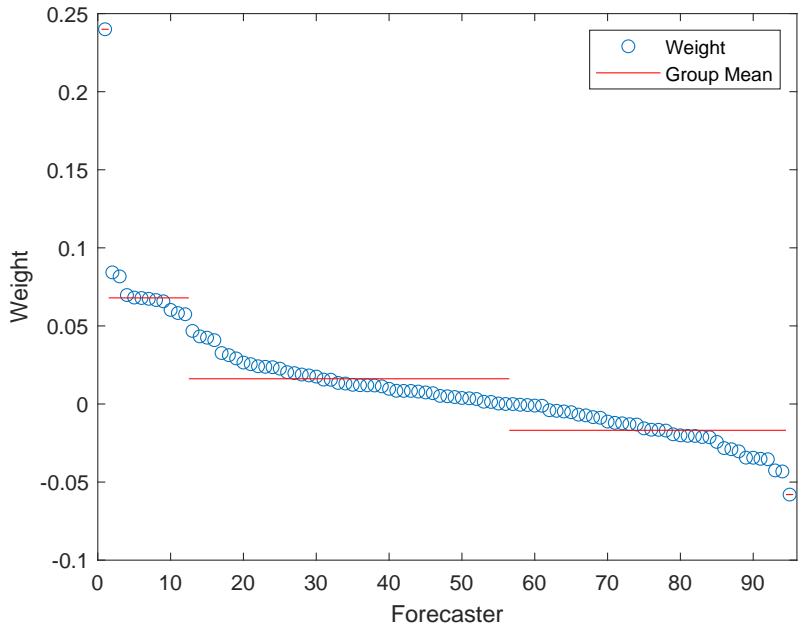}
\caption{Estimated Weights by $\ell_2$-relaxation ($n_\text{ev}=10$ and $%
\protect\tau=1$) }
\label{fmovie}
\end{figure}

To demonstrate the potential grouping pattern in the data, we show the
estimated weights of a typical replication on $n_{\mathrm{ev}}=10$ and $\tau
=1$ in Figure \ref{fmovie}. The pattern is similar for other values of $n_{%
\mathrm{ev}}$. The vertical axis of Figure \ref{fmovie} represents the
estimated weights, and the horizontal axis shows all the 95 forecasters
order by the weights from high to low. In addition, we divide the weights
into five groups according to the following manually selected intervals: $%
(-\infty ,-0.05)$, $(0.05,0)$, $(0,0.05)$, $(0.05,0.2)$, and $(0.2,+\infty )$%
. Circles and solid-lines represent the weights and group means,
respectively, in Figure \ref{fmovie}. The results demonstrate potential
latent grouping pattern. Interesting, more than 40\% of the individual
models receive negative weights.

\subsection{Inflation}

Firms, consumers, as well as monetary policy authorities count on the
outlook of inflation to make rational economic decisions. Besides
model-based inflation forecasts published by government and research
institutes, SPF reports experts' perceptions about the price level movement
in the future. A long-standing myth of forecast combination lies in the
robustness of the simple average which extract the mean or median as a
predictor in a simple linear regression, as documented by %
\citet{ang2007macro}. Recent research shows modern machine learning methods
can assist by assigning data-driven weights to individual forecasters to
gather disaggregate information; see, e.g., \citet{Diebold_Shin2019}.

The European Central Bank's SPF inquires many professional institutions for
their expectations of the euro-zone macroeconomic outlook. We revisit \cite%
{genre2013}'s harmonized index of consumer prices (HICP) dataset, which
covers 1999Q1--2018Q4. The experts were asked about their one-year- and
two-year-ahead predictions. The raw data record 119 forecasters in total,
but are highly unbalanced with plenty of missing values, mainly due to entry
and exit in the long time span. We follow \cite{genre2013} to obtain 30
qualified forecasters by first filtering out irregular respondents if he or
she missed more than 50\% of the observations, and then using a simple AR(1)
regression to interpolate the missing values in the middle.

\begin{table}[tph]
\caption{Relative MSFE of HICP Forecasting}
\label{t1}\centering{\ 
\begin{tabular}{lccccccc}
\hline
Horizon & SA & $\ell_2$-relax$^0$ & Lasso & Ridge & \multicolumn{3}{c}{$%
\ell_2$-relax} \\ 
&  &  &  &  & $\widehat{\boldsymbol{\Sigma}}_s$ & $\widehat{\boldsymbol{%
\Sigma}}_1$ & $\widehat{\boldsymbol{\Sigma}}_2$ \\ \hline
One-year-ahead & 1.000 & 2.029 & 0.844 & 0.869 & 0.908 & 0.777 & 0.824 \\ 
Two-year-ahead & 1.000 & 1.947 & 0.750 & 0.910 & 0.684 & 0.728 & 0.602 \\ 
\hline
\multicolumn{5}{l}{\scriptsize Note: The MSFE of SA is normalized as 1.} & 
&  & 
\end{tabular}%
}
\end{table}

Our benchmark is the simple average (SA) on all 30 forecasters. We compare
the forecast errors of SA, $\ell_2$-relax$^0$, Lasso, Ridge, and $\ell _{2}$%
-relaxation. We use a rolling window of 40 quarters for estimation. The
tuning parameters are selected by the OOS approach described in Box 2  with grid search from 0 to 5 with increment 0.01. The
results of relative risks are presented in Table \ref{t1}, with the MSFEs of
SA standardized as 1. $\ell_2$-relax$^0$ performs worse than SA. Lasso and
Ridge yield roughly 15\% improvement relative to SA. $\ell _{2}$-relaxation
exhibits robust performance under all choices of $\widehat{\boldsymbol{%
\Sigma }}$.

\begin{figure}[ht!]
\centering
\includegraphics[scale = 0.80]{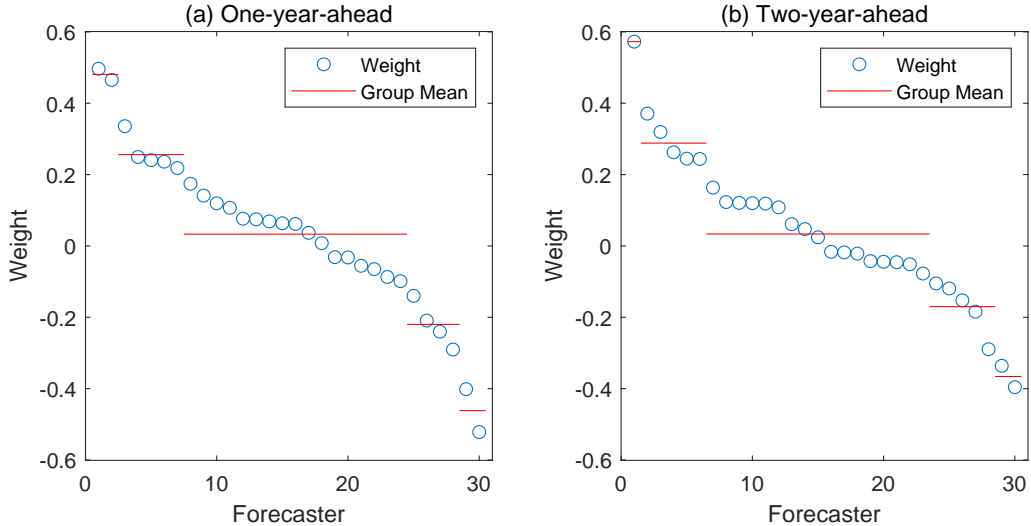}
\caption{$\ell_2$-relaxation's Estimated Weights under the Two Horizons }
\label{fhicp}
\end{figure}

Since we do not directly observe the underlying factors based upon which the
forecasters make decisions, we illustrate in Figure \ref{fhicp} the
estimated weights associated with the 30 forecasters of a typical roll from
1999Q1 to 2008Q4 and $\tau =0.02$. Sub-figures (a) and (b) are associated
with results of one-year-ahead and two-year-ahead forecasting, respectively.
The horizontal axis shows the forecasters and the vertical axis represents
the estimated weights. The weights can be roughly categorized into five
groups according to the following manually selected intervals: $(-\infty
,-0.3)$, $(-0.3,-0.1)$, $(-0.1,0.2)$, $(0.2,0.4)$, and $(0.4,+\infty )$. In
both sub-figures, the spread of the weights deviates the equal-weight SA
strategy. In sub-figure (b) the weights are more concentrated around 0 than
sub-figure (a), reflecting the challenges to forecast over a longer horizon.

\subsection{Fama and French 100 Portfolios}

\label{subsec:port}

Here we mimic the simulation design in Section \ref{subsec:sim_port} but
feed the algorithms with the real 2001--2020 Fama and French 100 monthly
portfolios on Size-BM, Size-INV, and Size-OP. The empirical results are
presented in Table \ref{psr}.

\begin{table}[ht!]
\caption{Sharp Ratio based on 100 Fama-French Portfolios}
\label{psr}\centering 
\begin{tabular}{cccccccccc}
\hline
$L$ & $N$ & SA &  & \multicolumn{2}{c}{GEC} &  & \multicolumn{3}{c}{$\ell_2$%
-relax} \\ 
&  &  &  & $c = 1$ & $c = 2$ &  & $\widehat{\boldsymbol{\Sigma}}_s$ & $%
\widehat{\boldsymbol{\Sigma}}_1$ & $\widehat{\boldsymbol{\Sigma}}_2$ \\ 
\hline
\multicolumn{5}{l}{\emph{Panel A: Size-BM}} &  &  &  &  &  \\ 
60 & 100 & 0.182 &  & 0.210 & 0.257 &  & 0.277 & 0.264 & 0.265 \\ 
120 & 100 & 0.249 &  & 0.362 & 0.428 &  & 0.441 & 0.428 & 0.445 \\ 
&  &  &  &  &  &  &  &  &  \\ 
\multicolumn{5}{l}{\emph{Panel B: Size-INV}} &  &  &  &  &  \\ 
60 & 100 & 0.191 &  & 0.296 & 0.255 &  & 0.192 & 0.215 & 0.232 \\ 
120 & 100 & 0.252 &  & 0.335 & 0.340 &  & 0.414 & 0.426 & 0.407 \\ 
&  &  &  &  &  &  &  &  &  \\ 
\multicolumn{5}{l}{\emph{Panel C: Size-OP}} &  &  &  &  &  \\ 
60 & 100 & 0.222 &  & 0.293 & 0.383 &  & 0.342 & 0.314 & 0.335 \\ 
120 & 100 & 0.287 &  & 0.431 & 0.513 &  & 0.472 & 0.548 & 0.541 \\ \hline
\end{tabular}%
\end{table}

The benchmark SA suggested by \citet{demiguel2009generalized} delivers
higher Sharpe ratios under the longer rolling window, indicating substantial
noise in the simple aggregation over the cross section when $L$ is small.
The performance of SA is eclipsed by GEC and $\ell _{2}$-relaxation in all
cases. GEC without short exposure ($c=1$) wins the Size-INV with $L=60$, and
that with 50\% short exposure ($c=2$) wins the Size-OP with $L=60$. In four
out of six cases, nevertheless, $\ell _{2}$-relaxation delivers the highest
Sharpe ratios.

\begin{figure}[tbp]
\centering
\subfloat{Sub-figure 4.1: Size-BM}{
             \includegraphics[scale = 0.70]{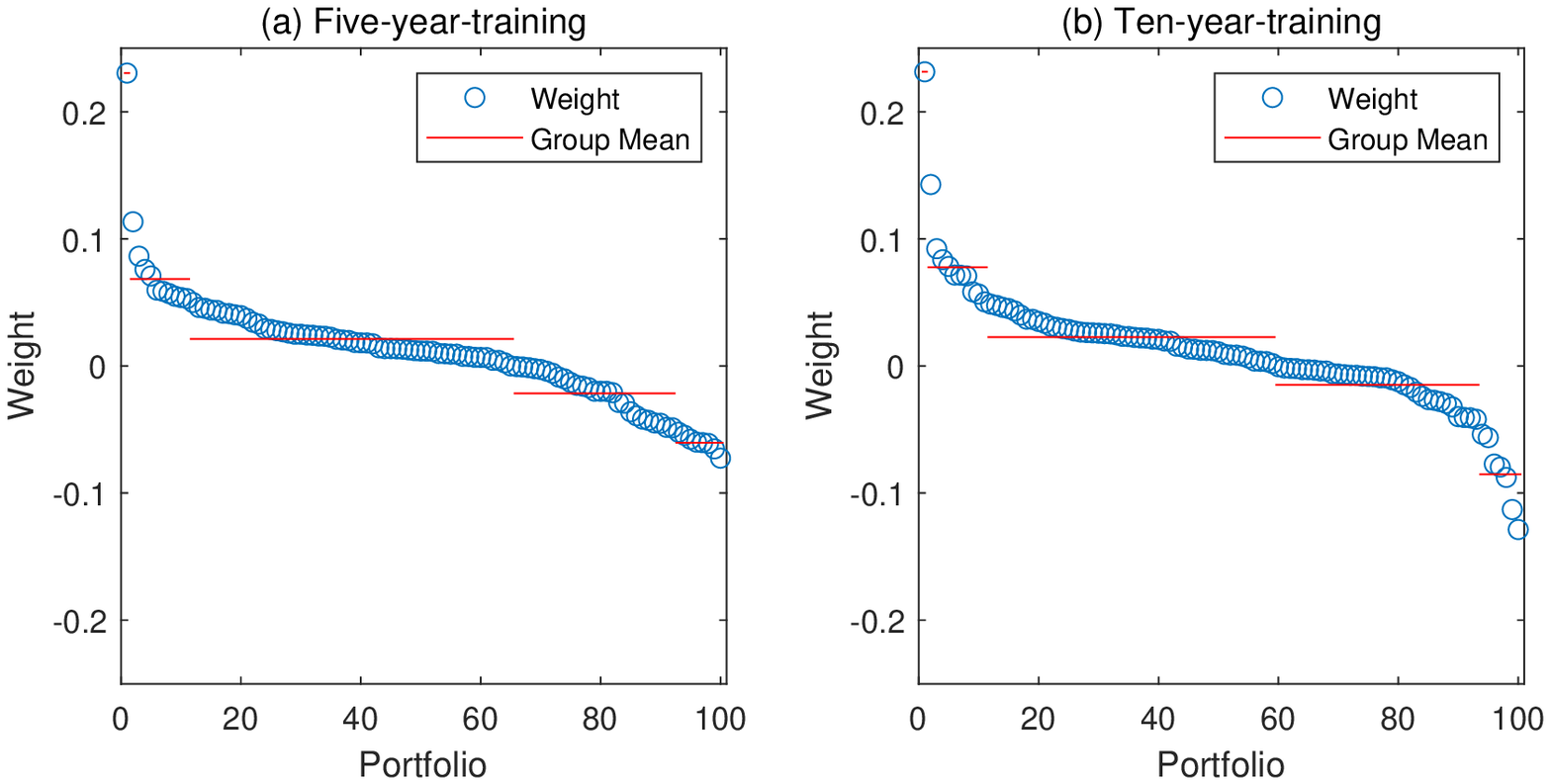}
          } \vspace{0.5cm} 
\subfloat{Sub-figure 4.2: Size-INV}{
             \includegraphics[scale = 0.70]{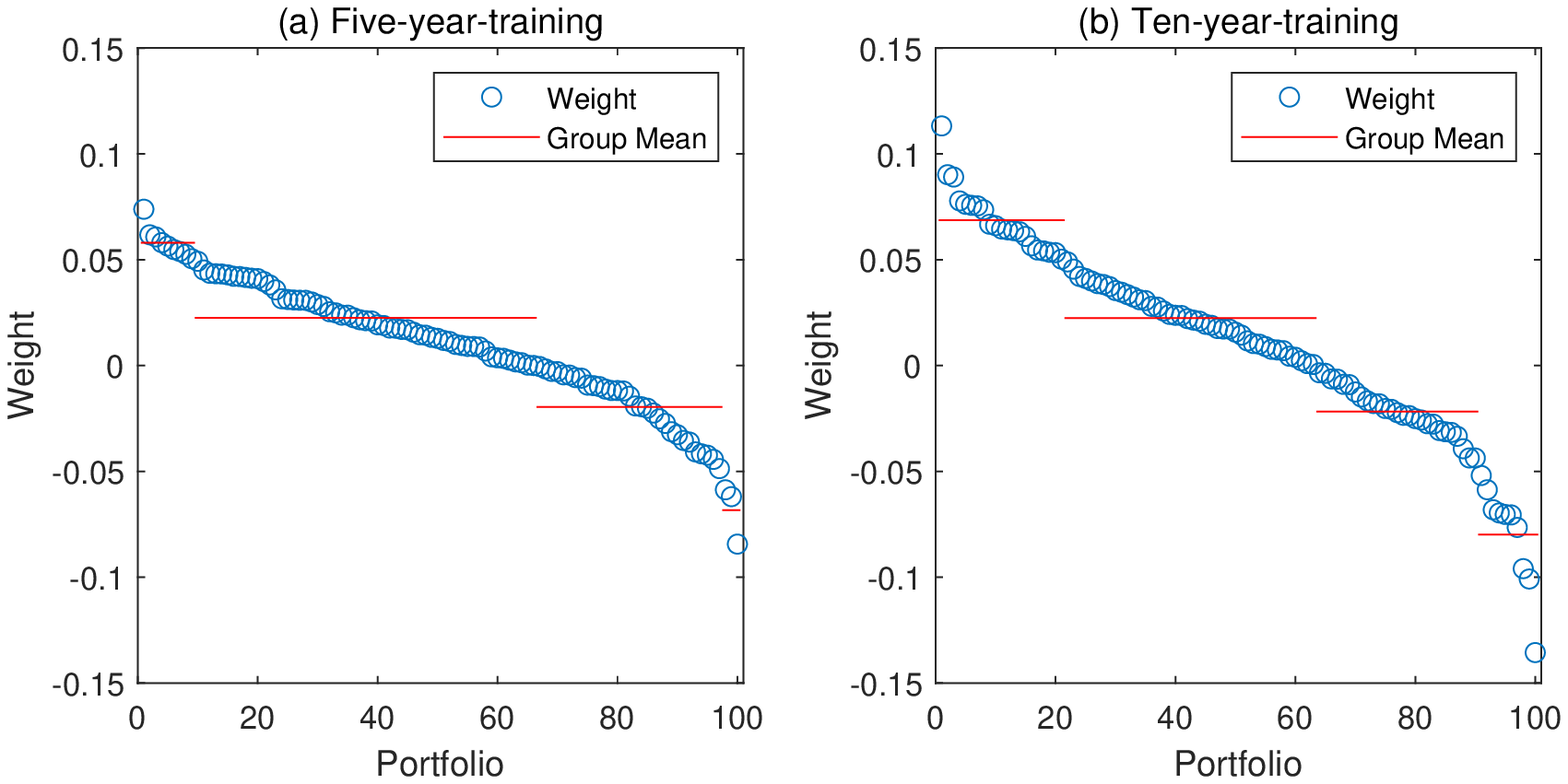}
           } 
\subfloat{Sub-figure 4.3: Size-OP}{
             \includegraphics[scale = 0.70]{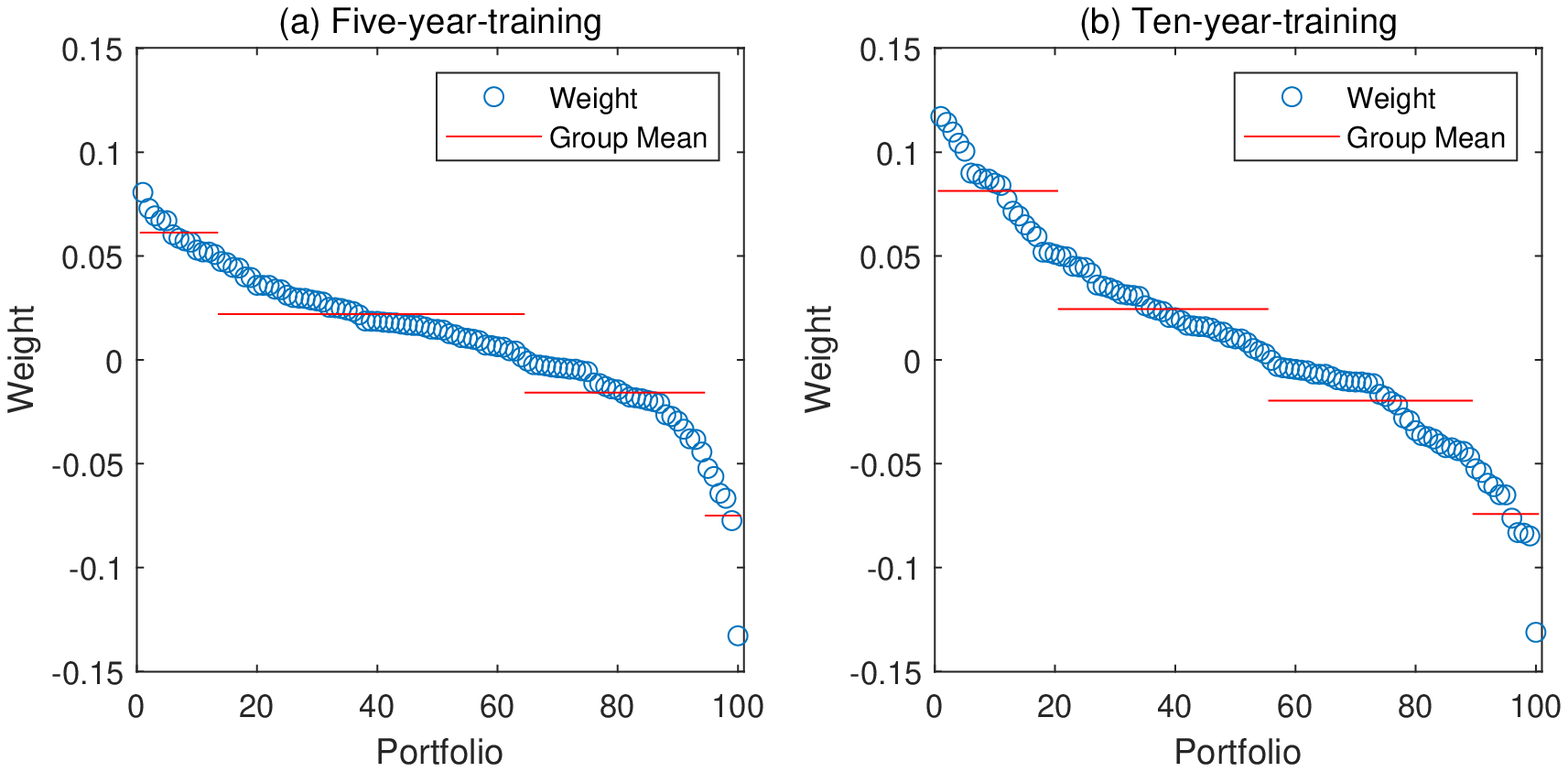}
           }
\caption{Estimated Weights by $\ell_2$-relaxation}
\label{fps}
\end{figure}

To better understand the behavior of $\ell _{2}$-relaxation, we plot in
Figure \ref{fps} its estimated weights of a typical estimation window with $%
L=60$ and 120, and the value of tuning parameter set to $\tau =1$. We assign
the 100 portfolios into 5 groups in each case and plot the group means by
the red horizontal lines. The weights are manually categorized into five
intervals: $(-\infty ,-0.05)$, $(-0.05,0)$, $(0,0.05)$, $(0.05,0.15)$, and $%
(0.15,+\infty )$. The distribution of the weights are similar across $L$ in
the sub-figure for the Size-BM sorted portfolios. Under the Size-INV
sorting, however, the weights for $L=60$ are less spread and many are close
to zero, and the weights under the Size-OP portfolios share similar
patterns. This phenomenon helps to explain the high Sharpe ratio of GEC
under $L=60$: intuitively, its $\ell _{1}$-norm restriction $\Vert \mathbf{w}%
\Vert _{1}\leq c$ would shrink all weights toward zero and moreover push
many small weights to be exactly zero. 

Acknowledging that the theoretical setup of parsimonious factors and group
structure is an approximation of the real financial world at best, in future
studies we are interested in investigating an enhanced $\ell _{2}$%
-relaxation with the exposure constraint $\Vert \mathbf{w}\Vert _{1}\leq c$.

\section{Conclusion}

This paper presents a new machine learning algorithm, namely, $\ell _{2}$%
-relaxation. When the forecast error VC or the portfolio VC can be
approximated by a block equicorrelation structure, we establish its
consistency and asymptotic optimality in the high dimensional context.
Simulations and real data applications demonstrate excellent performance of
the $\ell _{2}$-relaxation method.

Our work raises several interesting issues for further research. First, we
have not studied the optimal choice of the tuning parameter $\tau $ or
provided a formal justification for the use of the cross-validated $\tau . $
Recently, \citet{wu2020survey} have reviewed the literature of tuning
parameter selection for high dimensional regressions and discussed various
strategies to choose the tuning parameter to achieve either prediction
accuracy or support recovery such as the $L$-fold cross-validation, $m$%
-out-of-$n$ bootstrap and extended Bayesian information criterion (BIC). %
\citet{chetverikov2021cross} have studied the theoretical properties of
Lasso based on cross-validated choice of the tuning parameter. It will be
interesting to study whether we can draw support from these papers to
provide theoretical guidance concerning the choice of $\tau $ in our context.

Second, additional restrictions can be imposed to accompany the $\ell _{2}$%
-relaxation problem. For example, if sparsity is desirable, we may consider
adding the exposure constraint $\left\Vert \mathbf{w}\right\Vert _{1}\leq c$
for some tuning parameter $c$, which echoes the idea of mixed $\ell _{1}$-
and $\ell _{2}$-penalty of the elastic net method by \citet{Zou_Hastie2005}.
Another example is to incorporate the constraints $w_{i}\geq 0$ for all $i$
if non-negative weights are desirable \citep{jagannathan2003risk}. Third,
our $\ell _{2}$-relaxation is motivated from the MSFE loss function, it is
possible to consider other forms of relaxation if the other forms of loss
functions (e.g., MAFE) are under investigation. Last but not least, the $%
\ell _{2}$-relaxation theory in this paper requires the dominant component $%
\mathbf{\hat{\Sigma}}^{\ast }$ of the VC matrix to have a latent group
structure. It is desirable to extend the theory to the case where $\mathbf{%
\hat{\Sigma}}^{\ast }$ has only a low-rank structure instead of a latent
group structure. We shall explore some of these topics in future works.

{\small 
\bibliographystyle{chicagoa}
\bibliography{regressorComb}
}

\newpage

\newpage

{{%
%TCIMACRO{\TeXButton{appendix}{\appendix}}%
%BeginExpansion
\appendix%
%EndExpansion
}}%
%TCIMACRO{\TeXButton{setcounter}{\setcounter{page}{1}}}%
%BeginExpansion
\setcounter{page}{1}%
%EndExpansion
\setcounter{footnote}{0}\setcounter{section}{0} \setcounter{table}{0} %
\setcounter{figure}{0} %\renewcommand{\thesection}{S\arabic{section}} %
\numberwithin{equation}{section} 
\renewcommand{\thefigure}{S\arabic{figure}} \renewcommand{\thetable}{S%
\arabic{table}} \renewcommand{\thelemma}{S\arabic{lemma}}{{%
%TCIMACRO{\TeXButton{linespread}{\linespread{1.3}}}%
%BeginExpansion
\linespread{1.3}%
%EndExpansion
\setlength{\baselineskip}{39pt} }}

\begin{center}
{\Large Online Supplement for \medskip }

\textquotedblleft $\ell _{2}$\textbf{-Relaxation: With Applications to
Forecast Combination \\[0pt]
and Portfolio Analysis}\textquotedblright\ 

Zhentao Shi$^{a,b}$, Liangjun Su$^{c}$, Tian Xie$^{d}$ \textit{\medskip }

$^{a}$ School of Economics, Georgia Institute of Technology\\[0pt]
$^{b}$ Department of Economics, Chinese University of Hong Kong

$^{c}$ School of Economics and Management, Tsinghua University\\[0pt]
$^{d}$ College of Business, Shanghai University of Finance and Economics%
\textit{\medskip }
\end{center}

This online appendix is composed of two sections. Appendix A contains the
proofs of the theoretical results in the paper. Appendix B contains some
additional results on the simulation and empirical exercises.

\textbf{Additional Notation.} The notations in the appendix are consistent
with those in the main text. Here we introduce a few additional expressions.
For a generic $n\times m$ matrix $\mathbf{B}$, we denote $\mathbf{B}_{i\cdot
}$ as the $i$-th row ($1\times m$ vector), and define the spectral norm as $%
\left\Vert \mathbf{B}\right\Vert _{\mathrm{sp}}:=\phi _{\max }^{1/2}\left( 
\mathbf{B}^{\prime }\mathbf{B}\right) $. \textquotedblleft
w.p.a.1\textquotedblright\ is short for \textquotedblleft with probability
approaching one\textquotedblright\ in asymptotic statements.

\section{Technical Appendix}

\subsection{Optimization Formulation}

\label{subsec:bg_w}

While the original paper of \citet{bates1969combination} consider unbiased
forecasts in that $E\left[ y_{t+1}-f_{it}\right] =0$ for all $i$, %
\citet{granger1984improved} generalize it to accommodate biased forecasts.
We start with the latter. Besides the $N\times 1$ weight vector $\mathbf{w}$%
, we seek an additional intercept $\mu $, which is an unknown location
parameter, to correct the bias of the combined forecasts. The optimization
problem can be written as 
\begin{equation*}
\min_{\left( \mu ,\mathbf{w}\right) \in \mathbb{R}^{N+1}}\ \frac{1}{2T}%
\sum_{t=1}^{T}\left( y_{t+1}-\mu -\mathbf{w}^{\prime }\mathbf{f}_{t}\right)
^{2}\ \text{subject to\ }\boldsymbol{1}_{N}^{\prime }\mathbf{w}=1.
\end{equation*}%
Its Lagrangian is 
\begin{align*}
L\left( \mu ,\mathbf{w},\gamma \right) & =\frac{1}{2T}\sum_{t=1}^{T}\left(
y_{t+1}-\mu -\mathbf{w}^{\prime }\mathbf{f}_{t}\right) ^{2}+\gamma \left( 
\boldsymbol{1}_{N}^{\prime }\mathbf{w}-1\right) \\
& =\frac{1}{2T}\sum_{t=1}^{T}\left( \mathbf{w}^{\prime }\left[ \left(
y_{t+1}-\mu \right) \boldsymbol{1}_{N}-\mathbf{f}_{t}\right] \right)
^{2}+\gamma \left( \boldsymbol{1}_{N}^{\prime }\mathbf{w}-1\right) \\
& =\frac{1}{2T}\mathbf{w}^{\prime }\left[ \sum_{t=1}^{T}\left( \mathbf{e}%
_{t}-\mu \boldsymbol{1}_{N}\right) \left( \mathbf{e}_{t}-\mu \boldsymbol{1}%
_{N}\right) ^{\prime }\right] \mathbf{w}+\gamma \left( \boldsymbol{1}%
_{N}^{\prime }\mathbf{w}-1\right) ,
\end{align*}%
where $\gamma $ is the Lagrangian multiplier. The above formulation includes %
\citet{bates1969combination}'s problem as a special case when $\mu =0$.

{\ When $\mu $ is unconstrained, given any $\mathbf{w}$ its minimizer is $%
\widehat{\mu }=\widehat{\mu }\left( \mathbf{w}\right) =\mathbf{w}^{\prime
-1}\sum_{t=1}^{T}\mathbf{e}_{t}=\mathbf{w}^{\prime }\bar{\mathbf{e}}$.
Substituting $\widehat{\mu }$ back to the Lagrangian to profile out $\mu $
yields 
\begin{align}
\widehat{L}\left( \mathbf{w},\gamma \right) & :=L\left( \widehat{\mu }\left( 
\mathbf{w}\right) ,\mathbf{w},\gamma \right)  \notag \\
& =\frac{1}{2}\mathbf{w}^{\prime }\left[ \frac{1}{T}\sum_{t=1}^{T}\left( 
\mathbf{e}_{t}-\bar{\mathbf{e}}\right) \left( \mathbf{e}_{t}-\bar{\mathbf{e}}%
\right) ^{\prime }\right] \mathbf{w}+\gamma \left( \boldsymbol{1}%
_{N}^{\prime }\mathbf{w}-1\right)  \notag \\
& =\frac{1}{2}\mathbf{w}^{\prime }\widehat{\boldsymbol{\Sigma }}\mathbf{w}%
+\gamma \left( \boldsymbol{1}_{N}^{\prime }\mathbf{w}-1\right) ,
\label{eq:biased_Lag}
\end{align}%
which is exactly the Lagrangian of the problem in \eqref{eq:bates-granger}. }

\subsection{Finite Sample Numerical Properties}

The primal problem in (\ref{eq:relax}) induces the dual problem as stated in
the following lemma. \eqref{eq:sim_dual} below is a constrained $\ell _{1}$%
-penalized optimization where the criterion function is the summation of a
quadratic form of $\boldsymbol{\alpha }$, a linear combination of $%
\boldsymbol{\alpha }$, and the $\ell _{1} $-norm of $\boldsymbol{\alpha }$,
while the constraint is linear in $\boldsymbol{\alpha }$. The dual problem
is instrumental in our theoretical analyses due to its similarity to Lasso %
\citep{tibshirani1996regression}.

\begin{lemma}
{\ \label{lem:dual} The dual problem of (\ref{eq:relax}) is 
\begin{equation}
\min_{\boldsymbol{\alpha }\in \mathbb{R}^{N}}\left\{ \frac{1}{2}\boldsymbol{%
\alpha }^{\prime }\widehat{\mathbf{A}}^{\prime }\widehat{\mathbf{A}}%
\boldsymbol{\alpha }+\frac{1}{N}\boldsymbol{1}_{N}^{\prime }\widehat{%
\boldsymbol{\Sigma }}\boldsymbol{\alpha }+\tau \left\Vert \boldsymbol{\alpha 
}\right\Vert _{1}-\frac{1}{2N}\right\} \text{\ \ subject to \ }\boldsymbol{1}%
_{N}^{\prime }\boldsymbol{\alpha }=0,  \label{eq:sim_dual}
\end{equation}%
where $\widehat{\mathbf{A}}=\left( \mathbf{\mathbf{I}}_{N}-N^{-1}\boldsymbol{%
1}_{N}\boldsymbol{1}_{N}^{\prime }\right) \widehat{\boldsymbol{\Sigma }}$ is
the columnwise demeaned version of $\widehat{\boldsymbol{\Sigma }}$. Denote $%
\widehat{\boldsymbol{\alpha }}=\widehat{\boldsymbol{\alpha }}_{\tau }$ as a
solution to the dual problem in (\ref{eq:sim_dual}), and it is connected
with the solution to the primal problem in (\ref{eq:relax}) via 
\begin{equation}
\widehat{\mathbf{w}}=\widehat{\mathbf{A}}\widehat{\boldsymbol{\alpha }}+%
\frac{\boldsymbol{1}_{N}}{N}.  \label{eq:w_A_alpha}
\end{equation}%
}
\end{lemma}

{\ \noindent \textbf{Proof of Lemma \ref{lem:dual}}. First, we can rewrite
the minimization problem in (\ref{eq:relax}) in terms of linear constraints: 
\begin{gather}
\min_{\left( \mathbf{w},\gamma \right) \in \mathbb{R}^{N+1}}\frac{1}{2}%
\left\Vert \mathbf{w}\right\Vert _{2}^{2}  \notag \\
\text{s.t. }\mathbf{w}^{\prime }\boldsymbol{1}_{N}-1=0,\ 
\begin{pmatrix}
\widehat{\boldsymbol{\Sigma }} & \boldsymbol{1}_{N}%
\end{pmatrix}%
\begin{pmatrix}
\mathbf{w} \\ 
\gamma%
\end{pmatrix}%
\leq \tau \boldsymbol{1}_{N},\ \text{and }-%
\begin{pmatrix}
\widehat{\boldsymbol{\Sigma }} & \boldsymbol{1}_{N}%
\end{pmatrix}%
\begin{pmatrix}
\mathbf{w} \\ 
\gamma%
\end{pmatrix}%
\leq \tau \boldsymbol{1}_{N}  \label{eq:relax_linear}
\end{gather}%
where \textquotedblleft $\leq $\textquotedblright\ holds elementwise
hereafter. Define the Lagrangian function as 
\begin{eqnarray}
\mathcal{L}\left( \mathbf{w},\gamma ;\boldsymbol{\alpha }_{1},\boldsymbol{%
\alpha }_{2},\alpha _{3}\right) &=&\frac{1}{2}\mathbf{w}^{\prime }\mathbf{w}+%
\boldsymbol{\alpha }_{1}^{\prime }\left( 
\begin{pmatrix}
\widehat{\boldsymbol{\Sigma }} & \boldsymbol{1}_{N}%
\end{pmatrix}%
\begin{pmatrix}
\mathbf{w} \\ 
\gamma%
\end{pmatrix}%
-\tau \boldsymbol{1}_{N}\right)  \notag \\
&&-\boldsymbol{\alpha }_{2}^{\prime }\left( 
\begin{pmatrix}
\widehat{\boldsymbol{\Sigma }} & \boldsymbol{1}_{N}%
\end{pmatrix}%
\begin{pmatrix}
\mathbf{w} \\ 
\gamma%
\end{pmatrix}%
+\tau \boldsymbol{1}_{N}\right) +\alpha _{3}\left( \mathbf{w}^{\prime }%
\boldsymbol{1}_{N}-1\right)  \label{eq:Lagrangian2}
\end{eqnarray}%
and the associated Lagrangian dual function as $g\left( \boldsymbol{\alpha }%
_{1},\boldsymbol{\alpha }_{2},\alpha _{3}\right) =\inf_{\mathbf{w},\gamma }%
\mathcal{L}\left( \mathbf{w},\gamma ;\boldsymbol{\alpha }_{1},\boldsymbol{%
\alpha }_{2},\alpha _{3}\right) ,$ where $\boldsymbol{\alpha }_{1}\geq 0$, $%
\boldsymbol{\alpha }_{2}\geq 0$, and $\alpha _{3}$ are the Lagrangian
multipliers for the three constraints in (\ref{eq:relax_linear}),
respectively. }

{\ Let $\varphi\left(\mathbf{w,}\gamma\right)=\frac{1}{2}\left\Vert \mathbf{w%
}\right\Vert _{2}^{2}$, the objective function in (\ref{eq:relax_linear}).
Define its conjugate function as 
\begin{equation*}
\varphi^{\ast}\left(\mathbf{a,}b\right)=\sup_{\mathbf{w},\gamma}\left\{ 
\mathbf{a}^{\prime}\mathbf{w}+b\gamma-\frac{1}{2}\left\Vert \mathbf{w}%
\right\Vert _{2}^{2}\right\} =\left\{ 
\begin{array}{ll}
\frac{1}{2}\left\Vert \mathbf{a}\right\Vert _{2}^{2}\text{ } & \text{if }b=0
\\ 
\infty & \text{otherwise}%
\end{array}%
.\right.
\end{equation*}
The linear constraints indicate an explicit dual function (See %
\citet[p.221]{boyd2004convex}): 
\begin{eqnarray*}
& & g\left(\boldsymbol{\alpha}_{1},\boldsymbol{\alpha}_{2},\alpha_{3}\right)
\\
& = & -\tau\boldsymbol{1}_{N}^{\prime}\left(\boldsymbol{\alpha}_{1}+%
\boldsymbol{\alpha}_{2}\right)-\alpha_{3}-\varphi^{\ast}\left(\widehat{%
\boldsymbol{\Sigma}}\left(\boldsymbol{\alpha}_{2}-\boldsymbol{\alpha}%
_{1}\right)-\alpha_{3}\boldsymbol{1}_{N},\text{ }\boldsymbol{1}%
_{N}^{\prime}\left(\boldsymbol{\alpha}_{2}-\boldsymbol{\alpha}%
_{1}\right)\right) \\
& = & \left\{ 
\begin{array}{ll}
-\tau\boldsymbol{1}_{N}^{\prime}\left(\boldsymbol{\alpha}_{1}+\boldsymbol{%
\alpha}_{2}\right)-\alpha_{3}-\frac{1}{2}\left\Vert \widehat{\boldsymbol{%
\Sigma}}\left(\boldsymbol{\alpha}_{2}-\boldsymbol{\alpha}_{1}\right)-%
\alpha_{3}\boldsymbol{1}_{N}\right\Vert _{2}^{2},\text{ } & \text{if }%
\boldsymbol{1}_{N}^{\prime}\left(\boldsymbol{\alpha}_{2}-\boldsymbol{\alpha}%
_{1}\right)=0 \\ 
\infty, & \text{otherwise}%
\end{array}%
.\right.
\end{eqnarray*}
}

{\ Let $\boldsymbol{\alpha }=\boldsymbol{\alpha }_{2}-\boldsymbol{\alpha }%
_{1}$. When $\tau >0$, the two inequalities $\widehat{\boldsymbol{\Sigma }}%
_{i\cdot }\mathbf{w}+\gamma \leq \tau $ and $-\widehat{\boldsymbol{\Sigma }}%
_{i\cdot }\mathbf{w}-\gamma \leq \tau $ cannot be binding simultaneously.
The associated Lagrangian multipliers $\alpha _{1i}$ and $\alpha _{2i}$ must
satisfy $\alpha _{1i}\cdot \alpha _{2i}=0$ for all $i\in \left[ N\right] $.
This implies that $\left\Vert \boldsymbol{\alpha }\right\Vert _{1}=%
\boldsymbol{1}_{N}^{\prime }\boldsymbol{\alpha }_{1}+\boldsymbol{1}%
_{N}^{\prime }\boldsymbol{\alpha }_{2}$ so that the dual problem can be
simplified as 
\begin{equation}
\max_{\boldsymbol{\alpha },\alpha _{3}}\left\{ -\frac{1}{2}\left\Vert 
\widehat{\boldsymbol{\Sigma }}\boldsymbol{\alpha }-\alpha _{3}\boldsymbol{1}%
_{N}\right\Vert _{2}^{2}-\alpha _{3}-\tau \left\Vert \boldsymbol{\alpha }%
\right\Vert _{1}\right\} \ \text{ s.t. }\boldsymbol{1}_{N}^{\prime }%
\boldsymbol{\alpha }=0.  \label{eq:dual}
\end{equation}%
Taking the partial derivative of the above criterion function with respect
to $\alpha _{3}$ yields 
\begin{equation*}
(\widehat{\boldsymbol{\Sigma }}\boldsymbol{\alpha }-\alpha _{3}\boldsymbol{1}%
_{N})^{\prime }\boldsymbol{1}_{N}-1=0,
\end{equation*}%
or equivalently, $\alpha _{3}=\frac{1}{N}(\boldsymbol{1}_{N}^{\prime }%
\widehat{\boldsymbol{\Sigma }}\boldsymbol{\alpha }-1).$ Then 
\begin{equation*}
\left\Vert \widehat{\boldsymbol{\Sigma }}\boldsymbol{\alpha }-\alpha _{3}%
\boldsymbol{1}_{N}\right\Vert _{2}^{2}=\left\Vert \widehat{\mathbf{A}}%
\boldsymbol{\alpha }+\frac{\boldsymbol{1}_{N}}{N}\right\Vert _{2}^{2}=%
\boldsymbol{\alpha }^{\prime }\widehat{\mathbf{A}}^{\prime }\widehat{\mathbf{%
A}}\boldsymbol{\alpha }+\frac{1}{N}
\end{equation*}%
where $\widehat{\mathbf{A}}=\left( \mathbf{\mathbf{I}}_{N}-N^{-1}\boldsymbol{%
1}_{N}\boldsymbol{1}_{N}^{\prime }\right) \widehat{\boldsymbol{\Sigma }}$.
We conclude that the dual problem in (\ref{eq:dual}) is equivalent to 
\begin{equation}
\min_{\boldsymbol{\alpha }\in \mathbb{R}^{N}}\left\{ \frac{1}{2}\boldsymbol{%
\alpha }^{\prime }\widehat{\mathbf{A}}^{\prime }\widehat{\mathbf{A}}%
\boldsymbol{\alpha }+\frac{1}{N}\boldsymbol{1}_{N}^{\prime }\widehat{%
\boldsymbol{\Sigma }}\boldsymbol{\alpha }+\tau \left\Vert \boldsymbol{\alpha 
}\right\Vert _{1}-\frac{1}{2N}\right\} \text{ subject to }\boldsymbol{1}%
_{N}^{\prime }\boldsymbol{\alpha }=0,  \label{eq:sim_dual1}
\end{equation}%
where we keep the constant $-\frac{1}{2N}$ which is irrelevant to the
optimization. }

{\ When $\widehat{\boldsymbol{\alpha}}=\widehat{\boldsymbol{\alpha}}_{2}-%
\widehat{\boldsymbol{\alpha}}_{1}$ is the solution to (\ref{eq:sim_dual1}),
the solution of $\alpha_{3}$ in (\ref{eq:dual}) is $\widehat{\alpha}_{3}=%
\frac{1}{N}(\boldsymbol{1}_{N}^{\prime}\widehat{\boldsymbol{\Sigma}}\widehat{%
\boldsymbol{\alpha}}-1).$ The first order condition of (\ref{eq:Lagrangian2}%
) with respect to $\mathbf{w}$ evaluated at the solution gives 
\begin{equation*}
\boldsymbol{0}_{N}=\widehat{\mathbf{w}}+\widehat{\boldsymbol{\Sigma}}(%
\widehat{\boldsymbol{\alpha}}_{1}-\widehat{\boldsymbol{\alpha}}_{2})+%
\widehat{\alpha}_{3}\mathbf{1}_{N}=\widehat{\mathbf{w}}-\widehat{\boldsymbol{%
\Sigma}}\widehat{\boldsymbol{\alpha}}+\frac{1}{N}(\boldsymbol{1}_{N}^{\prime}%
\widehat{\boldsymbol{\Sigma}}\widehat{\boldsymbol{\alpha}}-1)\mathbf{1}_{N}=%
\widehat{\mathbf{w}}-\widehat{\boldsymbol{\Sigma}}\widehat{\boldsymbol{\alpha%
}}-\frac{1}{N}\mathbf{1}_{N}.
\end{equation*}
as $\boldsymbol{1}^{\prime }_{N}\widehat{\boldsymbol{\alpha}}=0$. The result
in (\ref{eq:w_A_alpha}) follows. \hfill ${\ \blacksquare}$ }

\begin{remark}
{\ Since $\mathrm{rank}(\widehat{\mathbf{A}})\leq \mathrm{rank}(\mathbf{%
\mathbf{I}}_{N}-N^{-1}\boldsymbol{1}_{N}\boldsymbol{1}_{N}^{\prime })=N-1$,
the singularity of\ $\widehat{\mathbf{A}}$ may induce multiple solutions to
the dual problem in (\ref{eq:sim_dual}). Despite this, the uniqueness of $%
\widehat{\mathbf{w}}$ as a solution to the primal problem in (\ref{eq:relax}%
) implies $\widehat{\mathbf{A}}\widehat{\boldsymbol{\alpha }}^{\left(
1\right) }=\widehat{\mathbf{A}}\widehat{\boldsymbol{\alpha }}^{\left(
2\right) }$ for any $\widehat{\boldsymbol{\alpha }}^{\left( 1\right) }$ and $%
\widehat{\boldsymbol{\alpha }}^{\left( 2\right) }$ that solve (\ref%
{eq:sim_dual}). It is sufficient to find any solution to the dual problem to
recover the same $\widehat{\mathbf{w}}$ in the primal. }
\end{remark}

{\ Next, we prove Lemma \ref{lem:w0} that the oracle primal problem induces
a within-group equal weight solution. }

{\ \noindent \textbf{Proof of Lemma \ref{lem:w0}.} The restriction of (\ref%
{eq:oracle_primal_0}) can be written as 
\begin{equation*}
\begin{pmatrix}
\widehat{\boldsymbol{\Sigma }}^{\ast } & \boldsymbol{1}_{N} \\ 
\boldsymbol{1}_{N}^{\prime } & 0%
\end{pmatrix}%
\begin{pmatrix}
\mathbf{w} \\ 
\gamma%
\end{pmatrix}%
=%
\begin{pmatrix}
\boldsymbol{0}_{N} \\ 
1%
\end{pmatrix}%
.
\end{equation*}%
Since the rank of $\widehat{\boldsymbol{\Sigma }}^{\ast }$ is $K$ and there
are an infinite number of solutions of $\left( \mathbf{w},\gamma \right) $
to the above system of $N+1$ equations. Since the $i$th equation and the $j$%
th equation are exactly the same if $i$ and $j$ are the in the same group,
the $\left( N+1\right) $-equation system can be reduced to a system of $K+1$
equations: 
\begin{equation*}
\begin{pmatrix}
\widehat{\boldsymbol{\Sigma }}^{\mathrm{co}} & \boldsymbol{1}_{K} \\ 
\boldsymbol{1}_{K}^{\prime } & 0%
\end{pmatrix}%
\left( \sum_{i\in \mathcal{G}_{1}}w_{i},\ldots ,\sum_{i\in \mathcal{G}%
_{K}}w_{i},\gamma \right) ^{\prime }=%
\begin{pmatrix}
\boldsymbol{0}_{K} \\ 
1%
\end{pmatrix}%
.
\end{equation*}%
Due to the $\ell _{2}$ criterion, the within-group weight takes equal value.
Define $r_{k}:=N_{k}/N$ as the fraction of the $k$-th group members on the
cross section. Let $\mathbf{r}=\left( r_{k}\right) _{k\in \lbrack K]}$ and $%
\mathbf{r}^{-1}=\left( r_{k}^{-1}\right) _{k\in \lbrack K]}$. Let
\textquotedblleft $\circ $\textquotedblright\ denote the Hadamard product.
As $\widehat{\boldsymbol{\Sigma }}^{\mathrm{co}}$ is of full rank, the
solution to the above $\left( K+1\right) $-equation system is unique: 
\begin{equation*}
\mathbf{b}_{0}^{\ast }\circ \mathbf{r}=\left( \sum_{i\in \mathcal{G}%
_{1}}w_{i}^{\ast },\ldots ,\sum_{i\in \mathcal{G}_{K}}w_{i}^{\ast }\right)
^{\prime }=\frac{(\widehat{\boldsymbol{\Sigma }}^{\mathrm{co}})^{-1}%
\boldsymbol{1}_{K}}{\boldsymbol{1}_{K}^{\prime }(\widehat{\boldsymbol{\Sigma 
}}^{\mathrm{co}})^{-1}\boldsymbol{1}_{K}}.
\end{equation*}%
%
%
%
%
%
%
%
%
% where we use the fact that $\sum_{i\in \mathcal{G}_{k}}w_{i}^{\ast
% }=b_{k}^{\ast }r_{k}$ for each $k\in \left[ K\right]$ 
% $\mathbf{b}_{0}^{\ast }=%
% \mathbf{r}^{-1}\circ \frac{(\widehat{\boldsymbol{\Sigma }}^{\mathrm{co}%
% })^{-1}\boldsymbol{1}_{K}}{\boldsymbol{1}_{K}^{\prime }(\widehat{\boldsymbol{%
% \Sigma }}^{\mathrm{co}})^{-1}\boldsymbol{1}_{K}}$.$
The explicit solution 
\begin{equation}
\mathbf{b}_{0}^{\ast }:=\left( b_{0k}^{\ast }\right) _{k\in \lbrack K]}=%
\mathbf{r}^{-1}\circ \frac{(\widehat{\boldsymbol{\Sigma }}^{\mathrm{co}%
})^{-1}\boldsymbol{1}_{K}}{\boldsymbol{1}_{K}^{\prime }(\widehat{\boldsymbol{%
\Sigma }}^{\mathrm{co}})^{-1}\boldsymbol{1}_{K}}  \label{eq:b0star}
\end{equation}%
follows immediately. \hfill ${\ \blacksquare }$}

\begin{remark}
{\ \label{rem:markowitz_oracle} In the portfolio analysis %
\eqref{eq:markowitz} the corresponding oracle problem can be written as 
\begin{equation}
\min_{\left( \mathbf{w},\gamma \right) \in \mathbb{R}^{N+1}}\frac{1}{2}%
\left\Vert \mathbf{w}\right\Vert _{2}^{2}\text{\ \ subject to \ } 
\boldsymbol{1}_{N}^{\prime } \mathbf{w} = 1,\ \ \bar{\mathbf{r}}^{\prime} 
\mathbf{w} \geq r^*, \text{ and } \widehat{\boldsymbol{\Sigma }}^{\ast }%
\mathbf{w}+\gamma \mathbf{1}_{N} = 0.
\end{equation}%
The solution to the above problem will be affected by the patterns in both $%
\widehat{\boldsymbol{\Sigma }}^{\ast }$ and $\bar{\mathbf{r}}$. Unlikely the
within-group equal weight solution in Lemma \ref{lem:w0}, if the elements in 
$\bar{\mathbf{r}}$ take distinctive values, the corresponding $\widehat{%
\mathbf{w}}^{\ast}$ will not share within-group equal weights in view of the
group membership defined solely by $\widehat{\boldsymbol{\Sigma }}^{\ast }$.
This observation motivates us to focus on MVP, instead of the one with the
mean return constraint. }
\end{remark}

{\ To proceed, we define some notations about group patterns. We call a
vector of \textit{similar sign} if no pair of its elements takes opposite
signs. Formally, an $n$-vector $\mathbf{b}$ is of similar sign if $\mathbf{b}%
\in \mathbb{S}^{n}:=\left\{ \mathbf{b}\in \mathbb{R}^{n}:b_{i}b_{j}\geq 0\ 
\text{ for all }i,j\in \left[ N\right] \right\} $. Let $\mathcal{G}%
_{1},\ldots ,\mathcal{G}_{K}$ be a partition of $\left[ N\right] $, and
denote $N_{k}=\left\vert \mathcal{G}_{k}\right\vert $. We call a generic $N$%
-vector $\mathbf{b}$ of \textit{similar sign within all groups} if $\mathbf{b%
}\in \mathbb{S}^{\mathrm{all}}:=\left\{ \mathbf{b}\in \mathbb{R}^{N}:\mathbf{%
b}_{\mathcal{G}_{k}}\in \mathbb{S}^{N_{k}}\text{ for all }k\in \lbrack
K]\right\} $, and define $\tilde{\mathbb{S}}^{\mathrm{all}}:=\left\{ \mathbf{%
b}\in \mathbb{S}^{\mathrm{all}}:\boldsymbol{1}_{N}^{\prime }\mathbf{b}%
=0\right\} $ with a further restriction that the elements in $\mathbf{b}$
add up to 0. }

{\ Let $\phi _{e}:=\Vert \widehat{\boldsymbol{\Sigma }}^{e}\Vert _{c2}$ be a
measurement of the noise level or contamination level of $\widehat{%
\boldsymbol{\Sigma }}^{\ast }$ in the model. Theorem \ref{thm:same_sign}
below characterizes the numerical properties of the sample estimator, where
the condition $\tau >\phi _{e}\left\Vert \mathbf{b}_{0}^{\ast }\right\Vert
_{\infty }/\sqrt{N}$ will be satisfied w.p.a.1 in the asymptotic analysis. }

\begin{theorem}
{\ \label{thm:same_sign} Suppose that $\tau >\phi _{e} \left\Vert \mathbf{b}%
_{0}^{\ast }\right\Vert _{\infty }/\sqrt{N}$. Then }

\begin{enumerate}
\item {\ $\left\Vert \widehat{\mathbf{w}}\right\Vert _{2}\leq\left\Vert 
\mathbf{w}^{*}\right\Vert _{2}\leq\left\Vert \mathbf{b}_{0}^{*}\right\Vert
_{\infty}/\sqrt{N}$; }

\item {\ $\widehat{\boldsymbol{\alpha}}\in\mathbb{S}^{\mathrm{all}} $. }
\end{enumerate}
\end{theorem}

{\ \noindent \textbf{Proof of Theorem \ref{thm:same_sign}}. \textbf{Part (a)}%
. Substituting $\left( \mathbf{w}^{\ast },\gamma ^{\ast }\right) $ into the
constraint in (\ref{eq:relax}), we obtain 
\begin{align}
\left\Vert \widehat{\boldsymbol{\Sigma }}\mathbf{w}^{\ast }+\gamma ^{\ast }%
\boldsymbol{1}_{N}\right\Vert _{\infty }& =\Vert \widehat{\boldsymbol{\Sigma 
}}^{\ast }\mathbf{w}^{\ast }+\widehat{\boldsymbol{\Sigma }}^{e}\mathbf{w}%
^{\ast }+\gamma ^{\ast }\boldsymbol{1}_{N}\Vert _{\infty }=\Vert \widehat{%
\boldsymbol{\Sigma }}^{e}\mathbf{w}^{\ast }\Vert _{\infty }  \notag \\
& =\max_{i}\Vert \widehat{\boldsymbol{\Sigma }}_{i\cdot }^{e}\mathbf{w}%
^{\ast }\Vert _{\infty }\leq \Vert \widehat{\boldsymbol{\Sigma }}^{e}\Vert
_{c2}\Vert \mathbf{w}^{\ast }\Vert _{2}\leq \phi _{e}\Vert \mathbf{b}%
_{0}^{\ast }\Vert _{\infty }/\sqrt{N}.  \label{eq:Sigma_w}
\end{align}%
where the second equality follows by the KKT condition $\widehat{\boldsymbol{%
\Sigma }}^{\ast }\mathbf{w}^{\ast }+\gamma ^{\ast }\boldsymbol{1}_{N}=0$.
The presumption $\phi _{e}\left\Vert \mathbf{b}_{0}^{\ast }\right\Vert
_{\infty }/\sqrt{N}<\tau $ in the statement makes sure that $\Vert \widehat{%
\boldsymbol{\Sigma }}\mathbf{w}^{\ast }+\gamma ^{\ast }\boldsymbol{1}%
_{N}\Vert _{\infty }<\tau $ holds with strict inequality. This strict
inequality means that $\left( \mathbf{w}^{\ast },\gamma ^{\ast }\right) $
lies in the interior of the feasible set of (\ref{eq:relax}). Because $%
\widehat{\mathbf{w}}$ is the minimizer of the problem in (\ref{eq:relax}),
its $\ell _{2}$-norm is no greater than any other feasible solution. Thus $%
\left\Vert \widehat{\mathbf{w}}\right\Vert _{2}\leq \left\Vert \mathbf{w}%
^{\ast }\right\Vert _{2}$ and furthermore $\left\Vert \mathbf{w}^{\ast
}\right\Vert _{2}$ is bounded by $\left\Vert \mathbf{b}_{0}^{\ast
}\right\Vert _{\infty }/\sqrt{N}$ by Lemma \ref{lem:w0}. }

{\ \textbf{Part (b)}. The Lagrangian of (\ref{eq:sim_dual}) can be written
as 
\begin{equation*}
\mathcal{L}\left( \boldsymbol{\alpha },\gamma \right) =\frac{1}{2}%
\boldsymbol{\alpha }^{\prime }\widehat{\mathbf{A}}^{\prime }\widehat{\mathbf{%
A}}\boldsymbol{\alpha }+\frac{1}{N}\boldsymbol{1}_{N}^{\prime }\widehat{%
\boldsymbol{\Sigma }}\boldsymbol{\alpha }+\tau \left\Vert \boldsymbol{\alpha 
}\right\Vert _{1}+\gamma \boldsymbol{1}_{N}^{\prime }\boldsymbol{\alpha -}%
\frac{1}{2N},
\end{equation*}%
where $\gamma $ is the Lagrangian multiplier for the constraint $\boldsymbol{%
1}_{N}^{\prime }\boldsymbol{\alpha }=0$. Consider the subgradient of $%
\mathcal{L}\left( \widehat{\boldsymbol{\alpha }},\widehat{\gamma }\right) $
with respect to $\alpha _{i}$ for any $i\in \lbrack N]$, where $\left( 
\widehat{\boldsymbol{\alpha }},\widehat{\gamma }\right) $ is the optimizer.
Noting that $\widehat{\mathbf{A}}^{\prime }\widehat{\mathbf{A}}=\widehat{%
\mathbf{A}}^{\prime }\widehat{\boldsymbol{\Sigma }}$ due to the fact that $%
\mathbf{\mathbf{I}}_{N}-N^{-1}\boldsymbol{1}_{N}\boldsymbol{1}_{N}^{\prime }$
is a projection matrix, the KKT conditions imply that 
\begin{equation*}
\left\vert \widehat{\boldsymbol{\alpha }}^{\prime }\widehat{\mathbf{A}}%
^{\prime }\widehat{\mathbf{A}}_{\cdot i}+\frac{1}{N}\boldsymbol{1}%
_{N}^{\prime }\widehat{\boldsymbol{\Sigma }}_{\cdot i}+\widehat{\gamma }%
\right\vert =\left\vert (\widehat{\mathbf{A}}\widehat{\boldsymbol{\alpha }}+%
\frac{1}{N}\boldsymbol{1}_{N})^{\prime }\widehat{\boldsymbol{\Sigma }}%
_{\cdot i}+\widehat{\gamma }\right\vert =\left\vert \widehat{\mathbf{w}}%
^{\prime }\widehat{\boldsymbol{\Sigma }}_{\cdot i}+\widehat{\gamma }%
\right\vert \leq \tau \ \text{for all }i\in \lbrack N]
\end{equation*}%
and furthermore 
\begin{equation}
\widehat{\mathbf{w}}^{\prime }\widehat{\boldsymbol{\Sigma }}_{\cdot i}+%
\widehat{\gamma }=\tau \mathrm{sign}\left( \widehat{\alpha }_{i}\right) \ 
\text{for all }\widehat{\alpha }_{i}\neq 0.  \label{eq:same_sign_1}
\end{equation}
}

{\ Suppose $\widehat{\boldsymbol{\alpha}}\notin\mathbb{S}^{\mathrm{all}}$.
Without loss of generality, let $\widehat{\alpha}_{i}>0$ and $\widehat{\alpha%
}_{j}<0$ for some $i,j\in\mathcal{G}_{k}$, $i\neq j$. (\ref{eq:same_sign_1})
indicates $\widehat{\mathbf{w}}^{\prime}\widehat{\boldsymbol{\Sigma}}_{\cdot
i}+\widehat{\gamma}=\tau$ and $\widehat{\mathbf{w}}^{\prime}\widehat{%
\boldsymbol{\Sigma}}_{\cdot j}+\widehat{\gamma}=-\tau$. Subtracting these
two equations on both sides yields 
\begin{align}
2\tau & =\left\vert \widehat{\mathbf{w}}^{\prime}(\widehat{\boldsymbol{\Sigma%
}}_{\cdot i}-\widehat{\boldsymbol{\Sigma}}_{\cdot j})\right\vert =\left\vert 
\widehat{\mathbf{w}}^{\prime}[(\widehat{\boldsymbol{\Sigma}}_{\cdot i}^{e}+%
\widehat{\boldsymbol{\Sigma}}_{\cdot i}^{\ast})-(\widehat{\boldsymbol{\Sigma}%
}_{\cdot j}^{e}+\widehat{\boldsymbol{\Sigma}}_{\cdot j}^{\ast})]\right\vert
=\left\vert \widehat{\mathbf{w}}^{\prime}(\widehat{\boldsymbol{\Sigma}}%
_{\cdot i}^{e}-\widehat{\boldsymbol{\Sigma}}_{\cdot j}^{e})\right\vert 
\notag \\
& \leq\Vert\widehat{\boldsymbol{\Sigma}}_{\cdot i}^{e}-\widehat{\boldsymbol{%
\Sigma}}_{\cdot j}^{e}\Vert_{2}\left\Vert \widehat{\mathbf{w}}\right\Vert
_{2}\leq2\Vert\widehat{\boldsymbol{\Sigma}}^{e}\Vert_{c2}\left\Vert \widehat{%
\mathbf{w}}\right\Vert _{2}\leq2\phi_{e}\left\Vert \mathbf{b}%
_{0}^{\ast}\right\Vert _{\infty}/\sqrt{N},  \label{eq:violation}
\end{align}
where the third equality holds as $\widehat{\boldsymbol{\Sigma}}_{\cdot
i}^{\ast}=\widehat{\boldsymbol{\Sigma}}_{\cdot j}^{\ast}$ for $i$ and $j$ in
the same group $k$, and the last inequality by Part (a) which bounds $%
\left\Vert \widehat{\mathbf{w}}\right\Vert _{2}$. The above inequality (\ref%
{eq:violation}) violates the presumption $\tau>\phi_{e}\left\Vert \mathbf{b}%
_{0}^{\ast}\right\Vert _{\infty}/\sqrt{N}$. We thus conclude $\widehat{%
\boldsymbol{\alpha}}\in\mathbb{S}^{\mathrm{all}}$. \hfill ${\ \blacksquare}$ 
}

{\ Theorem \ref{thm:same_sign}(a) sets an upper bound for $\left\Vert 
\widehat{\mathbf{w}}\right\Vert _{2}$, which is used in establishing part
(b). If the ratio between the tolerance $\tau $ and the noise level $\phi
_{e}$ is sufficiently large in that it is larger than $\left\Vert \mathbf{b}%
_{0}^{\ast }\right\Vert _{\infty }/\sqrt{N} $, the estimator $\widehat{%
\boldsymbol{\alpha }}$ must be of similar sign within each group. This
result is proved by exploiting the KKT conditions associated with the
Lagrangian of (\ref{eq:sim_dual}). The intuition is that when the specified $%
\tau $ is large, for any $i,j\in \mathcal{G}_{k},$ the column-wise
difference in the noise, i.e., $\widehat{\boldsymbol{\Sigma }}_{\cdot i}^{e}-%
\widehat{\boldsymbol{\Sigma }}_{\cdot j}^{e},$ is unable to push the two
associated KKT conditions to be satisfied simultaneously for the pair of $%
\widehat{\alpha }_{i}$ and $\widehat{\alpha }_{j}$ of opposite signs. }

\begin{remark}
{\ Theorem \ref{thm:same_sign}(b) reminds us of the grouping effect of
elastic net \citep{Zou_Hastie2005}. A regression method exhibits the
grouping effect if the regression coefficients of a group of highly
correlated regressors in the design matrix $\mathbf{X}$ tend to be equal (up
to a change of sign if negatively correlated). It is well-known that while
Lasso yields sparse solutions in many cases, it does not have the grouping
effect. In contrast, the elastic net penalty, as a convex combination of the
Lasso ($\ell _{1}$) and ridge ($\ell _{2}$) penalties, encourages the
grouping effect and has the advantage of including highly correlated
variables automatically in the group. }
\end{remark}

{\ Consider the following primal problem with $\widehat{\boldsymbol{\Sigma }}%
^{\ast }$: 
\begin{equation}  \label{eq:oracle_primal}
\min_{\left( \mathbf{w},\gamma \right) \in \mathbb{R}^{N+1}}\frac{1}{2}%
\left\Vert \mathbf{w}\right\Vert _{2}^{2}\text{\ \ subject to \ }\mathbf{w}%
^{\prime }\boldsymbol{1}_{N}=1\text{ and } \Vert \widehat{\boldsymbol{\Sigma 
}}^{\ast } \mathbf{w}+\gamma \mathbf{1}_{N} \Vert _{\infty } \leq \tau
\end{equation}
and we denote its solution as $\mathbf{w}_{\tau}^{\ast}$. Its dual is 
\begin{equation}
\min_{\boldsymbol{\alpha }\in \mathbb{R}^{N}}\left\{ \frac{1}{2}\boldsymbol{%
\alpha }^{\prime }\widehat{\mathbf{A}}^{\ast \prime }\widehat{\mathbf{A}}%
^{\ast }\boldsymbol{\alpha }+\frac{1}{N}\boldsymbol{1}_{N}^{\prime }\widehat{%
\boldsymbol{\Sigma }}^{\ast }\boldsymbol{\alpha }+\tau \left\Vert 
\boldsymbol{\alpha }\right\Vert _{1}-\frac{1}{2N}\right\} \text{ \ \ subject
to \ }\boldsymbol{1}_{N}^{\prime }\boldsymbol{\alpha }=0,
\label{eq:oracle_dual}
\end{equation}%
where $\widehat{\mathbf{A}}^{\ast }=\left( \mathbf{\mathbf{I}}_{N}-N^{-1}%
\boldsymbol{1}_{N}\boldsymbol{1}_{N}^{\prime }\right) \widehat{\boldsymbol{%
\Sigma }}^{\ast }$. For any $\boldsymbol{\alpha }^{\ast }=\boldsymbol{\alpha 
}_{\tau }^{\ast }$ that solves (\ref{eq:oracle_dual}), Lemma \ref{lem:dual}
implies that the solution to (\ref{eq:oracle_dual}) is not unique, and the
unique $\mathbf{w}_{\tau }^{\ast }$ and the non-unique $\boldsymbol{\alpha }%
_{\tau }^{\ast }$ are connected via 
\begin{equation}
\mathbf{w}_{\tau }^{\ast }=\widehat{\mathbf{A}}^{\ast }\boldsymbol{\alpha }%
_{\tau }^{\ast }+\frac{\boldsymbol{1}_{N}}{N}.  \label{eq:w_tau}
\end{equation}
The oracle problem (\ref{eq:oracle_primal_0}) is a special case of %
\eqref{eq:oracle_primal} when $\tau = 0$. }

{\ To develop the counterpart of Lemma \ref{lem:w0} for the dual problem, we
need some extra notations. For a generic $N$-vector $\boldsymbol{\alpha }%
=(\alpha_{i})_{i\in [N]}$, denote $a_{k}=\sum_{i\in \mathcal{G}_{k}}\alpha
_{i}$ as the $k$-th within-group summation of $\boldsymbol{\alpha }$. Let $%
\mathbf{a}=\left( a_{k}\right)_{k\in[K]}$. Let $\widehat{\mathbf{A}}^{%
\mathrm{co}}=\mathbf{R}^{1/2}\left( \mathbf{\mathbf{I}}_{K}-\boldsymbol{1}%
_{K}\mathbf{r}^{\prime }\right) \widehat{\boldsymbol{\Sigma }}^{\mathrm{co}}$%
, where $\mathbf{R}=\mathrm{diag}$$\left( \mathbf{r}\right) $ is the $%
K\times K$ diagonal matrix that stacks the elements of $\mathbf{r}$ along
the diagonal line. Note that $\widehat{\mathbf{A}}^{\mathrm{co}}$ is the 
\textit{weighted demeaned core} of $\widehat{\mathbf{A}}^{\ast }$ with the
weights depending on the relative group size $\mathbf{r}$. The following
lemma characterizes the features of $\boldsymbol{\alpha }_{\tau }^{\ast }.$ }

\begin{lemma}
{\ \label{lem:core_dual} }

\begin{enumerate}
\item {\ If $\tau >0$, any solution $\boldsymbol{\alpha }^{\ast }=%
\boldsymbol{\alpha }_{\tau }^{\ast }$ to (\ref{eq:oracle_dual}) must satisfy 
$\boldsymbol{\alpha }^{\ast }\in \mathbb{S}^{\mathrm{all}}$. }

\item {\ The low dimensional core dual problem for (\ref{eq:oracle_dual}) is 
\begin{equation}
\min_{\mathbf{a}\in \mathbb{R}^{K}}\left\{ \frac{N}{2}\boldsymbol{\mathbf{a}}%
^{\prime }\widehat{\mathbf{A}}^{\mathrm{co}\prime }\widehat{\mathbf{A}}^{%
\mathrm{co}}\mathbf{a}+\mathbf{r}^{\prime }\widehat{\boldsymbol{\Sigma }}^{%
\mathrm{co}}\mathbf{a}+\tau \left\Vert \mathbf{a}\right\Vert _{1}-\frac{1}{2N%
}\right\} \text{ \ \ subject to \ }\boldsymbol{1}_{K}^{\prime }\mathbf{a}=0.
\label{eq:core_dual}
\end{equation}
}

\item {\ In the special case of $\tau =0$, a solution $\mathbf{a}_{0}^{\ast
}:=\mathbf{a}_{\tau =0}^{\ast }$ to (\ref{eq:core_dual}) is 
\begin{equation}
\mathbf{a}_{0}^{\ast }=N^{-1}(\tilde{\mathbf{A}}^{\mathrm{co}\prime }\tilde{%
\mathbf{A}}^{\mathrm{co}})^{-1}\tilde{\mathbf{A}}^{\mathrm{co}\prime }%
\begin{pmatrix}
\left( \mathbf{b}_{0}^{\ast }\circ \mathbf{r}-\mathbf{r}\right) ^{\prime }%
\mathbf{R}^{-1/2} & 0%
\end{pmatrix}%
^{\prime },  \label{eq:a_0star}
\end{equation}%
whereas $\tilde{\mathbf{A}}^{\mathrm{co}}:= (\widehat{\mathbf{A}}^{\mathrm{co%
}\prime }, \mathbf{1}_{K})^{\prime }$ is a $\left( K+1\right) \times K$
matrix of full column rank. }
\end{enumerate}
\end{lemma}

{\ \noindent \textbf{Proof of Lemma \ref{lem:core_dual}. Part (a).} We
suppress the dependence of $\mathbf{\alpha}_{\tau }^{\ast }\ $and $%
\alpha_{\tau ,i}^{\ast }$ on $\tau $ when no confusion arises. We prove the
result by contradiction. Suppose that there exists some $k\in \left[ K\right]
$ such that the $k$th group of an optimizer $\boldsymbol{\alpha }^{\ast
}=(\alpha _{1}^{\ast },...,\alpha _{N}^{\ast })^{\prime }$ has elements of
opposite signs, viz, there are $i,j\in \mathcal{G}_{k}$ such that $\alpha
_{i}^{\ast }\alpha _{j}^{\ast }<0$. Construct an alternative estimator $%
\check{\boldsymbol{\alpha }}^{\ast }=(\check{\alpha}_{1}^{\ast },...,\check{%
\alpha}_{N}^{\ast })^{\prime },$ where $\check{\alpha}_{i}^{\ast
}=N_{k}^{-1}a_{k}^{\ast }\ \text{for }i\in \mathcal{G}_{k}\text{ and all }%
k\in \left[ K\right]$ %\end{equation*}%
and $a_{k}^{\ast }=\sum_{j\in \mathcal{G}_{k}}\alpha _{j}^{\ast }.$ }

{\ By construction, $\check{\boldsymbol{\alpha }}^{\ast }\in \mathbb{S}^{%
\mathrm{all}}$ as it replaces each $\alpha _{i}^{\ast }$ with $i\in \mathcal{%
G}_{k}$ by the groupwise average. It is obvious that $\boldsymbol{\alpha }%
^{\ast \prime }\widehat{\mathbf{A}}^{\ast \prime }\widehat{\mathbf{A}}^{\ast
}\boldsymbol{\alpha }=\check{\boldsymbol{\alpha }}^{\ast \prime }\widehat{%
\mathbf{A}}^{\ast \prime }\widehat{\mathbf{A}}^{\ast }\check{\boldsymbol{%
\alpha }}^{\ast }$ and $\boldsymbol{1}_{N}^{\prime }\widehat{\boldsymbol{%
\Sigma }}^{\ast }\boldsymbol{\alpha }=\boldsymbol{1}_{N}^{\prime }\widehat{%
\boldsymbol{\Sigma }}^{\ast }\check{\boldsymbol{\alpha }}^{\ast }$. On the
other hand, by the triangle inequality 
\begin{equation*}
\left\Vert \check{\boldsymbol{\alpha }}^{\ast }\right\Vert
_{1}=\sum_{i=1}^{N}\left\vert \check{\alpha}_{i}^{\ast }\right\vert
=\sum_{k=1}^{K}N_{k}\left\vert N_{k}^{-1}\sum_{j\in \mathcal{G}_{k}}\alpha
_{j}^{\ast }\right\vert <\sum_{i=1}^{N}\left\vert \alpha _{j}^{\ast
}\right\vert =\left\Vert \mathbf{\alpha }^{\ast }\right\Vert _{1},
\end{equation*}%
where the strict inequality follows from the fact that the elements in $%
\{\alpha _{j}^{\ast },$ $j\in \mathcal{G}_{k}\}$ change signs for some $k\in %
\left[ K\right] .$ As a result, the objective function of the dual problem
in (\ref{eq:oracle_dual}) is strictly larger when evaluated at $\boldsymbol{%
\alpha }^{\ast }$ than that at $\check{\boldsymbol{\alpha }}^{\ast }$. This
contradicts the presumption that $\boldsymbol{\alpha }^{\ast }$ is an
optimizer of (\ref{eq:oracle_dual}). }

{\ \textbf{Part (b).} For any $\boldsymbol{\alpha }\in \mathbb{R}^{N}$, we
have 
\begin{equation}
\left\Vert \widehat{\mathbf{A}}^{\ast }\boldsymbol{\alpha }\right\Vert
_{2}^{2}=\boldsymbol{\alpha }^{\prime }\widehat{\boldsymbol{\Sigma }}^{\ast
}\left( \mathbf{\mathbf{I}}_{N}-\frac{1}{N}\boldsymbol{1}_{N}\boldsymbol{1}%
_{N}^{\prime }\right) \widehat{\boldsymbol{\Sigma }}^{\ast }\boldsymbol{%
\alpha }=\boldsymbol{\alpha }^{\prime }\widehat{\boldsymbol{\Sigma }}^{\ast }%
\widehat{\boldsymbol{\Sigma }}^{\ast }\boldsymbol{\alpha }-\frac{1}{N}\left( 
\boldsymbol{1}_{N}^{\prime }\widehat{\boldsymbol{\Sigma }}^{\ast }%
\boldsymbol{\alpha }\right) ^{2}.  \label{eq:verify1-1}
\end{equation}%
The group structure in $\widehat{\boldsymbol{\Sigma }}^{\ast }$ implies $%
\widehat{\boldsymbol{\Sigma }}^{\ast }\boldsymbol{\alpha }=\left( \widehat{%
\boldsymbol{\Sigma }}_{1\cdot }^{\mathrm{co}}\mathbf{a}\cdot \boldsymbol{1}%
_{N_{1}}^{\prime },\ldots ,\widehat{\boldsymbol{\Sigma }}_{K\cdot }^{\mathrm{%
co}}\mathbf{a}\cdot \boldsymbol{1}_{N_{K}}^{\prime }\right) ^{\prime }$.
Therefore, we have 
\begin{align}  \label{eq:verify1-2}
\boldsymbol{\alpha }^{\prime }\widehat{\boldsymbol{\Sigma }}^{\ast }\widehat{%
\boldsymbol{\Sigma }}^{\ast }\boldsymbol{\alpha }&
=\sum_{k=1}^{K}N_{k}\left( \widehat{\boldsymbol{\Sigma }}_{k\cdot }^{\mathrm{%
co}}\mathbf{a}\right) ^{2}=N\sum_{k=1}^{K}r_{k}\left( \widehat{\boldsymbol{%
\Sigma }}_{k\cdot }^{\mathrm{co}}\mathbf{a}\right) ^{2}=N\mathbf{a}^{\prime }%
\widehat{\boldsymbol{\Sigma }}^{\mathrm{co}}\mathbf{R}\widehat{\boldsymbol{%
\Sigma }}^{\mathrm{co}}\mathbf{a,}  \notag \\
\boldsymbol{1}_{N}^{\prime }\widehat{\boldsymbol{\Sigma }}^{\ast }%
\boldsymbol{\alpha }& =\sum_{k=1}^{K}N_{k}\widehat{\boldsymbol{\Sigma }}%
_{k\cdot }^{\mathrm{co}}\mathbf{a}=N\sum_{k=1}^{K}r_{k}\widehat{\boldsymbol{%
\Sigma }}_{k\cdot }^{\mathrm{co}}\mathbf{a}=N\mathbf{r}^{\prime }\widehat{%
\boldsymbol{\Sigma }}^{\mathrm{co}}\mathbf{a}.
\end{align}%
Substituting the above two equations into (\ref{eq:verify1-1}) yields 
\begin{equation*}
\left\Vert \widehat{\mathbf{A}}^{\ast }\boldsymbol{\alpha }\right\Vert
_{2}^{2}=N\mathbf{a}^{\prime }\widehat{\boldsymbol{\Sigma }}^{\mathrm{co}}%
\mathbf{R}\widehat{\boldsymbol{\Sigma }}^{\mathrm{co}}\mathbf{a}-N\left( 
\mathbf{r}^{\prime }\widehat{\boldsymbol{\Sigma }}^{\mathrm{co}}\mathbf{a}%
\right) ^{2}=N\mathbf{a}^{\prime }\widehat{\boldsymbol{\Sigma }}^{\mathrm{co}%
}\left( \mathbf{R}-\mathbf{rr}^{\prime }\right) \widehat{\boldsymbol{\Sigma }%
}^{\mathrm{co}}\mathbf{a}.
\end{equation*}%
On the other hand, noticing $\mathbf{R}\boldsymbol{1}_{K}=\mathbf{r}$ and $%
\boldsymbol{1}_{K}^{\prime }\mathbf{R}\boldsymbol{1}_{K}=\boldsymbol{1}%
_{K}^{\prime }\mathbf{r}=1$, we have 
\begin{align*}
N\left\Vert \widehat{\mathbf{A}}^{\mathrm{co}}\mathbf{a}\right\Vert
_{2}^{2}& =N\mathbf{a}^{\prime }\widehat{\boldsymbol{\Sigma }}^{\mathrm{co}%
}\left( \mathbf{\mathbf{I}}_{K}-\mathbf{r}\boldsymbol{1}_{K}^{\prime
}\right) \mathbf{R}\left( \mathbf{\mathbf{I}}_{K}-\boldsymbol{1}_{K}\mathbf{r%
}^{\prime }\right) \widehat{\boldsymbol{\Sigma }}^{\mathrm{co}}\mathbf{a} \\
& =N\mathbf{a}^{\prime }\widehat{\boldsymbol{\Sigma }}^{\mathrm{co}}\left( 
\mathbf{R}-\mathbf{R}\boldsymbol{1}_{K}\mathbf{r}^{\prime }-\mathbf{r}%
\boldsymbol{1}_{K}^{\prime }\mathbf{R}+\mathbf{r}\boldsymbol{1}_{K}^{\prime }%
\mathbf{R}\boldsymbol{1}_{K}\mathbf{r}^{\prime }\right) \widehat{\boldsymbol{%
\Sigma }}^{\mathrm{co}}\mathbf{a} \\
& =N\mathbf{a}^{\prime }\widehat{\boldsymbol{\Sigma }}^{\mathrm{co}}\left( 
\mathbf{R}-\mathbf{rr}^{\prime }\right) \widehat{\boldsymbol{\Sigma }}^{%
\mathrm{co}}\mathbf{a}.
\end{align*}%
Therefore we obtain 
\begin{equation}
\left\Vert \widehat{\mathbf{A}}^{\ast }\boldsymbol{\alpha }\right\Vert
_{2}^{2}=N\left\Vert \widehat{\mathbf{A}}^{\mathrm{co}}\mathbf{a}\right\Vert
_{2}^{2}=N\boldsymbol{\mathbf{a}}^{\prime }\widehat{\mathbf{A}}^{\mathrm{co}%
\prime }\widehat{\mathbf{A}}^{\mathrm{co}}\mathbf{a}.  \label{eq:verify1-3}
\end{equation}
}

{\ In the objective function (\ref{eq:oracle_dual}), by (\ref{eq:verify1-3})
we can use $N\boldsymbol{\mathbf{a}}^{\prime }\widehat{\mathbf{A}}^{\mathrm{%
co}\prime }\widehat{\mathbf{A}}^{\mathrm{co}}\mathbf{a}$ to replace $%
\boldsymbol{\alpha }^{\prime }\widehat{\mathbf{A}}^{\ast \prime }\widehat{%
\mathbf{A}}^{\ast }\boldsymbol{\alpha }$, by (\ref{eq:verify1-2}) we can use 
$\mathbf{r}^{\prime }\widehat{\boldsymbol{\Sigma }}^{\mathrm{co}}\mathbf{a}$
to replace $\boldsymbol{1}_{N}^{\prime }\widehat{\boldsymbol{\Sigma }}^{\ast
}\boldsymbol{\alpha }/N$, and by Part (a) its solution must be of similar
sign in that $\Vert \boldsymbol{\alpha }^{\ast }\Vert _{1}=\Vert \mathbf{a}%
^{\ast }\Vert _{1}$. Consequently, the problem in (\ref{eq:oracle_dual}) is
equivalent to that in (\ref{eq:core_dual}). }

{\ \textbf{Part (c)}. We first show $\tilde{\mathbf{A}}^{\mathrm{co}}$ is of
full column rank. The first $K$ rows of $\tilde{\mathbf{A}}^{\mathrm{co}}$
is $\widehat{\mathbf{A}}^{\mathrm{co}}=\mathbf{R}^{1/2}\left( \mathbf{%
\mathbf{I}}_{K}-\boldsymbol{1}_{K}\mathbf{r}^{\prime }\right) \widehat{%
\boldsymbol{\Sigma }}^{\mathrm{co}}$. Notice that $(\mathbf{\mathbf{I}}_{K}-%
\mathbf{r}\boldsymbol{1}_{K}^{\prime })$ is idempotent and $\mathbf{R}^{1/2}$
and $\widehat{\boldsymbol{\Sigma }}^{\mathrm{co}}$ are both of full rank, $%
\mathrm{rank}(\widehat{\mathbf{A}}^{\mathrm{co}})=\mathrm{rank}(\mathbf{%
\mathbf{I}}_{K}-\mathbf{r}\boldsymbol{1}_{K}^{\prime })=\mathrm{trace}(%
\mathbf{\mathbf{I}}_{K}-\mathbf{r}\boldsymbol{1}_{K}^{\prime })=K-1$. In
other words, $\widehat{\mathbf{A}}^{\mathrm{co}}$ is rank deficient and its
null space is one-dimensional. The null space of $\widehat{\mathbf{A}}^{%
\mathrm{co}}$ is %\begin{equation*}
$\mathrm{ker}(\widehat{\mathbf{A}}^{\mathrm{co}})=\left\{ c(\widehat{%
\boldsymbol{\Sigma }}^{\mathrm{co}})^{-1}\boldsymbol{1}_{K}:c\in \mathbb{R}%
\backslash \left\{ 0\right\} \right\} ,$ %\end{equation*}%
as $\widehat{\mathbf{A}}^{\mathrm{co}}(\widehat{\boldsymbol{\Sigma }}^{%
\mathrm{co}})^{-1}\boldsymbol{1}_{K}=\mathbf{R}^{1/2}\left( \mathbf{\mathbf{I%
}}_{K}-\boldsymbol{1}_{K}\mathbf{r}^{\prime }\right) \boldsymbol{1}_{K}=%
\boldsymbol{0}_{N}$. Moreover, since $\boldsymbol{1}_{K}^{\prime }(\widehat{%
\boldsymbol{\Sigma }}^{\mathrm{co}})^{-1}\boldsymbol{1}_{K}\neq 0$, $(%
\widehat{\boldsymbol{\Sigma }}^{\mathrm{co}})^{-1}\boldsymbol{1}_{K}$ is not
in the null space of $\boldsymbol{1}_{K}.$ In other words, $\mathrm{ker}(%
\widehat{\mathbf{A}}^{\mathrm{co}})\cap \mathrm{ker}(\boldsymbol{1}%
_{K}^{\prime })$ is empty and we must have rank$(\tilde{\mathbf{A}}^{\mathrm{%
co}})=$rank$(\widehat{\mathbf{A}}^{\mathrm{co}})+$rank$(\boldsymbol{1}%
_{K}^{\prime })=(K-1)+1=K.$ }

{\ Setting $\tau =0$ in (\ref{eq:w_tau}), we have %\begin{equation}
$\widehat{\mathbf{A}}^{\ast }\boldsymbol{\alpha }_{0}^{\ast }=\mathbf{w}%
_{0}^{\ast } - N^{-1} \boldsymbol{1}_{N}.$ %\label{eq:w_0}
%\end{equation}%
Premultiplying both sides of the above equation by the $K\times N$ block
diagonal matrix $\mathrm{diag}(r_{1}^{-1/2}\boldsymbol{1}_{N_{1}}^{\prime
},...,$ $r_{K}^{-1/2}\boldsymbol{1}_{N_{K}}^{\prime }),$ we obtain 
%\begin{equation}
$N\widehat{\mathbf{A}}^{\mathrm{co}}\mathbf{a}_{0}^{\ast }=\mathbf{R}%
^{-1/2}\left( \mathbf{b}_{0}^{\ast }\circ \mathbf{r}-\mathbf{r}\right).$ 
%\label{eq:b01}
%\end{equation}%
As $\widehat{\mathbf{A}}^{\mathrm{co}}$ is a submatrix of $\tilde{\mathbf{A}}%
^{\mathrm{co}}$, the above equation implies 
\begin{equation*}
N\tilde{\mathbf{A}}^{\mathrm{co}}\mathbf{a}_{0}^{\ast }=\left( 
\begin{array}{c}
N\widehat{\mathbf{A}}^{\mathrm{co}}\mathbf{a}_{0}^{\ast } \\ 
N\mathbf{1}_{K}^{\prime }\mathbf{a}_{0}^{\ast }%
\end{array}%
\right) =\left( 
\begin{array}{c}
\mathbf{R}^{-1/2}\left( \mathbf{b}_{0}^{\ast }\circ \mathbf{r}-\mathbf{r}%
\right) \\ 
0%
\end{array}%
\right) ,
\end{equation*}%
where we use the restriction $N\boldsymbol{1}_{K}^{^{\prime }}\mathbf{a}%
_{0}^{\ast }=0.$ Since $\tilde{\mathbf{A}}^{\mathrm{co}}$ is of full column
rank, we explicitly solve $\mathbf{a}_{0}^{\ast }=\left( \tilde{\mathbf{A}}^{%
\mathrm{co}\prime }\tilde{\mathbf{A}}^{\mathrm{co}}\right) ^{-1}\tilde{%
\mathbf{A}}^{\mathrm{co}\prime }\tilde{\mathbf{b}}^{\mathrm{co}}/N.$ \hfill $%
{\blacksquare }$ }

{\ Lemma \ref{lem:core_dual}(a) shows that the $\ell _{1}$-norm penalization
in (\ref{eq:oracle_dual}) precludes opposite signs of the estimates $%
\boldsymbol{\alpha }_{\tau }^{\ast }$ within a group, which implies $\Vert 
\boldsymbol{\alpha }_{\tau }^{\ast }\Vert _{1}=\Vert \mathbf{a}_{\tau
}^{\ast }\Vert _{1}$ for any $\tau >0$. Lemma \ref{lem:core_dual}(b) reduces
the high dimensional oracle dual problem in $\boldsymbol{\alpha }\in \mathbb{%
R}^{N}$ to the low dimensional oracle dual one in $\boldsymbol{a}\in \mathbb{%
R}^{K}$. Lemma \ref{lem:core_dual}(c) is the counterpart of (\ref{eq:b0star}%
) for the dual, which involves the \textit{augmented (by a row of 1's)
weighted demeaned core} $\tilde{\mathbf{A}}^{\mathrm{co}}$. A numerical
lower bound and a stochastic lower bound of $\tilde{\mathbf{A}}^{\mathrm{co}%
} $ will be established in Lemma \ref{lem:res_eigen}(b) and Lemma \ref%
{lem:phi_e}(a).% in Section \ref{subsec:Lemmas-Prepared} of the Appendix.
}

\subsection{Convergence of the Dual Problem}

The convergence of the weight $\widehat{\mathbf{w}}$ will be established via
the convergence of $\widehat{\boldsymbol{\alpha }}$ in view of their
connection in \eqref{eq:w_A_alpha}. We start with the dual problem (\ref%
{eq:sim_dual}) under Assumption \ref{assu:idiosyn}. There are multiple
solutions to the oracle dual in (\ref{eq:oracle_dual}) due to the rank
deficiency of $\boldsymbol{\Sigma }^{\ast }$. But if we want to establish
convergence in probability, we must declare a target to which the estimator
will converge. We construct such a desirable $\boldsymbol{\alpha }_{0}^{\ast
}$ in (\ref{eq:alpha_hat_star}) below, denoted as $\widehat{\boldsymbol{%
\alpha }}^{\ast }$ where the \textquotedblleft hat\textquotedblright\
signifies its dependence on the realization of $\widehat{\boldsymbol{\alpha }%
}$ and \textquotedblleft star\textquotedblright\ indicates its validity as
an oracle estimator. Define $\widehat{\boldsymbol{\alpha }}^{\ast }=(%
\widehat{\boldsymbol{\alpha }}_{\mathcal{G}_{k}}^{\ast })_{k\in \lbrack K]}$%
, where 
\begin{equation}
\widehat{\boldsymbol{\alpha }}_{\mathcal{G}_{k}}^{\ast }=a_{0k}^{\ast
}\left( \frac{\widehat{\boldsymbol{\alpha }}_{\mathcal{G}_{k}}}{\widehat{a}%
_{k}}\cdot 1\left\{ \widehat{a}_{k}a_{0k}^{\ast }>0\right\} +\frac{%
\boldsymbol{1}_{N_{k}}}{N_{k}}\cdot 1\left\{ \widehat{a}_{k}a_{0k}^{\ast
}\leq 0\right\} \right) ,  \label{eq:alpha_hat_star}
\end{equation}%
where $\widehat{a}_{k}=\sum_{i\in \mathcal{G}_{k}}\widehat{\alpha }_{i}$ is
the sum of the $\widehat{\alpha }_{i}$ in the $k$-th group and $1\left\{
\cdot \right\} $ is the usual indicator function. The above $\widehat{%
\mathbf{\alpha }}_{\mathcal{G}_{k}}^{\ast }$ is designed such that the $k$th
oracle group weight $a_{0k}^{\ast }$ is distributed across the $k$th group
members proportionally to $\widehat{\boldsymbol{\alpha }}_{\mathcal{G}_{k}}/%
\widehat{a}_{k}$ when $\widehat{a}_{k}$ and $a_{0k}^{\ast }$ share the same
sign, whereas $a_{0k}^{\ast }$ is distributed equally across the $k$th group
members when they take opposite signs. When $\widehat{\boldsymbol{\alpha }}%
\in \tilde{\mathbb{S}}^{\mathrm{all}}$, which holds w.p.a.1 in view of
Theorem \ref{thm:same_sign} and Lemma \ref{lem:phi_e}(b), it is easy to
verify that 
\begin{equation}
(i)\text{ }\widehat{\boldsymbol{\alpha }}^{\ast }\in \tilde{\mathbb{S}}^{%
\mathrm{all}},\ (ii)\text{ }\left\Vert \boldsymbol{\alpha }_{0}^{\ast
}\right\Vert _{1}=\left\Vert \widehat{\boldsymbol{\alpha }}^{\ast
}\right\Vert _{1}\ \text{ and }(iii)\text{ }\widehat{\boldsymbol{\alpha }}-%
\widehat{\boldsymbol{\alpha }}^{\ast }\in \tilde{\mathbb{S}}^{\mathrm{all}}.
\label{eq:alpha_star}
\end{equation}%
For example, ($i$) in (\ref{eq:alpha_star}) holds because by construction $%
\widehat{\boldsymbol{\alpha }}^{\ast }\in \mathbb{S}^{\mathrm{all}}$ as long
as $\widehat{\boldsymbol{\alpha }}\in \tilde{\mathbb{S}}^{\mathrm{all}},$
and $\mathbf{1}_{N}^{\prime }\widehat{\boldsymbol{\alpha }}^{\ast
}=\sum_{k=1}^{K}\mathbf{1}_{N_{k}}^{\prime }\widehat{\boldsymbol{\alpha }}_{%
\mathcal{G}_{k}}^{\ast }=\sum_{k=1}^{K}a_{0k}^{\ast }=\mathbf{1}_{K}^{\prime
}\mathbf{a}_{0}^{\ast }=0.$ The following theorem shows that the solution to
the $\ell _{1}$-penalization problem (\ref{eq:sim_dual}) is close to the
desirable oracle estimator $\widehat{\boldsymbol{\alpha }}^{\ast }$.

\begin{theorem}
\label{thm:alpha_converg} {\ Suppose that Assumptions \ref{assu:idiosyn} and %
\ref{assu:rate} hold. Then 
\begin{equation*}
\left\Vert \widehat{\boldsymbol{\alpha}}-\widehat{\boldsymbol{\alpha}}%
^{\ast}\right\Vert _{1}=O_{p}\left(N^{-1}K^{3}\tau\right)\ \mbox{and }\Vert 
\widehat{\mathbf{A}}\left(\widehat{\boldsymbol{\alpha}}-\widehat{\boldsymbol{%
\alpha}}^{\ast}\right)\Vert_{2} =O_{p}(N^{-1/2}K^{2}\tau).
\end{equation*}
}
\end{theorem}

Theorem \ref{thm:alpha_converg} is a key result that characterizes the
convergence rate of the high-dimensional parameter $\widehat{\boldsymbol{%
\alpha}}$ in the dual problem to its oracle group counterpart $\boldsymbol{%
\alpha}_{0}^{\ast}$, represented by the constructed unique solution $%
\widehat{\boldsymbol{\alpha}}^{\ast}$. Although our ultimate interest lies
in the weight estimate $\widehat{\mathbf{w}}$ in the primal problem, in
theoretical analysis we first work with $\widehat{\boldsymbol{\alpha}}$ in
the dual problem instead. This detour is taken because the dual is an $%
\ell_{1}$-penalized optimization which resembles Lasso. The intensive study
of Lasso in statistics and econometrics offers a set of inequalities
involving the $\ell_{1}$-norms of $\widehat{\boldsymbol{\alpha}}$, $\widehat{%
\boldsymbol{\alpha}}^{\ast}$ and their difference $\left(\widehat{%
\boldsymbol{\alpha}}-\widehat{\boldsymbol{\alpha}}^{\ast}\right)$ at our
disposal.

\begin{remark}
{\ All high dimensional estimation problems require certain notion of
sparsity to reduce dimensionality. It is helpful to compare our setting of
latent group structures with the sparse regression estimated by Lasso. For
Lasso estimation, the complexity of the problem is governed by the total
number of regressors %($p$ in Example \ref{exa:OLS})
while under sparsity those non-zero coefficients control the effective
number of parameters, which is assumed to be far fewer than the sample size.
For the $\ell _{1}$-penalized (\ref{eq:sim_dual}), the complexity is the
number of forecasters $N$ whereas under the group structures the number of
groups $K$ determines the effective number of parameters. }
\end{remark}

\begin{remark}
A critical technical step in proving the consistency of high-dimensional
Lasso problems is the \emph{compatibility condition} 
\citep[Ch
6.13]{buhlmann2011statistics} in conjunction with the \emph{restricted
eigenvalue condition} \citep{Bickel2009,belloni2012sparse}. The consistency
of Lasso requires that the correlation among the columns of the design
matrix should not be too strong; otherwise, various versions of restricted
eigenvalue conditions break down. In our paper, the correlation of
individuals in the same group are indeed strong. We must design a
compatibility condition tailored for the group structure, which is to be
establish in Lemma \ref{lem:res_eigen}(a) and the restricted eigenvalue to
be represented by $\phi _{A}:=1\wedge \phi _{\min }(\tilde{\mathbf{A}}^{%
\mathrm{co}\prime }\tilde{\mathbf{A}}^{\mathrm{co}})$. In particular,
instead of \emph{assuming} a lower bound for the restricted eigenvalues as
most high dimensional Lasso papers do, we \emph{derive} a finite sample
lower bound for $\phi _{A}$ in Lemma \ref{lem:res_eigen}(b) below as well as
its convergence rate in Lemma \ref{lem:phi_e}(a) under our latent group
structures and primitive Assumptions \ref{assu:idiosyn} and \ref{assu:rate}.
These developments are original contributions to the literature, though they
appear here in the appendix due to the technical nature.
\end{remark}

{\ We will first establish Lemmas \ref{lem:res_eigen} and \ref{lem:phi_e}
before we prove Theorem \ref{thm:alpha_converg}. Lemma \ref{lem:res_eigen}%
(a) provides a \textit{compatibility inequality} that links $\left\Vert 
\boldsymbol{\delta }\right\Vert _{1}$ and $\Vert \widehat{\mathbf{A}}%
\boldsymbol{\delta }\Vert _{2}^{2}$ for any $\boldsymbol{\delta }\in \tilde{%
\mathbb{S}}^{\mathrm{all}}$. Recall 
%that $\phi _{A}:=\phi _{\min }(\tilde{\mathbf{A}}^{\mathrm{co}\prime }%
%\tilde{\mathbf{A}}^{\mathrm{co}})\wedge 1$ and
$\underline{r}:=\min_{k\in \left[ K\right] }r_{k}$. }

\begin{lemma}
{\ \label{lem:res_eigen} }

\begin{enumerate}
\item {\ If $\phi_{e}\leq\frac{1}{2}\sqrt{N\phi_{A}/K}$, we have $\left\Vert 
\boldsymbol{\delta}\right\Vert _{1}\leq2\sqrt{K/(N\phi_{A})}\Vert\widehat{%
\mathbf{A}}\boldsymbol{\delta}\Vert_{2}$ for any $\boldsymbol{\delta}\in%
\tilde{\mathbb{S}}^{\mathrm{all}}$. }

\item {\ $\phi_{A}^{-1}\leq2\underline{r}^{-1}\phi_{\min}^{-2}(\widehat{%
\boldsymbol{\Sigma}}^{\mathrm{co}})+K^{-1}\phi_{\max}(\widehat{\boldsymbol{%
\Sigma}}^{\mathrm{co}})/\phi_{\min}(\widehat{\boldsymbol{\Sigma}}^{\mathrm{co%
}}).$ }

\item {\ $\left\Vert \mathbf{a}_{0}^{\ast}\right\Vert _{1}\leq N^{-1}\sqrt{%
K/\phi_{A}}\left(\left\Vert \mathbf{b}_{0}^{\ast}\right\Vert
_{\infty}+1\right)$. }
\end{enumerate}
\end{lemma}

\begin{remark}
{\ The constants $1/2$ and 2 in Lemma \ref{lem:res_eigen}(a) are not
important in the asymptotic analysis. It means if the magnitude of the
idiosyncratic shock, represented by $\phi_{e}$, is controlled by the order $%
\sqrt{N\phi_{A}/K}$, then the $\ell_{1}$-norm of $\boldsymbol{\delta}$ can
be controlled by the $\ell_{2}$-norm of $\Vert\widehat{\mathbf{A}}%
\boldsymbol{\delta}\Vert_{2}$ multiplied by a factor involving $K/\phi_{A}$,
which is the ratio between the number of groups $K$ and the square of the
minimum non-trivial singular value of the augmented weighted demeaned core $%
\tilde{\mathbf{A}}^{\mathrm{co}}$. In the proof of Lemma \ref{lem:res_eigen}%
, we introduce an original self-defined semi-norm (\ref{eq:semi-norm}) to
take advantage of the group pattern. Another necessary condition for Lasso
to achieve reasonable performance is that the $\ell _{1}$-norm of the true
coefficients cannot be too large. In our context, since Lemma \ref%
{lem:core_dual} implies $\left\Vert \boldsymbol{\alpha }^{\ast }\right\Vert
_{1}=\left\Vert \mathbf{a}^{\ast }\right\Vert _{1}$, Part (c) sets an upper
bound for the $\ell _{1}$-norm of the true coefficient value for (\ref%
{eq:oracle_dual}). }
\end{remark}

{\ \noindent \textbf{Proof of Lemma \ref{lem:res_eigen}. Part (a)}. For a
generic vector $\boldsymbol{\delta }\in \tilde{\mathbb{S}}^{\mathrm{all}}$,
we have 
\begin{equation}
\Vert \widehat{\mathbf{A}}\boldsymbol{\delta }\Vert _{2}\geq \Vert \widehat{%
\mathbf{A}}^{\ast }\boldsymbol{\delta }\Vert _{2}-\Vert \left( \mathbf{%
\mathbf{I}}_{N}-N^{-1}\boldsymbol{1}_{N}\boldsymbol{1}_{N}^{\prime }\right) 
\widehat{\boldsymbol{\Sigma }}^{e}\boldsymbol{\delta }\Vert _{2}\geq \Vert 
\widehat{\mathbf{A}}^{\ast }\boldsymbol{\delta }\Vert _{2}-\Vert \widehat{%
\boldsymbol{\Sigma }}^{e}\boldsymbol{\delta }\Vert _{2},
\label{eq:eig_ineq1}
\end{equation}%
where the first inequality holds by the the triangle inequality, and the
second follows because $(\mathbf{\mathbf{I}}_{N}-N^{-1}\boldsymbol{1}_{N}%
\boldsymbol{1}_{N}^{\prime })$ is a projection matrix. We will bound the two
terms on the right hand side. }

{\ To take advantage of the group structure to handle collinearity, we
introduce a novel groupwise semi-norm and establish a corresponding version
of compatibility condition. Let $d_{k}=\sum_{i\in \mathcal{G}_{k}}\delta
_{i} $ for $k\in \left[ K\right] $ and $\mathbf{d}=\left( d_{1},\ldots
,d_{K}\right) ^{\prime }$. Define a groupwise $\ell _{2} $ semi-norm $%
\left\Vert \cdot \right\Vert _{2\mathcal{G}}:\mathbb{R}^{N}\mapsto \mathbb{R}%
^{+}$ as 
\begin{equation}
\left\Vert \boldsymbol{\delta }\right\Vert _{2\mathcal{G}}=\left\Vert 
\mathbf{d}\right\Vert _{2}.  \label{eq:semi-norm}
\end{equation}%
The definition of the semi-norm depends on the true group membership, which
is infeasible in reality. We introduce this semi-norm only for theoretical
development. In the estimation we do not need to know the true group
membership. This semi-norm $\left\Vert \boldsymbol{\delta }\right\Vert _{2%
\mathcal{G}}$ allows $\left\Vert \boldsymbol{\delta }\right\Vert _{2\mathcal{%
G}}=0$ even if $\boldsymbol{\delta }\neq \boldsymbol{0}_{N}$, while it
remains homogeneous, sub-additive, and non-negative---all other desirable
properties of a norm. Moreover, if $\boldsymbol{\delta }\in \mathbb{S}^{%
\mathrm{all}}$ it is obvious 
\begin{equation}
\left\Vert \boldsymbol{\delta }\right\Vert _{1}=\sum_{k\in \left[ K\right] }%
\bigg|\sum_{i\in \mathcal{G}_{k}}\delta _{i}\bigg|=\sum_{k\in \left[ K\right]
}\left\vert d_{k}\right\vert \leq \sqrt{K}\left\Vert \boldsymbol{\delta }%
\right\Vert _{2\mathcal{G}}  \label{eq:delta_L1}
\end{equation}%
by either the Cauchy-Schwarz or Jensen's inequality. }

{\ For any $\boldsymbol{\delta }\in \text{$\tilde{\mathbb{S}}^{\mathrm{all}}$%
}$, we have 
\begin{align}
\Vert \widehat{\mathbf{A}}^{\ast }\boldsymbol{\delta }\Vert _{2}& =\sqrt{N}%
\Vert \widehat{\mathbf{A}}^{\mathrm{co}}\mathbf{d}\Vert _{2}=\sqrt{N}%
\left\Vert 
\begin{pmatrix}
\widehat{\mathbf{A}}^{\mathrm{co}}\mathbf{d} \\ 
0%
\end{pmatrix}%
\right\Vert _{2}=\sqrt{N}\Vert \tilde{\mathbf{A}}^{\mathrm{co}}\mathbf{d}%
\Vert _{2}  \notag \\
& \geq \sqrt{N\phi _{A}}\left\Vert \mathbf{d}\right\Vert _{2}=\sqrt{N\phi
_{A}}\left\Vert \boldsymbol{\delta }\right\Vert _{2\mathcal{G}}\geq \sqrt{%
\frac{N\phi _{A}}{K}}\left\Vert \boldsymbol{\delta }\right\Vert _{1},
\label{eq:bound1}
\end{align}%
where the first equality follows by (\ref{eq:verify1-3}), the third equality
by $\boldsymbol{1}_{K}^{\prime }\mathbf{d}=\boldsymbol{1}_{N}^{\prime }%
\boldsymbol{\delta }=0$, and the last inequality by (\ref{eq:delta_L1}). We
have found a lower bound for the first term on the right hand side of (\ref%
{eq:eig_ineq1}). For the second term on the right hand side of (\ref%
{eq:eig_ineq1}), we have 
\begin{equation}
\Vert \widehat{\boldsymbol{\Sigma }}^{e}\boldsymbol{\delta }\Vert _{2}\leq
\Vert \widehat{\boldsymbol{\Sigma }}||_{c2}\left\Vert \boldsymbol{\delta }%
\right\Vert _{1}\leq \phi _{e}\left\Vert \boldsymbol{\delta }\right\Vert _{1}
\label{eq:bound2}
\end{equation}%
by (\ref{eq:holder_CS1}) and the definition of $\phi _{e}$. Under the
presumption $\phi _{e}\leq \frac{1}{2}\sqrt{N\phi _{A}/K}$, (\ref%
{eq:eig_ineq1}), (\ref{eq:bound1}), and (\ref{eq:bound2}) together imply $%
\Vert \widehat{\mathbf{A}}\boldsymbol{\delta }\Vert _{2}\geq (\sqrt{N\phi
_{A}/K}-\phi _{e})\left\Vert \boldsymbol{\delta }\right\Vert _{1}\geq \frac{1%
}{2}\sqrt{N\phi _{A}/K}\left\Vert \boldsymbol{\delta }\right\Vert _{1}.$ 
%\end{equation*}%
Then the result in part (a) follows. }

{\ \noindent \textbf{Part (b)}. Notice 
\begin{eqnarray*}
\tilde{\mathbf{A}}^{\mathrm{co}\prime }\tilde{\mathbf{A}}^{\mathrm{co}} &=&%
\widehat{\mathbf{A}}^{\mathrm{co}\prime }\widehat{\mathbf{A}}^{\mathrm{co}}+%
\boldsymbol{1}_{K}\boldsymbol{1}_{K}^{\prime }=\widehat{\boldsymbol{\Sigma }}%
^{\mathrm{co}}\left( \mathbf{\mathbf{I}}_{K}-\mathbf{r}\cdot \boldsymbol{1}%
_{K}^{\prime }\right) \mathbf{R}\left( \mathbf{\mathbf{I}}_{K}-\boldsymbol{1}%
_{K}\cdot \mathbf{r}^{\prime }\right) \widehat{\boldsymbol{\Sigma }}^{%
\mathrm{co}}+\boldsymbol{1}_{K}\boldsymbol{1}_{K}^{\prime } \\
&=&\widehat{\boldsymbol{\Sigma }}^{\mathrm{co}}\mathbf{R}\widehat{%
\boldsymbol{\Sigma }}^{\mathrm{co}}+\boldsymbol{1}_{K}\boldsymbol{1}%
_{K}^{\prime }-\widehat{\boldsymbol{\Sigma }}^{\mathrm{co}}\mathbf{rr}%
^{\prime }\widehat{\boldsymbol{\Sigma }}^{\mathrm{co}}.
\end{eqnarray*}%
By the Sherman-Morrison formula in (\ref{eq:sherman2}), 
\begin{equation}
(\tilde{\mathbf{A}}^{\mathrm{co}\prime }\tilde{\mathbf{A}}^{\mathrm{co}%
})^{-1}=\mathbf{A}_{1}^{-1}+\frac{\mathbf{A}_{1}^{-1}\widehat{\boldsymbol{%
\Sigma }}^{\mathrm{co}}\mathbf{rr}^{\prime }\widehat{\boldsymbol{\Sigma }}^{%
\mathrm{co}}\mathbf{A}_{1}^{-1}}{1-\mathbf{r}^{\prime }\widehat{\boldsymbol{%
\Sigma }}^{\mathrm{co}}\mathbf{A}_{1}^{-1}\widehat{\boldsymbol{\Sigma }}^{%
\mathrm{co}}\mathbf{r}},  \label{eq:Aco1}
\end{equation}%
where $\mathbf{A}_{1}=\widehat{\boldsymbol{\Sigma }}^{\mathrm{co}}\mathbf{R}%
\widehat{\boldsymbol{\Sigma }}^{\mathrm{co}}+\boldsymbol{1}_{K}\boldsymbol{1}%
_{K}^{\prime }$ and moreover $\mathbf{A}_{1}^{-1}=\mathbf{A}_{2}^{-1}-\frac{%
\mathbf{A}_{2}^{-1}\boldsymbol{1}_{K}\boldsymbol{1}_{K}^{\prime }\mathbf{A}%
_{2}^{-1}}{1+\boldsymbol{1}_{K}^{\prime }\mathbf{A}_{2}^{-1}\boldsymbol{1}%
_{K}}$ %\label{eq:Aco4}
by (\ref{eq:sherman1}), where $\mathbf{A}_{2}=\widehat{\boldsymbol{\Sigma }}%
^{\mathrm{co}}\mathbf{R}\widehat{\boldsymbol{\Sigma }}^{\mathrm{co}}.$
Obviously 
\begin{equation}
\phi _{\max }\left( \mathbf{A}_{1}^{-1}\right) \leq \lbrack \phi _{\min
}\left( \mathbf{A}_{2}\right) ]^{-1}\leq \underline{r}^{-1}\phi _{\min
}^{-2}(\widehat{\boldsymbol{\Sigma }}^{\mathrm{co}})  \label{eq:Aco7}
\end{equation}%
and 
\begin{equation}
\boldsymbol{1}_{K}^{\prime }\mathbf{A}_{2}^{-1}\boldsymbol{1}_{K}\leq \phi
_{\min }^{-1}(\mathbf{R})\phi _{\min }^{-1}(\widehat{\boldsymbol{\Sigma }}^{%
\mathrm{co}})\boldsymbol{1}_{K}^{\prime }(\widehat{\boldsymbol{\Sigma }}^{%
\mathrm{co}})^{-1}\boldsymbol{1}_{K}.  \label{eq:Aco5}
\end{equation}
}

{\ The denominator of the second term on the right hand side of (\ref%
{eq:Aco1}) is 
\begin{align}
1-\mathbf{r}^{\prime }\widehat{\boldsymbol{\Sigma }}^{\mathrm{co}}\mathbf{A}%
_{1}^{-1}\widehat{\boldsymbol{\Sigma }}^{\mathrm{co}}\mathbf{r}& =1-\mathbf{r%
}^{\prime }\widehat{\boldsymbol{\Sigma }}^{\mathrm{co}}\left( \mathbf{A}%
_{2}^{-1}-\frac{\mathbf{A}_{2}^{-1}\boldsymbol{1}_{K}\boldsymbol{1}%
_{K}^{\prime }\mathbf{A}_{2}^{-1}}{1+\boldsymbol{1}_{K}^{\prime }\mathbf{A}%
_{2}^{-1}\boldsymbol{1}_{K}}\right) \widehat{\boldsymbol{\Sigma }}^{\mathrm{%
co}}\mathbf{r}  \notag \\
& =\mathbf{r}^{\prime }\widehat{\boldsymbol{\Sigma }}^{\mathrm{co}}\frac{%
\mathbf{A}_{2}^{-1}\boldsymbol{1}_{K}\boldsymbol{1}_{K}^{\prime }\mathbf{A}%
_{2}^{-1}}{1+\boldsymbol{1}_{K}^{\prime }\mathbf{A}_{2}^{-1}\boldsymbol{1}%
_{K}}\widehat{\boldsymbol{\Sigma }}^{\mathrm{co}}\mathbf{r}=\frac{\left[ 
\boldsymbol{1}_{K}^{\prime }(\widehat{\boldsymbol{\Sigma }}^{\mathrm{co}%
})^{-1}\boldsymbol{1}_{K}\right] ^{2}}{1+\boldsymbol{1}_{K}^{\prime }\mathbf{%
A}_{2}^{-1}\boldsymbol{1}_{K}}>0,  \label{eq:Aco2}
\end{align}%
where the second equality follows by $\mathbf{r}^{\prime }\widehat{%
\boldsymbol{\Sigma }}^{\mathrm{co}}\mathbf{A}_{2}^{-1}\widehat{\boldsymbol{%
\Sigma }}^{\mathrm{co}}\mathbf{r=r}^{\prime }\mathbf{R}^{-1}\mathbf{r=}%
\boldsymbol{1}_{K}^{\prime }\mathbf{r}=1,$ and the third equality by $%
\mathbf{r}^{\prime }\widehat{\boldsymbol{\Sigma }}^{\mathrm{co}}\mathbf{A}%
_{2}^{-1}\boldsymbol{1}_{K}=\boldsymbol{1}_{K}^{\prime }(\widehat{%
\boldsymbol{\Sigma }}^{\mathrm{co}})^{-1}\boldsymbol{1}_{K}$. The numerator
of the second term on the right hand side of (\ref{eq:Aco1}) has rank 1, and
thus 
\begin{eqnarray}
& & \phi _{\max }\left( \mathbf{A}_{1}^{-1}\widehat{\boldsymbol{\Sigma }}^{%
\mathrm{co}}\mathbf{rr}^{\prime }\widehat{\boldsymbol{\Sigma }}^{\mathrm{co}}%
\mathbf{A}_{1}^{-1}\right)  \notag \\
&=&\text{trace}\left( \mathbf{A}_{1}^{-1}\widehat{\boldsymbol{\Sigma }}^{%
\mathrm{co}}\mathbf{rr}^{\prime }\widehat{\boldsymbol{\Sigma }}^{\mathrm{co}}%
\mathbf{A}_{1}^{-1}\right) =\mathbf{r}^{\prime }\widehat{\boldsymbol{\Sigma }%
}^{\mathrm{co}}\mathbf{A}_{1}^{-2}\widehat{\boldsymbol{\Sigma }}^{\mathrm{co}%
}\mathbf{r}  \notag \\
&\leq &\mathbf{r}^{\prime }\widehat{\boldsymbol{\Sigma }}^{\mathrm{co}}%
\mathbf{A}_{2}^{-2}\widehat{\boldsymbol{\Sigma }}^{\mathrm{co}}\mathbf{r}=%
\mathbf{r}^{\prime }\mathbf{R}^{-1}(\widehat{\boldsymbol{\Sigma }}^{\mathrm{%
co}})^{-2}\mathbf{R}^{-1}\mathbf{r}=\boldsymbol{1}_{K}^{\prime }(\widehat{%
\boldsymbol{\Sigma }}^{\mathrm{co}})^{-2}\boldsymbol{1}_{K}.\text{ \ }
\label{eq:Aco3}
\end{eqnarray}%
Combine (\ref{eq:Aco2}) and (\ref{eq:Aco3}), 
\begin{align}
\frac{\phi _{\max }\left( \mathbf{A}_{1}^{-1}\widehat{\boldsymbol{\Sigma }}^{%
\mathrm{co}}\mathbf{rr}^{\prime }\widehat{\boldsymbol{\Sigma }}^{\mathrm{co}}%
\mathbf{A}_{1}^{-1}\right) }{1-\mathbf{r}^{\prime }\widehat{\boldsymbol{%
\Sigma }}^{\mathrm{co}}\mathbf{A}_{1}^{-1}\widehat{\boldsymbol{\Sigma }}^{%
\mathrm{co}}\mathbf{r}}& \leq \boldsymbol{1}_{K}^{\prime }(\widehat{%
\boldsymbol{\Sigma }}^{\mathrm{co}})^{-2}\boldsymbol{1}_{K}\times \frac{1+%
\boldsymbol{1}_{K}^{\prime }\mathbf{A}_{2}^{-1}\boldsymbol{1}_{K}}{\left[ 
\boldsymbol{1}_{K}^{\prime }(\widehat{\boldsymbol{\Sigma }}^{\mathrm{co}%
})^{-1}\boldsymbol{1}_{K}\right] ^{2}}  \notag \\
& \leq \phi _{\min }^{-1}(\widehat{\boldsymbol{\Sigma }}^{\mathrm{co}})\left[
\left( \boldsymbol{1}_{K}^{\prime }(\widehat{\boldsymbol{\Sigma }}^{\mathrm{%
co}})^{-1}\boldsymbol{1}_{K}\right) ^{-1}+\phi _{\min }^{-1}(\mathbf{R})\phi
_{\min }^{-1}(\widehat{\boldsymbol{\Sigma }}^{\mathrm{co}})\right]  \notag \\
& \leq \frac{\phi _{\min }^{-1}(\widehat{\boldsymbol{\Sigma }}^{\mathrm{co}})%
}{K\phi _{\max }^{-1}(\widehat{\boldsymbol{\Sigma }}^{\mathrm{co}})}+%
\underline{r}^{-1}\phi _{\min }^{-2}(\widehat{\boldsymbol{\Sigma }}^{\mathrm{%
co}}),  \label{eq:Aco6}
\end{align}%
where the second inequality holds by (\ref{eq:Aco5}). }

{\ Applying the spectral norm to (\ref{eq:Aco1}) yields 
\begin{eqnarray*}
\phi _{A}^{-1} &=&\phi _{\max }\left( (\tilde{\mathbf{A}}^{\mathrm{co}\prime
}\tilde{\mathbf{A}}^{\mathrm{co}})^{-1}\right) \leq \phi _{\max }\left( 
\mathbf{A}_{1}^{-1}\right) +\frac{\phi _{\max }\left( \mathbf{A}_{1}^{-1}%
\widehat{\boldsymbol{\Sigma }}^{\mathrm{co}}\mathbf{rr}^{\prime }\widehat{%
\boldsymbol{\Sigma }}^{\mathrm{co}}\mathbf{A}_{1}^{-1}\right) }{1-\mathbf{r}%
^{\prime }\widehat{\boldsymbol{\Sigma }}^{\mathrm{co}}\mathbf{A}_{1}^{-1}%
\widehat{\boldsymbol{\Sigma }}^{\mathrm{co}}\mathbf{r}} \\
&\leq &\underline{r}^{-1}\phi _{\min }^{-2}(\widehat{\boldsymbol{\Sigma }}^{%
\mathrm{co}})+\frac{\phi _{\max }(\widehat{\boldsymbol{\Sigma }}^{\mathrm{co}%
})}{K\phi _{\min }(\widehat{\boldsymbol{\Sigma }}^{\mathrm{co}})}+\underline{%
r}^{-1}\phi _{\min }^{-2}(\widehat{\boldsymbol{\Sigma }}^{\mathrm{co}})
\end{eqnarray*}%
by (\ref{eq:Aco7}) and (\ref{eq:Aco6}). }

{\ \noindent \textbf{Part (c)}. Given the expression of $\mathbf{a}%
_{0}^{\ast }$ in Lemma \ref{lem:core_dual}, its $\ell _{2}$-norm is bounded
by 
\begin{align}
\left\Vert \mathbf{a}_{0}^{\ast }\right\Vert _{2}& \leq \Vert (\tilde{%
\mathbf{A}}^{\mathrm{co}\prime }\tilde{\mathbf{A}}^{\mathrm{co}})^{-1}\tilde{%
\mathbf{A}}^{\mathrm{co}\prime }\Vert _{\text{sp}}\Vert \left( \mathbf{b}%
_{0}^{\ast }\circ \mathbf{r}-\mathbf{r}\right) ^{\prime }\mathbf{R}%
^{-1/2}\Vert _{2}/N  \notag \\
& \leq \frac{1}{N\sqrt{\phi _{A}}}\left( \Vert \mathbf{\mathbf{b}}_{0}^{\ast
}\circ \mathbf{r}^{1/2}\Vert _{2}+\Vert \mathbf{r}^{1/2}\Vert _{2}\right)
\leq \frac{1}{N\sqrt{\phi _{A}}}\left( \Vert \mathbf{b}_{0}^{\ast }\Vert
_{\infty }+1\right) ,  \label{eq:a0_star_L2}
\end{align}%
where $\mathbf{r}^{1/2}=(r_{1}^{1/2},\ldots ,r_{k}^{1/2})^{\prime }$. In
addition, the Cauchy-Schwarz inequality entails 
\begin{equation}
\left\Vert \mathbf{a}_{0}^{\ast }\right\Vert _{1}\leq \sqrt{K}\left\Vert 
\mathbf{a}_{0}^{\ast }\right\Vert _{2}=N^{-1}\sqrt{K/\phi _{A}}\left(
\left\Vert \mathbf{b}_{0}^{\ast }\right\Vert _{\infty }+1\right)
\label{eq:a0_star_L1}
\end{equation}%
as stated in the lemma. \hfill ${\ \blacksquare }$ }

{\ \bigskip{} }

{\ Lemma \ref{lem:phi_e} collects the implications of Assumptions \ref%
{assu:idiosyn} and \ref{assu:rate} for some building blocks of our
asymptotic theory. Lemma \ref{lem:phi_e}(a) provides the magnitude $\phi_{e}$%
, $\phi_{A}^{-1}$ and $\left\Vert \mathbf{b}_{0}^{\ast}\right\Vert _{\infty}$%
, and part (b) shows that the key condition in the numerical properties in
Theorem \ref{thm:same_sign} is satisfied. }

\begin{lemma}
{\ \label{lem:phi_e} Under Assumptions \ref{assu:idiosyn} and \ref{assu:rate}%
, we have }

\begin{enumerate}
\item {\ $\phi_{e}=O_{p}(\sqrt{N}\phi_{NT})$ , $\phi_{A}^{-1}=O_{p}\left(K%
\right)$, and $\left\Vert \mathbf{b}_{0}^{\ast}\right\Vert _{\infty}=O_{p}(%
\sqrt{K})$; }

\item {\ The event $\left\{ \phi_{e}\left\Vert \mathbf{b}_{0}^{\ast}\right%
\Vert _{\infty}/\sqrt{N}<\tau\right\} $ occurs w.p.a.1. }
\end{enumerate}
\end{lemma}

{\ \noindent \textbf{Proof of Lemma \ref{lem:phi_e}.} \textbf{Part (a).} By
the definition of $\phi_{e}$ and the triangle inequality, 
\begin{eqnarray}
\phi_{e} & = & \Vert\widehat{\boldsymbol{\Sigma}}^{e}\Vert_{c2}\leq\Vert%
\boldsymbol{\Sigma}_{0}^{e}\Vert_{c2}+\Vert\boldsymbol{\Delta}^{e}\Vert_{c2}
\notag \\
& \leq & C_{e0}\phi_{\max}\left(\boldsymbol{\Sigma}_{0}^{e}\right)+\sqrt{N}%
\left\Vert \boldsymbol{\Delta}^{e}\right\Vert _{\infty}=O_{p}(\sqrt{N}%
\phi_{NT}),  \label{eq:phi_e}
\end{eqnarray}
where the second inequality and the last equality follow by Assumption \ref%
{assu:idiosyn}(a). }

{\ Noting that $\widehat{\boldsymbol{\Sigma}}^{\mathrm{co}}=\boldsymbol{%
\Sigma}_{0}^{\mathrm{co}}+\boldsymbol{\Delta}^{\mathrm{co}},$ and then
w.p.a.1 
\begin{eqnarray}
\phi_{\min}(\widehat{\boldsymbol{\Sigma}}^{\mathrm{co}}) & \geq &
\phi_{\min}(\boldsymbol{\Sigma}_{0}^{\mathrm{co}})-\left\Vert \boldsymbol{%
\Delta}^{\mathrm{co}}\right\Vert _{\text{sp}}\geq\phi_{\min}(\boldsymbol{%
\Sigma}_{0}^{\mathrm{co}})-K\left\Vert \boldsymbol{\Delta}^{\mathrm{co}%
}\right\Vert _{\infty}  \notag \\
& \geq & \underline{c}-O_{p}(K(T/\log N)^{-1/2})\geq\underline{c}/2
\label{eq:phi_min}
\end{eqnarray}
where the first inequality follows by the Weyl inequality, the second
inequality by the Gershgorin circle theorem, the third inequality by
Assumption \ref{assu:idiosyn}(b), and the last inequality holds when the
sample size is sufficiently large. Similarly, 
\begin{equation}
\phi_{\max}(\widehat{\boldsymbol{\Sigma}}^{\mathrm{co}})\leq\phi_{\max}(%
\boldsymbol{\Sigma}_{0}^{\mathrm{co}})+\left\Vert \boldsymbol{\Delta}^{%
\mathrm{co}}\right\Vert _{\text{sp}}\leq2\overline{c}  \label{eq:phi_max}
\end{equation}
w.p.a.1. Suppose (\ref{eq:phi_min}) and (\ref{eq:phi_max}) occur. Given
Assumption \ref{assu:rate}(b) about the rate of $\underline{r}$, Lemma \ref%
{lem:res_eigen}(b) implies $\phi_{A}^{-1}\leq8\underline{r}^{-1}\underline{c}%
^{-2}+ 4\overline{c} / (K \underline{c} ) =O_{p}\left(K\right)$ and (\ref%
{eq:b0star}) implies 
\begin{align}
\left\Vert \mathbf{b}_{0}^{\ast}\right\Vert _{\infty} & \leq\left\Vert (%
\widehat{\boldsymbol{\Sigma}}^{\mathrm{co}})^{-1}\boldsymbol{1}%
_{K}\right\Vert _{\infty}/\left[\underline{r}\cdot\boldsymbol{1}%
_{K}^{\prime}(\widehat{\boldsymbol{\Sigma}}^{\mathrm{co}})^{-1}\boldsymbol{1}%
_{K}\right]\leq\left(\boldsymbol{1}_{K}^{\prime }(\widehat{\boldsymbol{\Sigma%
}}^{\mathrm{co}})^{-2}\boldsymbol{1}_{K}\right)^{1/2}/\left[\underline{r}%
\cdot\boldsymbol{1}_{K}^{\prime}(\widehat{\boldsymbol{\Sigma}}^{\mathrm{co}%
})^{-1}\boldsymbol{1}_{K}\right]  \notag \\
& \leq\underline{r}^{-1}\phi_{\min}^{-1}\left(\widehat{\boldsymbol{\Sigma}}^{%
\mathrm{co}}\right)\left(\boldsymbol{1}_{K}^{\prime }(\widehat{\boldsymbol{%
\Sigma}}^{\mathrm{co}})^{-1}\boldsymbol{1}_{K}\right)^{-1/2}\leq\underline{r}%
^{-1}K^{-1/2}\phi_{\max}^{1/2}\left(\widehat{\boldsymbol{\Sigma}}^{\mathrm{co%
}}\right)/\phi_{\min}\left(\widehat{\boldsymbol{\Sigma}}^{\mathrm{co}}\right)
\notag \\
& \leq\underline{r}^{-1}K^{-1/2}\cdot O_{p}(\overline{c}^{1/2}/\underline{c}%
)=O_{p}(\sqrt{K}).  \label{eq:b0star_bound}
\end{align}
}

{\ \noindent \textbf{Part (b).} Part (a) has given $\phi _{e}\left\Vert 
\mathbf{b}_{0}^{\ast }\right\Vert _{\infty }/\sqrt{N}=O_{p}(K^{1/2}\phi
_{NT}).$ Since $K^{1/2}\phi _{NT}/\tau \rightarrow 0$ in Assumption \ref%
{assu:rate}(a), we have $\tau >\phi _{e}\left\Vert \mathbf{b}_{0}^{\ast
}\right\Vert _{\infty }/\sqrt{N}$ when $\left( N,T\right) $ are sufficiently
large and thus the event $\left\{ \phi _{e}\left\Vert \mathbf{b}_{0}^{\ast
}\right\Vert _{\infty }/\sqrt{N}<\tau \right\} $ occurs w.p.a.1. $\hfill{%
\blacksquare }$ }

{\ \bigskip }

{\ Given the above results, we are ready to proceed with proving Theorem \ref%
{thm:alpha_converg}. }

{\ %\subsection{Proofs of the Results in Section \protect\ref%
%{sec:Asymptotic-Theory}}
}

{\ \noindent \textbf{Proof of Theorem \ref{thm:alpha_converg}.} When the
sample size is sufficiently large, w.p.a.1 the event $\left\{ \phi
_{e}\left\Vert \mathbf{b}_{0}^{\ast }\right\Vert _{\infty }/\sqrt{N}<\tau
\right\} $ occurs according to Lemma \ref{lem:phi_e}(b), which allows us to
construct the desirable $\widehat{\boldsymbol{\alpha }}^{\ast }$ in (\ref%
{eq:alpha_hat_star}). Since $\widehat{\boldsymbol{\alpha }}$ is the solution
to (\ref{eq:sim_dual1}), we have 
\begin{equation*}
\frac{1}{2}\widehat{\boldsymbol{\alpha }}^{\prime }\widehat{\mathbf{A}}%
^{\prime }\widehat{\mathbf{A}}\widehat{\boldsymbol{\alpha }}+\frac{1}{N}%
\boldsymbol{1}_{N}^{\prime }\widehat{\boldsymbol{\Sigma }}\widehat{%
\boldsymbol{\alpha }}+\tau \left\Vert \boldsymbol{\alpha }\right\Vert
_{1}\leq \frac{1}{2}\widehat{\boldsymbol{\alpha }}^{\ast \prime }\widehat{%
\mathbf{A}}^{\prime }\widehat{\mathbf{A}}\widehat{\boldsymbol{\alpha }}%
^{\ast }+\frac{1}{N}\boldsymbol{1}_{N}^{\prime }\widehat{\boldsymbol{\Sigma }%
}\widehat{\boldsymbol{\alpha }}^{\ast }+\tau \left\Vert \widehat{\boldsymbol{%
\alpha }}^{\ast }\right\Vert _{1}.
\end{equation*}%
Define $\boldsymbol{\psi }=\widehat{\boldsymbol{\alpha }}-\widehat{%
\boldsymbol{\alpha }}^{\ast }$. Rearranging the above inequality yields 
\begin{equation}
\boldsymbol{\psi }^{\prime }\widehat{\mathbf{A}}^{\prime }\widehat{\mathbf{A}%
}\boldsymbol{\psi }+2\tau \left\Vert \widehat{\boldsymbol{\alpha }}%
\right\Vert _{1}\leq -2\boldsymbol{\psi }^{\prime }\widehat{\boldsymbol{%
\Sigma }}(\widehat{\mathbf{A}}\widehat{\boldsymbol{\alpha }}^{\ast }+\frac{%
\boldsymbol{1}_{N}}{N})+2\tau \left\Vert \widehat{\boldsymbol{\alpha }}%
^{\ast }\right\Vert _{1}.  \label{eq:basic1}
\end{equation}%
Notice that 
\begin{align}
& \boldsymbol{\psi }^{\prime }\widehat{\boldsymbol{\Sigma }}\left( \widehat{%
\mathbf{A}}\widehat{\boldsymbol{\alpha }}^{\ast }+\frac{\boldsymbol{1}_{N}}{N%
}\right)  \notag \\
& =\boldsymbol{\psi }^{\prime }\widehat{\boldsymbol{\Sigma }}(\widehat{%
\mathbf{A}}^{\ast }\widehat{\boldsymbol{\alpha }}^{\ast }+\frac{\boldsymbol{1%
}_{N}}{N})+\boldsymbol{\psi }^{\prime }\widehat{\boldsymbol{\Sigma }}(%
\widehat{\mathbf{A}}-\widehat{\mathbf{A}}^{\ast })\widehat{\boldsymbol{%
\alpha }}^{\ast }=\boldsymbol{\psi }^{\prime }\widehat{\boldsymbol{\Sigma }}%
\mathbf{w}^{\ast }+\boldsymbol{\psi }^{\prime }\widehat{\boldsymbol{\Sigma }}%
^{\prime }\left( \mathbf{I}_{N}-N^{-1}\boldsymbol{1}_{N}\boldsymbol{1}%
_{N}^{\prime }\right) \widehat{\boldsymbol{\Sigma }}^{e}\widehat{\boldsymbol{%
\alpha }}^{\ast }  \notag \\
& =(\boldsymbol{\psi }^{\prime }\widehat{\boldsymbol{\Sigma }}^{\ast }%
\mathbf{w}^{\ast }+\boldsymbol{\psi }^{\prime }\widehat{\boldsymbol{\Sigma }}%
^{e}\mathbf{w}^{\ast })+\boldsymbol{\psi }^{\prime }\widehat{\mathbf{A}}%
^{\prime }\widehat{\boldsymbol{\Sigma }}^{e}\widehat{\boldsymbol{\alpha }}%
^{\ast }=(-\gamma^{\ast }\boldsymbol{\psi }^{\prime }\boldsymbol{1}_{N}+%
\boldsymbol{\psi }^{\prime }\widehat{\boldsymbol{\Sigma }}^{e}\mathbf{w}%
^{\ast })+\boldsymbol{\psi }^{\prime }\widehat{\mathbf{A}}^{\prime }\widehat{%
\boldsymbol{\Sigma }}^{e}\widehat{\boldsymbol{\alpha }}^{\ast }  \notag \\
& =\boldsymbol{\psi }^{\prime }\widehat{\boldsymbol{\Sigma }}^{e}\mathbf{w}%
_{0}^{\ast }+\boldsymbol{\psi }^{\prime }\widehat{\mathbf{A}}^{\prime }%
\widehat{\boldsymbol{\Sigma }}^{e}\widehat{\boldsymbol{\alpha }}^{\ast },
\label{eq:cross-term}
\end{align}%
where the fourth equality follows by the fact that $\widehat{\boldsymbol{%
\Sigma }}^{\ast }\mathbf{w}^{\ast }=-\gamma^{\ast }\boldsymbol{1}_{N}$
implied by the KKT conditions in (\ref{eq:KKT}) with $\tau =0$, and the last
equality by $\boldsymbol{\psi }^{\prime }\boldsymbol{1}_{N}=\left( \widehat{%
\boldsymbol{\alpha }}-\widehat{\boldsymbol{\alpha }}^{\ast }\right) ^{\prime
}\boldsymbol{1}_{N}=0$ as the dual problems entail both $\widehat{%
\boldsymbol{\alpha }}$ and $\widehat{\boldsymbol{\alpha }}^{\ast }$ sum up
to 0. Plugging (\ref{eq:cross-term}) into (\ref{eq:basic1}) to bound the
right hand side of (\ref{eq:basic1}), we have 
\begin{eqnarray}  \label{eq:basic2}
\boldsymbol{\psi }^{\prime }\widehat{\mathbf{A}}^{\prime }\widehat{\mathbf{A}%
}\boldsymbol{\psi }+2\tau \left\Vert \widehat{\boldsymbol{\alpha }}%
\right\Vert _{1} & \leq & 2\left\vert \boldsymbol{\psi }^{\prime }\widehat{%
\boldsymbol{\Sigma }}^{e}\mathbf{w}^{\ast }+\boldsymbol{\psi }^{\prime }%
\widehat{\mathbf{A}}^{\prime }\widehat{\boldsymbol{\Sigma }}^{e}\widehat{%
\boldsymbol{\alpha }}^{\ast }\right\vert +2\tau \left\Vert \widehat{%
\boldsymbol{\alpha }}^{\ast }\right\Vert _{1}  \notag \\
& \leq & 2\left\Vert \boldsymbol{\psi }\right\Vert _{1}\left( \zeta
_{1}+\zeta _{2}\right) +2\tau \left\Vert \widehat{\boldsymbol{\alpha }}%
^{\ast }\right\Vert _{1},
\end{eqnarray}%
where $\zeta _{1}=\Vert \widehat{\boldsymbol{\Sigma }}^{e}\mathbf{w}%
_{0}^{\ast }\Vert _{\infty }$ and $\zeta _{2}=\Vert \widehat{\mathbf{A}}%
^{\prime }\widehat{\boldsymbol{\Sigma }}^{e}\widehat{\boldsymbol{\alpha }}%
^{\ast }\Vert _{\infty }$. }

{\ Now we bound $\zeta _{1}$ and $\zeta _{2}$ in turn. By (\ref{eq:Sigma_w}%
), (\ref{eq:phi_e}) and Lemma \ref{lem:phi_e}(a), we have 
\begin{equation*}
\zeta _{1}\leq \Vert \widehat{\boldsymbol{\Sigma }}^{e}\Vert _{c2}\left\Vert 
\mathbf{w}^{\ast }\right\Vert _{2}\leq \phi _{e}\left\Vert \mathbf{b}%
_{0}^{\ast }\right\Vert _{\infty }/\sqrt{N}=O_{p}(K^{1/2}\phi _{NT}).
\end{equation*}%
For $\zeta _{2}$, by (\ref{eq:holder_ABb}) we have 
\begin{equation}
\zeta _{2}\leq \Vert \widehat{\mathbf{A}}\Vert _{c2}\Vert \widehat{%
\boldsymbol{\Sigma }}^{e}\Vert _{c2}\left\Vert \widehat{\boldsymbol{\alpha }}%
^{\ast }\right\Vert _{1}=\Vert \widehat{\mathbf{A}}\Vert _{c2}\cdot \phi
_{e}\left\Vert \widehat{\boldsymbol{\alpha }}^{\ast }\right\Vert _{1}.
\label{eq:cross_term_zeta_2}
\end{equation}%
By the fact $\left\Vert \mathbf{\mathbf{I}}_{N}-N^{-1}\boldsymbol{1}_{N}%
\boldsymbol{1}_{N}^{\prime }\right\Vert _{\text{sp}}=1$ and by the triangle
inequality, we have 
\begin{align*}
\Vert \widehat{\mathbf{A}}\Vert _{c2}& \leq \Vert \widehat{\boldsymbol{%
\Sigma }}\Vert _{c2}\leq \Vert \widehat{\boldsymbol{\Sigma }}^{\ast }\Vert
_{c2}+\Vert \widehat{\boldsymbol{\Sigma }}_{c2}^{e}\Vert \\
& \leq \sqrt{N}\left( \phi _{\max }\left( \boldsymbol{\Sigma }_{0}^{\mathrm{%
co}}\right) +\left\Vert \boldsymbol{\Delta }^{\mathrm{co}}\right\Vert
_{\infty }\right) =O_{p}(\sqrt{N}),
\end{align*}%
where the third inequality follows from Assumption \ref{assu:idiosyn}(b).
Noting that $\left\Vert \widehat{\boldsymbol{\alpha }}^{\ast }\right\Vert
_{1}=\left\Vert \mathbf{a}_{0}^{\ast }\right\Vert _{1}\leq \left( \left\Vert 
\mathbf{b}_{0}^{\ast }\right\Vert _{\infty }+1\right) $ $\times \sqrt{K/\phi
_{A}}/N$ by Lemma \ref{lem:res_eigen}(c), we continue (\ref%
{eq:cross_term_zeta_2}) to obtain 
\begin{equation*}
\zeta _{2}=O_{p}(\sqrt{N})O_{p}(\sqrt{N}\phi _{NT})N^{-1}\sqrt{K/\phi _{A}}%
\left( \left\Vert \mathbf{b}_{0}^{\ast }\right\Vert _{\infty }+1\right)
=O_{p}(K^{1/2}\phi _{NT}\sqrt{K/\phi _{A}})
\end{equation*}%
as $\left\Vert \mathbf{b}_{0}^{\ast }\right\Vert _{\infty }=O_{p}\left(
K^{1/2}\right) $ by Lemma \ref{lem:phi_e}(a). We thus obtain $\zeta
_{1}+\zeta _{2}=O_{p}(K^{1/2}\phi _{NT}\sqrt{K/\phi _{A}}))$ by the fact $%
K/\phi _{A}\geq 1$ according to the definition of $\phi _{A}$. }

{\ Now, suppose that the sample size is sufficiently large so that $\zeta
_{1}+\zeta _{2}\leq \tau \sqrt{K/\phi _{A}}/2$ in view of the rate of $\tau $
in Assumption \ref{assu:rate}(a). We push (\ref{eq:basic2}) further to
attain 
\begin{equation*}
\boldsymbol{\psi }^{\prime }\widehat{\mathbf{A}}^{\prime }\widehat{\mathbf{A}%
}\boldsymbol{\psi }+2\tau \left\Vert \widehat{\boldsymbol{\alpha }}%
\right\Vert _{1}\leq \tau \sqrt{K/\phi _{A}}\left\Vert \boldsymbol{\psi }%
\right\Vert _{1}+2\tau \left\Vert \widehat{\boldsymbol{\alpha }}^{\ast
}\right\Vert _{1}.
\end{equation*}%
Then 
\begin{equation*}
\boldsymbol{\psi }^{\prime }\widehat{\mathbf{A}}^{\prime }\widehat{\mathbf{A}%
}\boldsymbol{\psi }\leq \tau \sqrt{K/\phi _{A}}\left\Vert \boldsymbol{\psi }%
\right\Vert _{1}+2\tau \left( \left\Vert \widehat{\boldsymbol{\alpha }}%
^{\ast }\right\Vert _{1}-\left\Vert \widehat{\boldsymbol{\alpha }}%
\right\Vert _{1}\right) \leq \tau \left( \sqrt{K/\phi _{A}}+2\right)
\left\Vert \boldsymbol{\psi }\right\Vert _{1},
\end{equation*}%
where the last inequality follows by the triangle inequality: $\left\Vert 
\widehat{\boldsymbol{\alpha }}^{\ast }\right\Vert _{1}-\left\Vert \widehat{%
\boldsymbol{\alpha }}\right\Vert _{1}\leq \left\Vert \boldsymbol{\psi }%
\right\Vert _{1}$. Adding $\tau \sqrt{K/\phi _{A}}\left\Vert \boldsymbol{%
\psi }\right\Vert _{1}$ to both sides of the above inequality yields 
\begin{equation}
\boldsymbol{\psi }^{\prime }\widehat{\mathbf{A}}^{\prime }\widehat{\mathbf{A}%
}\boldsymbol{\psi }+\tau \sqrt{K/\phi _{A}}\left\Vert \boldsymbol{\psi }%
\right\Vert _{1}\leq 2\tau \left( \sqrt{K/\phi _{A}}+1\right) \left\Vert 
\boldsymbol{\psi }\right\Vert _{1}\leq 4\tau \sqrt{K/\phi _{A}}\left\Vert 
\boldsymbol{\psi }\right\Vert _{1}  \label{eq:4tau}
\end{equation}%
where the last inequality follows the fact $K/\phi _{A}\geq 1$. }

{\ By $\phi _{A}^{-1}=O_{p}\left( K\right) $ in Lemma \ref{lem:phi_e}(a), we
have 
\begin{eqnarray*}
\sqrt{\frac{K/\phi _{A}}{N}}\phi _{e} & = & \sqrt{\frac{K/\phi _{A}}{N}}%
O_{p}\left( \sqrt{N}\phi _{NT}\right) =O_{p}\left( \sqrt{K/\phi _{A}}\phi
_{NT}\right) \\
& =& O_{p}\left( K\phi _{NT}\right) =O_{p}(K^{1/2}\tau )=o_{p}\left(
1\right) .
\end{eqnarray*}%
where the last two equalities hold by Assumption \ref{assu:rate}(a). This
implies that the condition $\phi _{e}\leq \frac{1}{2}\sqrt{N\phi _{A}/K}$ in
Lemma \ref{lem:res_eigen} is satisfied w.p.a.1. Moreover, $\boldsymbol{\psi }%
\in \tilde{\mathbb{S}}^{\mathrm{all}}$ by construction of $\widehat{%
\boldsymbol{\alpha }}^{\ast }$ in (\ref{eq:alpha_hat_star}). We hence invoke
Lemma \ref{lem:res_eigen} to continue (\ref{eq:4tau}): 
\begin{equation}
4\tau \sqrt{K/\phi _{A}}\left\Vert \boldsymbol{\psi }\right\Vert _{1}\leq
8\tau \frac{K/\phi _{A}}{\sqrt{N}}\Vert \widehat{\mathbf{A}}\boldsymbol{\psi 
}\Vert _{2}\leq \frac{1}{2}\boldsymbol{\psi }^{\prime }\widehat{\mathbf{A}}%
^{\prime }\widehat{\mathbf{A}}\boldsymbol{\psi }+32\tau ^{2}\frac{(K/\phi
_{A})^{2}}{N},  \label{eq:4tau1}
\end{equation}%
where the last inequality follows by $8ab\leq \frac{1}{2}a^{2}+32b^{2}.$
Combining (\ref{eq:4tau}) and (\ref{eq:4tau1}), we have 
\begin{equation}
\frac{1}{2}\boldsymbol{\psi }^{\prime }\widehat{\mathbf{A}}^{\prime }%
\widehat{\mathbf{A}}\boldsymbol{\psi }+\tau \sqrt{K/\phi _{A}}\left\Vert 
\boldsymbol{\psi }\right\Vert _{1}\leq 32\tau ^{2}\frac{(K/\phi _{A})^{2}}{N}%
.  \label{eq:4tau3}
\end{equation}%
The above equality immediately implies 
\begin{equation*}
\left\Vert \boldsymbol{\psi }\right\Vert _{1}=\left\Vert \widehat{%
\boldsymbol{\alpha }}-\widehat{\boldsymbol{\alpha }}^{\ast }\right\Vert
_{1}\leq 32\tau \frac{(K/\phi _{A})^{3/2}}{N}=O_{p}\left( \frac{(K/\phi
_{A})^{3/2}\tau }{N}\right) =O_{p}\left( \frac{K^{3}\tau }{N}\right)
\end{equation*}%
and 
\begin{equation}
\sqrt{\boldsymbol{\psi }^{\prime }\widehat{\mathbf{A}}^{\prime }\widehat{%
\mathbf{A}}\boldsymbol{\psi }}=\left\Vert \widehat{\mathbf{A}}\left( 
\widehat{\boldsymbol{\alpha }}-\widehat{\boldsymbol{\alpha }}^{\ast }\right)
\right\Vert _{2}\leq 8\tau \frac{K/\phi _{A}}{\sqrt{N}}=O_{p}\left( \frac{%
(K/\phi _{A})\tau }{\sqrt{N}}\right) =O_{p}\left( \frac{K^{2}\tau }{\sqrt{N}}%
\right) ,  \label{eq:thm_quad}
\end{equation}%
given $K/\phi _{A}=O_{p}\left( K^{2}\right) $ by Lemma \ref{lem:phi_e}(a). 
% This completes the proof of the theorem. 
\hfill ${\ \blacksquare }$ }

{\ To summarize the theoretical development up to now, we have shown that
under the high dimensional asymptotic framework where $N/T\rightarrow\infty$
is allowed as $\left(N,T\right)\rightarrow\infty,$ we can construct a unique
oracle target $\widehat{\boldsymbol{\alpha}}^{\ast}$ that satisfies a set of
desirable properties. Since the dual problem (\ref{eq:sim_dual}) is an $%
\ell_{1}$-penalized optimization, we establish in Theorem \ref{thm:same_sign}
the convergence of $\widehat{\boldsymbol{\alpha}} $ to $\widehat{\boldsymbol{%
\alpha}}^{\ast}$ by statistical techniques that deal with the $\ell_{1}$%
-regularization, thanks to the amenable comparability condition and the
derived restricted eigenvalue. }

\subsection{Proofs of the Main Theorems}

The convergence of $\widehat{\boldsymbol{\alpha }}$ in Theorem \ref%
{thm:alpha_converg} implies the convergence of the weight $\widehat{\mathbf{w%
}}$ in Theorem \ref{thm:w_converg} and the convergence of the sample risk to
the oracle risks in Theorem \ref{thm:oracle}.

{\ \bigskip }

{\ \noindent \textbf{Proof of Theorem \ref{thm:w_converg}}. Recall $\widehat{%
\mathbf{A}}^{\ast }=(\mathbf{I}_{N}-N^{-1}\mathbf{1}_{N}\mathbf{1}%
_{N}^{\prime })\widehat{\boldsymbol{\Sigma }}^{\ast }\ $and $\widehat{%
\mathbf{A}}=(\mathbf{I}_{N}-N^{-1}\mathbf{1}_{N}\mathbf{1}_{N}^{\prime })%
\widehat{\boldsymbol{\Sigma }}.$ Let $\widehat{\mathbf{A}}^{e}:=(\mathbf{I}%
_{N}-N^{-1}\mathbf{1}_{N}\mathbf{1}_{N}^{\prime })\widehat{\boldsymbol{%
\Sigma }}^{e}.$ Then we have 
\begin{eqnarray}  \label{eq:w_conv_decomp}
\widehat{\mathbf{w}}-\mathbf{w}^{\ast } & = & \left( \widehat{\mathbf{A}}%
\widehat{\boldsymbol{\alpha }}+\frac{\boldsymbol{1}_{N}}{N}\right) -\left( 
\widehat{\mathbf{A}}^{\ast }\widehat{\boldsymbol{\alpha }}^{\ast }+\frac{%
\boldsymbol{1}_{N}}{N}\right)  \notag \\
& =& \widehat{\mathbf{A}}\widehat{\boldsymbol{\alpha }}-\left( \widehat{%
\mathbf{A}}-\widehat{\mathbf{A}}^{e}\right) \widehat{\boldsymbol{\alpha }}%
^{\ast }=\widehat{\mathbf{A}}\left( \widehat{\boldsymbol{\alpha }}-\widehat{%
\boldsymbol{\alpha }}^{\ast }\right) +\widehat{\mathbf{A}}^{e}\widehat{%
\boldsymbol{\alpha }}^{\ast }.
\end{eqnarray}%
For the first term in (\ref{eq:w_conv_decomp}), by (\ref{eq:thm_quad}) we
have 
\begin{equation}
\Vert \widehat{\mathbf{A}}\left( \widehat{\boldsymbol{\alpha }}-\widehat{%
\boldsymbol{\alpha }}^{\ast }\right) \Vert _{2}=O_{p}\left(
N^{-1/2}K^{2}\tau \right) .  \label{eq:w_conv1}
\end{equation}%
For the second term in (\ref{eq:w_conv_decomp}), we have $\Vert \widehat{%
\mathbf{A}}^{e}\widehat{\boldsymbol{\alpha }}^{\ast }\Vert _{2}\leq \Vert 
\widehat{\boldsymbol{\Sigma }}^{e}\widehat{\boldsymbol{\alpha }}^{\ast
}\Vert _{2}\leq \Vert \boldsymbol{\Sigma }_{0}^{e}\widehat{\boldsymbol{%
\alpha }}^{\ast }\Vert _{2}+\Vert \boldsymbol{\Delta }^{e}\widehat{%
\boldsymbol{\alpha }}^{\ast }\Vert _{2}:=I_{1}+I_{2}$ by the triangle
inequality. Notice that 
\begin{align}
I_{1}& \leq \phi _{\max }\left( \boldsymbol{\Sigma }_{0}^{e}\right)
\left\Vert \widehat{\boldsymbol{\alpha }}^{\ast }\right\Vert _{2}\leq \phi
_{\max }\left( \boldsymbol{\Sigma }_{0}^{e}\right) \left\Vert \mathbf{a}%
_{0}^{\ast }\right\Vert _{2}  \notag \\
& \leq O_{p}\left( \sqrt{N}\phi _{NT}\right) \frac{\left\Vert \mathbf{b}%
_{0}^{\ast }\right\Vert _{\infty }+1}{N\sqrt{\phi _{A}}}=O_{p}\left( \frac{%
K^{1/2}\phi _{NT}}{\sqrt{N\phi _{A}}}\right) ,  \label{eq:w_conv2}
\end{align}%
where the second inequality follows as $\widehat{\boldsymbol{\alpha }}^{\ast
}\in \tilde{\mathbb{S}}^{\mathrm{all}}\subset \mathbb{S}^{\mathrm{all}}$ by
construction, the third inequality by (\ref{eq:a0_star_L2}) and Assumption %
\ref{assu:idiosyn}(a), and the last equality by and Lemma \ref{lem:phi_e}%
(a). Moreover, 
\begin{align}
I_{2}& \leq \left\Vert \boldsymbol{\Delta }^{e}\right\Vert _{c2}\left\Vert 
\widehat{\boldsymbol{\alpha }}^{\ast }\right\Vert _{1}\leq \sqrt{N}%
\left\Vert \boldsymbol{\Delta }^{e}\right\Vert _{\infty }\left\Vert \widehat{%
\boldsymbol{\alpha }}^{\ast }\right\Vert _{1}  \notag \\
& =\sqrt{N}\left\Vert \boldsymbol{\Delta }^{e}\right\Vert _{\infty
}\left\Vert \mathbf{a}_{0}^{\ast }\right\Vert _{1}\leq \sqrt{N}O_{p}\left(
\left( T/\log N\right) ^{-1/2}\right) N^{-1}\sqrt{K/\phi _{A}}\left(
\left\Vert \mathbf{b}_{0}^{\ast }\right\Vert _{\infty }+1\right)  \notag \\
& =O_{p}\left( K\phi _{NT}/\sqrt{N\phi _{A}}\right) ,  \label{eq:w_conv3}
\end{align}%
where the first inequality follows by (\ref{eq:holder_CS1}), the first
equality holds by the fact that $\left\Vert \widehat{\boldsymbol{\alpha }}%
^{\ast }\right\Vert _{1}=\left\Vert \mathbf{\alpha }_{0}^{\ast }\right\Vert
_{1}=\left\Vert \mathbf{a}_{0}^{\ast }\right\Vert _{1}$ as in (\ref%
{eq:alpha_star}), the third inequality by Assumption \ref{assu:idiosyn}(a)
and Lemma \ref{lem:res_eigen}(c), and the last equality holds by Lemma \ref%
{lem:phi_e}(a). }

{\ Collecting (\ref{eq:w_conv_decomp}), (\ref{eq:w_conv1}), (\ref{eq:w_conv2}%
) and (\ref{eq:w_conv3}), we have 
\begin{equation*}
\left\Vert \widehat{\mathbf{w}}-\mathbf{w}^{\ast }\right\Vert
_{2}=O_{p}\left( N^{-1/2}\tau K^{2}\right) +O_{p}\left( K\phi _{NT}/\sqrt{%
N\phi _{A}}\right) =O_{p}\left( N^{-1/2}\tau K^{2}\right) =o_{p}\left(
N^{-1/2}\right)
\end{equation*}%
by Assumption \ref{assu:rate}(a) and Lemma \ref{lem:phi_e}(a). In addition,
the Cauchy-Schwarz inequality immediately implies $\left\Vert \widehat{%
\mathbf{w}}-\mathbf{w}^{\ast }\right\Vert _{1}\leq \sqrt{N}\left\Vert 
\widehat{\mathbf{w}}-\mathbf{w}^{\ast }\right\Vert _{2}=O_{p}\left(
K^{2}\tau \right) =o_{p}\left( 1\right) .\hfill \blacksquare$ }

{\ % \begin{remark}
% To connect the primal problem with the dual problem, we
% consider the decomposition: 
% \begin{equation*}
% \widehat{\mathbf{w}}-\mathbf{w}^{*}=\widehat{\mathbf{A}}(\widehat{%
% \boldsymbol{\alpha}}-\widehat{\boldsymbol{\alpha}}^{\ast})+(\widehat{\mathbf{%
% A}}-\mathbf{A}^{\ast})\widehat{\boldsymbol{\alpha}}^{\ast};
% \end{equation*}
% see (\ref{eq:w_conv_decomp}) in the appendix. The $\ell_{2}$-norm of the
% first term on the right-hand side is bounded by Theorem \ref%
% {thm:alpha_converg}. Moreover, the first term dominates the $\ell_{2}$-norm
% of the second term in which the magnitude of $(\widehat{\mathbf{A}}-\mathbf{A%
% }^{\ast} )$ is well controlled under Assumption \ref{assu:idiosyn}.
% \end{remark}
}

{\ \bigskip }

{\ \noindent \textbf{Proof of Theorem \ref{thm:oracle}}. \textbf{Part (a)}.
Denote $\boldsymbol{\psi }_{w}=\widehat{\mathbf{w}}-\mathbf{w}^{\ast }$. We
first show the in-sample oracle inequality. Decompose 
\begin{eqnarray*}
&&\widehat{\mathbf{w}}^{\prime }\widehat{\boldsymbol{\Sigma }}\widehat{%
\mathbf{w}}-\mathbf{w}^{\ast \prime }\widehat{\boldsymbol{\Sigma }}^{\ast }%
\mathbf{w}^{\ast } \\
&=&(\mathbf{w}^{\ast \prime }\widehat{\boldsymbol{\Sigma }}\mathbf{w}%
_{0}^{\ast }+2\boldsymbol{\psi }_{w}^{\prime }\widehat{\boldsymbol{\Sigma }}%
\mathbf{w}^{\ast }+\boldsymbol{\psi }_{w}\widehat{\boldsymbol{\Sigma }}%
\boldsymbol{\psi }_{w})-\mathbf{w}^{\ast \prime }\widehat{\boldsymbol{\Sigma 
}}^{\ast }\mathbf{w}^{\ast }=\mathbf{w}^{\ast \prime }\widehat{\boldsymbol{%
\Sigma }}^{e}\mathbf{w}^{\ast }+2\boldsymbol{\psi }_{w}^{\prime }\widehat{%
\boldsymbol{\Sigma }}\mathbf{w}^{\ast }+\boldsymbol{\psi }_{w}\widehat{%
\boldsymbol{\Sigma }}\boldsymbol{\psi }_{w} \\
&=&\mathbf{w}^{\ast \prime }\widehat{\boldsymbol{\Sigma }}^{e}\mathbf{w}%
_{0}^{\ast }+2\boldsymbol{\psi }_{w}^{\prime }(\widehat{\boldsymbol{\Sigma }}%
^{\ast }+\boldsymbol{\Delta }^{e})\mathbf{w}^{\ast }+2\boldsymbol{\psi }%
_{w}^{\prime }\boldsymbol{\Sigma }_{0}^{e}\mathbf{w}^{\ast }+\boldsymbol{%
\psi }_{w}^{\prime }(\widehat{\boldsymbol{\Sigma }}^{\ast }+\boldsymbol{%
\Delta }^{e})\boldsymbol{\psi }_{w}+\boldsymbol{\psi }_{w}^{\prime }%
\boldsymbol{\Sigma }_{0}^{e}\boldsymbol{\psi }_{w} \\
&=&:II_{1}+2II_{2}+2II_{3}+II_{4}+II_{5}.
\end{eqnarray*}%
We bound $II_{1}$ by 
\begin{align*}
\left\vert II_{1}\right\vert & \leq \phi _{\max }(\widehat{\boldsymbol{%
\Sigma }}^{e})\left\Vert \mathbf{w}^{\ast }\right\Vert _{2}^{2}\leq \left(
\phi _{\max }\left( \boldsymbol{\Sigma }_{0}^{e}\right) +\phi _{\max }\left( 
\boldsymbol{\Delta }^{e}\right) \right) \left\Vert \mathbf{w}_{0}^{\ast
}\right\Vert _{2}^{2} \\
& \leq \left( \phi _{\max }\left( \boldsymbol{\Sigma }_{0}^{e}\right) +N||%
\boldsymbol{\Delta }^{e}||_{\infty }\right) \left\Vert \mathbf{w}_{0}^{\ast
}\right\Vert _{2}^{2} \\
& \leq \left( O_{p}(\sqrt{N}\phi _{NT})+NO_{p}((T/\log N)^{-1/2})\right) 
\frac{\left\Vert \mathbf{b}_{0}^{\ast }\right\Vert _{\infty }^{2}}{N}%
=O_{p}(K\phi _{NT}),
\end{align*}%
where the third inequality holds by the Gershgorin circle theorem, and the
fourth by Assumption \ref{assu:idiosyn}, and the last by Lemma \ref%
{lem:phi_e}(a). The second term $II_{2}$ is bounded by 
\begin{align*}
\left\vert II_{2}\right\vert & \leq \Vert \widehat{\boldsymbol{\Sigma }}%
^{\ast }+\boldsymbol{\Delta }^{e}\Vert _{\infty }\left\Vert \boldsymbol{\psi 
}_{w}\right\Vert _{1}\left\Vert \mathbf{w}^{\ast }\right\Vert _{1}\leq
\left( \Vert \widehat{\boldsymbol{\Sigma }}^{\ast }\Vert _{\infty
}+\left\Vert \boldsymbol{\Delta }^{e}\right\Vert _{\infty }\right)
\left\Vert \boldsymbol{\psi }_{w}\right\Vert _{1}\left\Vert \mathbf{w}%
_{0}^{\ast }\right\Vert _{1} \\
& \leq \left( \Vert \widehat{\boldsymbol{\Sigma }}^{\mathrm{co}}\Vert
_{\infty }+\left\Vert \boldsymbol{\Delta }^{e}\right\Vert _{\infty }\right)
\left\Vert \boldsymbol{\psi }_{w}\right\Vert _{1}\sqrt{N}\left\Vert \mathbf{w%
}^{\ast }\right\Vert _{2} \\
& =O_{p}\left( \overline{c}+(T/\log N)^{-1/2}\right) O_{p}\left( \tau
K^{2}\right) \left\Vert \mathbf{b}_{0}^{\ast }\right\Vert _{\infty
}=O_{p}\left( \tau K^{5/2}\right) ,
\end{align*}%
where the first inequality follows by (\ref{eq:holder_aAb}), the third
inequality by the Cauchy-Schwarz inequality, the first equality holds by
Assumptions \ref{assu:idiosyn}, Theorem \ref{thm:w_converg}, and Theorem \ref%
{thm:same_sign}(a), and the last equality by Lemma \ref{lem:phi_e}(a). For $%
II_{3},$ we have 
\begin{align*}
\left\vert II_{3}\right\vert & \leq \phi _{\max }\left( \boldsymbol{\Sigma }%
_{0}^{e}\right) \left\Vert \boldsymbol{\psi }_{w}\right\Vert _{2}\left\Vert 
\mathbf{w}^{\ast }\right\Vert _{2} \\
& =O_{p}(\sqrt{N}\phi _{NT})O_{p}\left( N^{-1/2}\tau K^{2}\right) \left\Vert 
\mathbf{b}_{0}^{\ast }\right\Vert _{\infty }N^{-1/2}=O_{p}\left(
N^{-1/2}\phi _{NT}\tau K^{5/2}\right)
\end{align*}%
by (\ref{eq:holder_psd}), Assumptions \ref{assu:idiosyn}, and Theorem \ref%
{thm:w_converg}. Similarly, 
\begin{align*}
\left\vert II_{4}\right\vert & \leq \Vert \widehat{\boldsymbol{\Sigma }}%
^{\ast }+\boldsymbol{\Delta }^{e}\Vert _{\infty }\left\Vert \boldsymbol{\psi 
}_{w}\right\Vert _{1}^{2}=O_{p}(\overline{c}+(T/\log N)^{-1/2})O_{p}\left(
\tau ^{2}K^{4}\right) =O_{p}\left( \tau ^{2}K^{4}\right) ,\text{ and} \\
\left\vert II_{5}\right\vert & \leq \phi _{\max }\left( \boldsymbol{\Sigma }%
_{0}^{e}\right) \left\Vert \boldsymbol{\psi }_{w}\right\Vert _{2}^{2}=O_{p}(%
\sqrt{N}\phi _{NT})\left( N^{-1}\tau ^{2}K^{4}\right) =O_{p}\left(
N^{-1/2}\phi _{NT}\tau ^{2}K^{4}\right) .
\end{align*}%
Collecting all these five terms, and notice that $O_{p}\left( \tau
K^{5/2}\right) $ is the dominating order, we have 
\begin{equation*}
\left\vert \widehat{\mathbf{w}}^{\prime }\widehat{\boldsymbol{\Sigma }}%
\widehat{\mathbf{w}}-\mathbf{w}^{\ast \prime }\widehat{\boldsymbol{\Sigma }}%
^{\ast }\mathbf{w}^{\ast }\right\vert =O_{p}\left( \tau K^{5/2}\right)
=o_{p}\left( 1\right)
\end{equation*}%
under Assumption \ref{assu:rate}(a). }

{\ \textbf{Part (b)}. Given $T^{\mathrm{new}} \asymp T$, the same argument
goes through if we replace $\widehat{\boldsymbol{\Sigma }}$ with $\widehat{%
\boldsymbol{\Sigma }}^{\mathrm{new}}$, and replace $\widehat{\boldsymbol{%
\Sigma }}^{\ast }$ with $\widehat{\boldsymbol{\Sigma }}^{\ast \mathrm{new}}$%
, because in the above analysis of the in-sample oracle inequality we always
bound the various quantities by separating the norms of vectors and the
square matrices. We conclude the out-of-sample oracle equality. }

{\ \textbf{Part (c)}. This proof involves two steps: (i) Establish the
closeness between $\widehat{\mathbf{w}}^{\prime }\widehat{\boldsymbol{\Sigma 
}}^{\mathrm{new}}\widehat{\mathbf{w}}$ and $\widehat{\mathbf{w}}^{\prime }%
\widehat{\boldsymbol{\Sigma }}\widehat{\mathbf{w}}$ as shown in (\ref%
{eq:opt_5}) below; (ii) Establish the closeness between $\widehat{\mathbf{w}}%
^{\prime }\widehat{\boldsymbol{\Sigma }}\widehat{\mathbf{w}}$ and $Q(%
\boldsymbol{\Sigma }_{0})$ as shown in (\ref{eq:opt_6}) below, where $Q(%
\mathbf{S}):=\min_{\mathbf{w}\in \mathbb{R}^{N},\mathbf{w}^{\prime }%
\boldsymbol{1}_{N}=1}\,\mathbf{w}^{\prime }\mathbf{S}\mathbf{w}$ for a
generic $N\times N$ positive semi-definite matrix $\mathbf{S}$. }

{\ Obviously $\widehat{\mathbf{w}}^{\prime}\widehat{\boldsymbol{\Sigma}}^{%
\mathrm{new}}\widehat{\mathbf{w}}\ge\widehat{\mathbf{w}}^{\prime}\widehat{%
\boldsymbol{\Sigma}}\widehat{\mathbf{w}}=Q(\widehat{\boldsymbol{\Sigma}})$.
On the other hand, 
\begin{equation}
\widehat{\mathbf{w}}^{\prime}\widehat{\boldsymbol{\Sigma}}\widehat{\mathbf{w}%
}=\widehat{\mathbf{w}}^{\prime}\widehat{\boldsymbol{\Sigma}}^{\mathrm{new}}%
\widehat{\mathbf{w}}+\widehat{\mathbf{w}}^{\prime}(\widehat{\boldsymbol{%
\Sigma}}-\widehat{\boldsymbol{\Sigma}}^{\mathrm{new}})\widehat{\mathbf{w}}%
\geq\widehat{\mathbf{w}}^{\prime}\widehat{\boldsymbol{\Sigma}}^{\mathrm{new}}%
\widehat{\mathbf{w}}-\Vert\widehat{\boldsymbol{\Sigma}}-\widehat{\boldsymbol{%
\Sigma}}^{\mathrm{new}}\Vert_{\mathrm{sp}}\left\Vert \widehat{\mathbf{w}}%
\right\Vert _{2}^{2}  \label{eq:opt_1}
\end{equation}
by the triangle inequality and (\ref{eq:holder_psd}). We focus on the term $%
\Vert\widehat{\boldsymbol{\Sigma}}-\widehat{\boldsymbol{\Sigma}}^{\mathrm{new%
}}\Vert_{\mathrm{sp}}\left\Vert \widehat{\mathbf{w}}\right\Vert _{2}^{2}$. }

{\ For the first factor, notice 
\begin{equation*}
\widehat{\boldsymbol{\Sigma }}-\widehat{\boldsymbol{\Sigma }}^{\mathrm{new}}=%
\widehat{\boldsymbol{\Sigma }}^{\ast }-\widehat{\boldsymbol{\Sigma }}^{\ast ,%
\mathrm{new}}+\widehat{\boldsymbol{\Sigma }}^{e}-\widehat{\boldsymbol{\Sigma 
}}^{e,\mathrm{new}}=(\widehat{\boldsymbol{\Sigma }}^{\ast }-\boldsymbol{%
\Sigma }_{0}^{\ast })-(\widehat{\boldsymbol{\Sigma }}^{\ast ,\mathrm{new}}-%
\boldsymbol{\Sigma }_{0}^{\ast })+\boldsymbol{\Delta }^{e}-\boldsymbol{%
\Delta }^{e,\mathrm{new}}.
\end{equation*}%
Under Assumption \ref{assu:idiosyn}(b), $\Vert \widehat{\boldsymbol{\Sigma }}%
^{\ast }-\boldsymbol{\Sigma }_{0}^{\ast }\Vert _{\infty }=\Vert \widehat{%
\boldsymbol{\Sigma }}^{\mathrm{co}}-\boldsymbol{\Sigma }_{0}^{\mathrm{co}%
}\Vert _{\infty }=\left\Vert \Delta ^{\mathrm{co}}\right\Vert _{\infty
}=O_{p}((T/\log N)^{-1/2})$ and therefore by Gershgorin circle theorem the
spectral norm is bounded by 
\begin{equation}
\Vert \widehat{\boldsymbol{\Sigma }}^{\ast }-\boldsymbol{\Sigma }_{0}^{\ast
}\Vert _{\mathrm{sp}}\leq N\left\Vert \Delta ^{\mathrm{co}}\right\Vert
_{\infty }=O_{p}\left( N(T/\log N)^{-1/2}\right) .  \label{eq:opt_4}
\end{equation}%
Gershgorin circle theorem also implies $\left\Vert \boldsymbol{\Delta }%
^{e}\right\Vert _{\mathrm{sp}}\leq \phi _{e}=O_{p}(\sqrt{N}\phi _{NT}),$
where the stochastic order follows by Lemma \ref{lem:phi_e}(a). Since the
new testing data come from the same DGP as that of the training data, the
same stochastic bounds are applicable to the terms involving the new data,
and then 
\begin{align}
\Vert \widehat{\boldsymbol{\Sigma }}-\widehat{\boldsymbol{\Sigma }}^{\mathrm{%
new}}\Vert _{\mathrm{sp}}& \leq \Vert \widehat{\boldsymbol{\Sigma }}^{\ast }-%
\boldsymbol{\Sigma }_{0}^{\ast }\Vert _{\mathrm{sp}}+\Vert \widehat{%
\boldsymbol{\Sigma }}^{\ast ,\mathrm{new}}-\boldsymbol{\Sigma }_{0}^{\ast
}\Vert _{\mathrm{sp}}+\left\Vert \boldsymbol{\Delta }^{e}\right\Vert _{%
\mathrm{sp}}+\left\Vert \boldsymbol{\Delta }^{e,\mathrm{new}}\right\Vert _{%
\mathrm{sp}}  \notag \\
& =O_{p}\left( N(T/\log N)^{-1/2}\right) +O_{p}(\sqrt{N}\phi
_{NT})=O_{p}(N\phi _{NT}).  \label{eq:opt_3}
\end{align}%
The second factor is bounded by 
\begin{equation}
\left\Vert \widehat{\mathbf{w}}\right\Vert _{2}^{2}\leq \left\Vert \mathbf{b}%
_{0}^{\ast }\right\Vert _{\infty }^{2}/N=O_{p}\left( K/N\right)
\label{eq:opt_2}
\end{equation}%
according to Theorem \ref{thm:same_sign}(a) and Lemma \ref{lem:phi_e}(a).
Collecting (\ref{eq:opt_1}), (\ref{eq:opt_3}) and (\ref{eq:opt_2}), we have 
\begin{eqnarray}  \label{eq:opt_5}
0 & \leq & \widehat{\mathbf{w}}^{\prime }\widehat{\boldsymbol{\Sigma }}^{%
\mathrm{new}}\widehat{\mathbf{w}}-\widehat{\mathbf{w}}^{\prime }\widehat{%
\boldsymbol{\Sigma }}\widehat{\mathbf{w}}\leq \Vert \widehat{\boldsymbol{%
\Sigma }}-\widehat{\boldsymbol{\Sigma }}^{\mathrm{new}}\Vert _{\mathrm{sp}%
}\left\Vert \widehat{\mathbf{w}}\right\Vert _{2}^{2}  \notag \\
& = & O_{p}(N\phi _{NT})O_{p}\left( K/N\right) =O_{p}\left( K\phi
_{NT}\right).
\end{eqnarray}
}

{\ Next, consider the population matrices $\boldsymbol{\Sigma}_{0}$ and $%
\boldsymbol{\Sigma}_{0}^{*}$. Because $\boldsymbol{\Sigma}_{0}-\boldsymbol{%
\Sigma}_{0}^{*}=\boldsymbol{\Sigma}_{0}^{e}$ is positive semi-definite, 
\begin{equation}
Q(\boldsymbol{\Sigma}_{0})\geq Q\left(\boldsymbol{\Sigma}_{0}^{*}\right).
\label{eq:ineq_pop_1}
\end{equation}
Since $\mathrm{rank}\left(\boldsymbol{\Sigma}_{0}^{*}\right)=K\ll N$, the
solution to $\min_{\mathbf{w}\in\mathbb{R}^{N},\mathbf{w}^{\prime}%
\boldsymbol{1}_{N}=1}\,\mathbf{w}^{\prime}\boldsymbol{\Sigma}_{0}^{*}\mathbf{%
w}$ is not unique but all the solutions give the same minimum $Q\left(%
\boldsymbol{\Sigma}_{0}^{*}\right)$. Thus in order to evaluate $Q\left(%
\boldsymbol{\Sigma}_{0}^{*}\right)$ we can simply use the within-group equal
weight optimizer $\mathbf{w}^{\sharp}$ which solves 
\begin{equation*}
\min_{\left(\mathbf{w},\gamma\right)\in\mathbb{R}^{N+1}}\frac{1}{2}%
\left\Vert \mathbf{w}\right\Vert _{2}^{2}\text{\ \ subject to \ }\mathbf{w}%
^{\prime}\boldsymbol{1}_{N}=1,\text{ and }\boldsymbol{\Sigma}_{0}^{*}\mathbf{%
w}+\gamma=0.
\end{equation*}
}

{\ \noindent The only difference between $\mathbf{w}^{\sharp}$ and $\mathbf{w%
}_{0}^{*}$ is that the former is associated with the population $\boldsymbol{%
\Sigma}_{0}^{*}$ and the latter associated with the sample $\widehat{%
\boldsymbol{\Sigma}}^{*}$. Parallel to % (\ref{eq:equal_weight_star}), 
(\ref{eq:b0star}) and (\ref{eq:b0star_bound}), we can write $\mathbf{w}%
_{0}^{\sharp}=\left( b_{01}^{\sharp}\cdot\boldsymbol{1}_{N_{1}}^{\prime}/N,%
\cdots,b_{0K}^{\sharp} \cdot\boldsymbol{1}_{N_{K}}^{\prime}/N\right)^{%
\prime} $ where $\mathbf{b}_{0}^{\sharp}=\mathbf{r}\circ\frac{(\boldsymbol{%
\Sigma}_{0}^{\mathrm{co}})^{-1}\boldsymbol{1}_{K}}{\boldsymbol{1}%
_{K}^{\prime}(\boldsymbol{\Sigma}_{0}^{\mathrm{co}})^{-1}\boldsymbol{1}_{K}}$
and it is bounded by 
\begin{align*}
\left\Vert \mathbf{b}_{0}^{\sharp}\right\Vert _{\infty} & \leq\left\Vert (%
\boldsymbol{\Sigma}_{0}^{\mathrm{co}})^{-1}\boldsymbol{1}_{K}\right\Vert
_{\infty}/\left[\underline{r}\cdot\boldsymbol{1}_{K}^{\prime}(\boldsymbol{%
\Sigma}_{0}^{\mathrm{co}})^{-1}\boldsymbol{1}_{K}\right]\leq\underline{r}%
^{-1}K^{-1/2}\phi_{\max}^{1/2}\left(\boldsymbol{\Sigma}_{0}^{\mathrm{co}%
}\right)/\phi_{\min}\left(\boldsymbol{\Sigma}_{0}^{\mathrm{co}}\right) \\
& \leq\overline{c}^{1/2}/(\underline{r}\underline{c}K^{1/2})=O(\sqrt{K})
\end{align*}
under Assumption \ref{assu:idiosyn}(b) and furthermore 
\begin{equation}
\Vert\mathbf{w}^{\sharp}\Vert_{2}^{2}\leq N\Vert\mathbf{w}%
_{0}^{\sharp}\Vert_{\infty}^{2}=N(\Vert\mathbf{b}_{0}^{\sharp}\Vert_{%
\infty}/N)^{2}=O\left(K/N\right).  \label{eq:ineq_pop_4}
\end{equation}
}

We continue (\ref{eq:ineq_pop_1}): 
\begin{align}
Q(\boldsymbol{\Sigma }_{0}^{\ast })& =\mathbf{w}^{\sharp \prime }\widehat{%
\boldsymbol{\Sigma }}^{\ast }\mathbf{w}^{\sharp }+\mathbf{w}_{0}^{\sharp
\prime }\left( \boldsymbol{\Sigma }_{0}^{\ast }-\widehat{\boldsymbol{\Sigma }%
}^{\ast }\right) \mathbf{w}^{\sharp }\geq \mathbf{w}_{0}^{\ast \prime }%
\widehat{\Sigma }^{\ast }\mathbf{w}^{\ast }+\mathbf{w}_{0}^{\sharp \prime
}\left( \boldsymbol{\Sigma }_{0}^{\ast }-\widehat{\boldsymbol{\Sigma }}%
^{\ast }\right) \mathbf{w}^{\sharp }  \notag \\
& \geq \mathbf{w}^{\ast \prime }\widehat{\Sigma }^{\ast }\mathbf{w}%
_{0}^{\ast }-\Vert \boldsymbol{\Sigma }_{0}^{\ast }-\widehat{\boldsymbol{%
\Sigma }}^{\ast }\Vert _{\mathrm{sp}}\Vert \mathbf{w}^{\sharp }\Vert
_{2}^{2},  \label{eq:ineq_pop_2}
\end{align}%
where the first inequality follows as $\mathbf{w}^{\ast }$ is the optimizer
associated with $\widehat{\Sigma }^{\ast }$, and the second inequality is
derived by the same reasoning as used to obtain (\ref{eq:opt_1}). (\ref%
{eq:ineq_pop_2}) and (\ref{eq:ineq_pop_1}) imply 
\begin{equation}
\mathbf{w}^{\ast \prime }\widehat{\Sigma }^{\ast }\mathbf{w}^{\ast }\leq Q(%
\boldsymbol{\Sigma }_{0})+\Vert \boldsymbol{\Sigma }_{0}^{\ast }-\widehat{%
\boldsymbol{\Sigma }}^{\ast }\Vert _{\mathrm{sp}}\Vert \mathbf{w}%
_{0}^{\sharp }\Vert _{2}^{2}\leq Q(\boldsymbol{\Sigma }_{0})+O_{p}\left(
K(T/\log N)^{-1/2}\right)  \label{eq:ineq_pop_3}
\end{equation}
in view of (\ref{eq:opt_4}) and (\ref{eq:ineq_pop_4}). Combine Part (a) and (%
\ref{eq:ineq_pop_3}): 
\begin{equation}
\widehat{\mathbf{w}}^{\prime }\widehat{\boldsymbol{\Sigma }}\widehat{\mathbf{%
w}}\leq Q(\boldsymbol{\Sigma }_{0})+O_{p}(\tau K^{5/2}).  \label{eq:opt_6}
\end{equation}%
In conjunction with (\ref{eq:opt_5}) and notice $K\phi _{NT}=O\left( \tau
K^{1/2}\right) $ is of smaller order than $\tau K^{5/2}$, the conclusion
follows. \hfill ${\ \blacksquare }$

\subsection{Elementary Inequalities on Matrix Norms}

{\ We collect some elementary inequalities used in the proofs. }

\begin{lemma}
{\ \label{Lem:inequalities} Let $\mathbf{a}$ and $\mathbf{b}$ be two
vectors, and $\mathbf{A}$ and $\mathbf{B}$ be two matrices of compatible
dimensions. Then we have 
\begin{align}
\left\Vert \mathbf{A}\mathbf{b}\right\Vert _{\infty} & \leq\left\Vert 
\mathbf{A}\right\Vert _{\infty}\left\Vert \mathbf{b}\right\Vert _{1}
\label{eq:holder_Ab} \\
\left\vert \mathbf{a}^{\prime}\mathbf{A}\mathbf{b}\right\vert &
\leq\left\Vert \mathbf{A}\right\Vert _{\infty}\left\Vert \mathbf{a}%
\right\Vert _{1}\left\Vert \mathbf{b}\right\Vert _{1}  \label{eq:holder_aAb}
\\
\left\vert \mathbf{a}^{\prime}\mathbf{A}\mathbf{b}\right\vert &
\leq\left\Vert \mathbf{A}\right\Vert _{\mathrm{sp}}\left\Vert \mathbf{a}%
\right\Vert _{2}\left\Vert \mathbf{b}\right\Vert _{2}  \label{eq:holder_psd}
\\
\left\Vert \mathbf{ABb}\right\Vert _{\infty} & \leq\left\Vert \mathbf{A}%
^{\prime }\right\Vert _{c2}\left\Vert \mathbf{B}\right\Vert _{c2}\left\Vert 
\mathbf{b}\right\Vert _{1}  \label{eq:holder_ABb}
\end{align}
If $\mathbf{S}$ is a symmetric matrix, 
\begin{equation}
\left\Vert \mathbf{S}\mathbf{b}\right\Vert _{2}\leq\left\Vert \mathbf{S}%
\right\Vert _{c2}\left\Vert \mathbf{b}\right\Vert _{1}.
\label{eq:holder_CS1}
\end{equation}
If $\boldsymbol{\Sigma}$ is positive definite, 
\begin{align}
\left(\boldsymbol{\Sigma}+\mathbf{a}\mathbf{a}^{\prime }\right)^{-1} & =%
\boldsymbol{\Sigma}^{-1}- \boldsymbol{\Sigma}^{-1}\mathbf{a}\mathbf{a}%
^{\prime }\boldsymbol{\Sigma}^{-1}/(1+\mathbf{a}^{\prime }\boldsymbol{\Sigma}%
^{-1}\mathbf{a})  \label{eq:sherman1} \\
\left(\boldsymbol{\Sigma}-\mathbf{a}\mathbf{a}^{\prime }\right)^{-1} & =%
\boldsymbol{\Sigma}^{-1} +\boldsymbol{\Sigma}^{-1}\mathbf{a}\mathbf{a}%
^{\prime }\boldsymbol{\Sigma}^{-1}/(1-\mathbf{a}^{\prime }\boldsymbol{\Sigma}%
^{-1}\mathbf{a}).  \label{eq:sherman2}
\end{align}
}
\end{lemma}

{\ \noindent \textbf{Proof of Lemma \ref{Lem:inequalities}.} The first
inequality follows because 
\begin{equation*}
\left\Vert \mathbf{A}\mathbf{b}\right\Vert _{\infty}\leq\max_{i}\left\vert 
\mathbf{A}_{i\cdot}\mathbf{b}\right\vert \leq\max_{i}\left\Vert \mathbf{A}%
_{i\cdot}\right\Vert _{\infty}\left\Vert \mathbf{b}\right\Vert
_{1}=\left\Vert \mathbf{A}\right\Vert _{\infty}\left\Vert \mathbf{b}%
\right\Vert _{1}.
\end{equation*}
It implies the second inequality $\left\vert \mathbf{a}^{\prime}\mathbf{A}%
\mathbf{b}\right\vert \leq\left\Vert \mathbf{a}\right\Vert _{1}\left\Vert 
\mathbf{A}\mathbf{b}\right\Vert _{\infty}\leq\left\Vert \mathbf{A}%
\right\Vert _{\infty}\left\Vert \mathbf{a}\right\Vert _{1}\left\Vert \mathbf{%
b}\right\Vert _{1}$ and the fourth inequality 
\begin{equation*}
\left\Vert \mathbf{ABb}\right\Vert _{\infty}\leq\left\Vert \mathbf{AB}%
\right\Vert _{\infty}\left\Vert \mathbf{b}\right\Vert
_{1}=\max_{i,j}\left\vert \mathbf{A}_{i\cdot}\mathbf{B}_{\cdot j}\right\vert
\left\Vert \mathbf{b}\right\Vert _{1}\leq\left\Vert \mathbf{A}^{\prime
}\right\Vert _{c2}\left\Vert \mathbf{B}\right\Vert _{c2}\left\Vert \mathbf{b}%
\right\Vert _{1}.
\end{equation*}
The third inequality follows by the Cauchy-Schwarz inequality $\left\vert 
\mathbf{a}^{\prime}\mathbf{A}\mathbf{b}\right\vert \leq\left\Vert \mathbf{A}%
^{\prime}\mathbf{a}\right\Vert _{2}\left\Vert \mathbf{b}\right\Vert
_{2}\leq\left\Vert \mathbf{A}\right\Vert _{\mathrm{sp}}\left\Vert \mathbf{a}%
\right\Vert _{2}\left\Vert \mathbf{b}\right\Vert _{2}.$ For the symmetric
matrix $\mathbf{S}$, 
\begin{equation*}
\left\Vert \mathbf{S}\mathbf{b}\right\Vert _{2}=\sqrt{\mathbf{b}^{\prime}%
\mathbf{S}\mathbf{S}\mathbf{b}}\leq\sqrt{\left\Vert \mathbf{S}\mathbf{S}%
\right\Vert _{\infty}}\left\Vert \mathbf{b}\right\Vert _{1}\leq\sqrt{%
\max_{i}\left(\mathbf{S}\mathbf{S}\right)_{ii}}\left\Vert \mathbf{b}%
\right\Vert _{1}=\left\Vert \mathbf{S}\right\Vert _{c2}\left\Vert \mathbf{b}%
\right\Vert _{1}
\end{equation*}
where the first inequality follows by (\ref{eq:holder_aAb}), and the second
inequality and the last equality are due to the symmetry of $\mathbf{Q}$. }

{\ \noindent The Sherman-Morrison formula gives $\left(\boldsymbol{\Sigma}+%
\mathbf{a}\mathbf{b}^{\prime }\right)^{-1}=\boldsymbol{\Sigma}^{-1}-%
\boldsymbol{\Sigma}^{-1}\mathbf{a}\mathbf{b}^{\prime }\boldsymbol{\Sigma}%
^{-1}/\left(1+\mathbf{a}^{\prime }\boldsymbol{\Sigma}^{-1}\mathbf{b}\right)$
for any compatible vector $\mathbf{a}$ and $\mathbf{b}$. (\ref{eq:sherman1})
and (\ref{eq:sherman2}) follow by setting $\mathbf{b}=\mathbf{a}$ and $%
\mathbf{b}=-\mathbf{a}$, respectively. \hfill ${\ \blacksquare}$ }

\section{Additional Results for the Numerical Work}

\label{subsec:num_demo}

In this appendix, we report some additional designs and results for the
numerical work in the main text of the paper.

\subsection{Simulation Example on Bias-Variance Trade-off}

\label{subsec:demo}

The solution $\widehat{\mathbf{w}}^{\mathrm{C}}$ in (\ref%
{eq:bates-granger-explicit}) often performs unsatisfactorily in practice
because of the presence of parameter estimation error in the estimation of
the optimal weights. $\ell _{2}$-relaxation intends to achieve a balance
between the optimal weighting and the simple average by exploring the
bias-variance trade-off via the tuning parameter $\tau $.

\begin{figure}[ht!]
{\ \centering
\includegraphics[scale = 0.65]{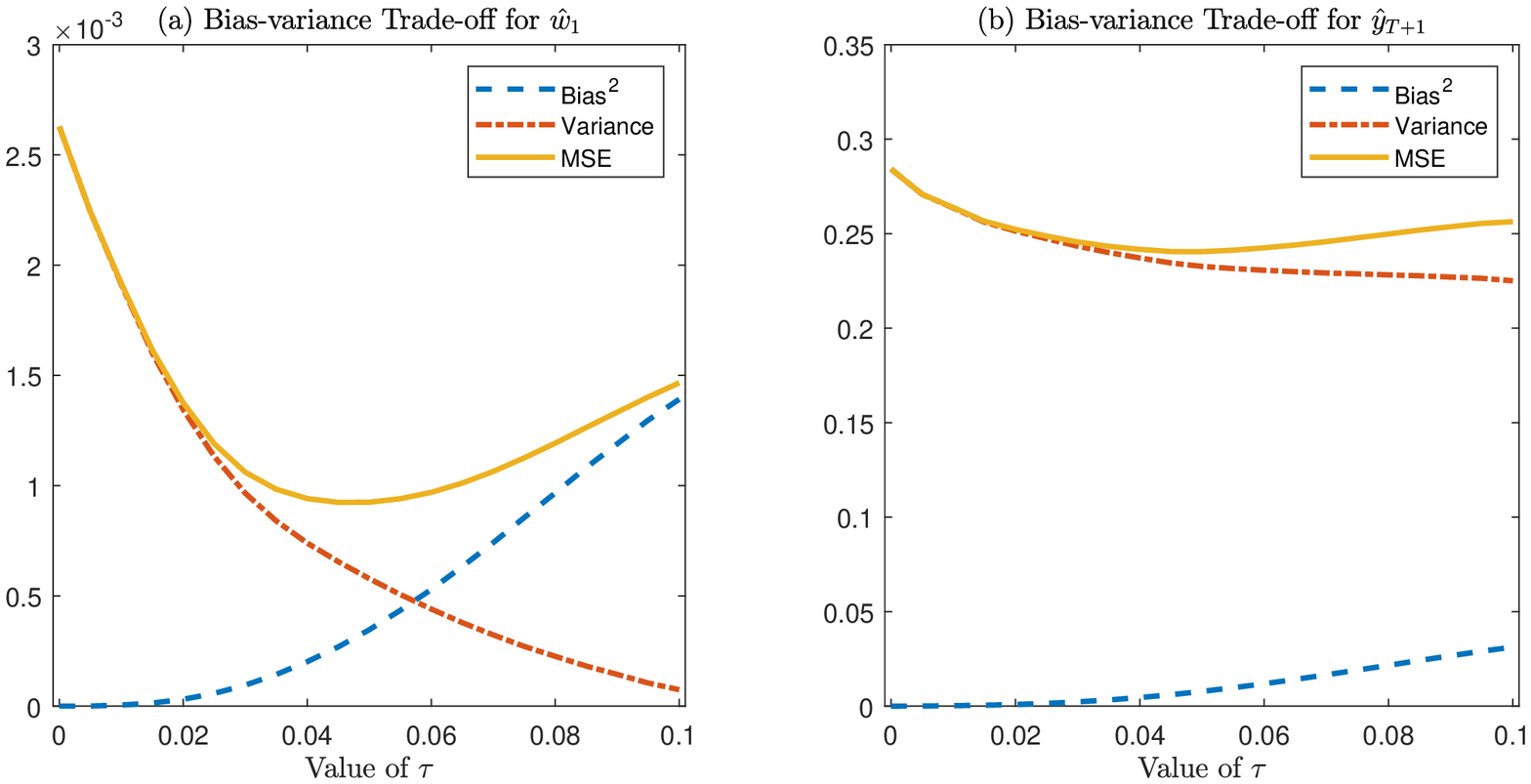} }
\caption{The Bias and Variance Trade-off under Different $\protect\tau $
Values}
\label{Fig1}
\end{figure}

Figure \ref{Fig1} demonstrates the bias and variance trade-off under a range
of $\tau $ in a simulated numerical example. Let 
\begin{equation*}
y_{t+1}=\sum_{i=1}^{20}w_{i}f_{it}+u_{t+1},\ \ t=1,\ldots ,100.
\end{equation*}%
The dependent variable $y_{t+1}$ is a linear combination of two groups of
input variables $\{f_{it}\}_{i=1}^{10}$ and $\{f_{it}\}_{i=11}^{20}$ with
group weights 
\begin{equation*}
w_{i}=0.09\cdot 1\left\{ 1\leq i\leq 10\right\} +0.01\cdot 1\left\{ 11\leq
i\leq 20\right\}
\end{equation*}
so that $\sum_{i=1}^{20}w_{i}=1$. We set $f_{it}\sim \mbox{i.i.d.\,}N(1,1)$
for $1\leq i\leq 10$, $f_{it}\sim \mbox{i.i.d.\,}N(0,1)$ for $11\leq i\leq
20 $, and $u_{t+1}\sim \mbox{i.i.d.\,}N(0,0.25)$. We estimate the weights by 
$\ell _{2}$-relaxation under different values of $\tau =0,0.005,0.01,\ldots
,0.1$. Let $\widehat{w}_{1}=\widehat{w}_{1\tau } $ denote the first element
of the $\ell _{2}$-relaxation estimator $\mathbf{\widehat{w}}_{\tau }.$

We report in Figure \ref{Fig1} the empirical squared bias, variance and MSE
of $(\widehat{w}_{1}=\widehat{w}_{1\tau })$ and those of the one-step-ahead
forecast $\widehat{y}_{102}$ over 1000 replications as a function of $\tau $%
. The figure shows that both the $\ell _{2}$-relaxation weight estimator and
the one-step-ahead forecast associated with a small value of $\tau $ have
small biases but large variances whereas those associated with a large value
of $\tau $ have large biases but small variances. In the middle, there is a
wide range of values of $\tau $ where the combined forecast yields smaller
MSFE than either that of the simple average estimator (when $\tau $ is
sufficiently large) or that of the classical optimal weight $\widehat{%
\mathbf{w}}^{\mathrm{C}}$ (when $\tau =0$). % \end{enumerate}

\subsection{SNR in the Section \protect\ref{sec:simulations}}

\label{signal}

To calculate the signal-to-noise ratio (SNR) in the simulations, we
decompose $y_{t+1}=\mathbf{w}_{\psi }^{\ast \prime }\left( \mathbf{f}_{t}-%
\mathbf{u}_{t}\right) +u_{y,t+1}$ into 
\begin{equation*}
y_{t+1}=\underbrace{\mathbf{w}_{\psi }^{\ast \prime }\left( \mathbf{f}_{t}-%
\mathcal{P}\left[ \mathbf{u}_{t}|\mathbf{f}_{t}\right] \right) }_{\mathrm{%
signal}}-\underbrace{\mathbf{w}_{\psi }^{\ast \prime }\left( \mathbf{u}_{t}-%
\mathcal{P}\left[ \mathbf{u}_{t}|\mathbf{f}_{t}\right] \right) +u_{y,t+1}}_{%
\mathrm{noise}},
\end{equation*}%
where $\mathcal{P}\left[ \mathbf{u}_{t}|\mathbf{f}_{t}\right] :=E\left[ 
\mathbf{u}_{t}\mathbf{f}_{t}^{\prime }\right] \left( E\left[ \mathbf{f}_{t}%
\mathbf{f}_{t}^{\prime }\right] \right) ^{-1}\mathbf{f}_{t}=\mathbf{\Omega }%
_{u}\left( \mathbf{\Psi }+\mathbf{\Omega }_{u}\right) ^{-1}\mathbf{f}_{t}$
is the linear projection of $\mathbf{u}_{t}$ onto the linear space spanned
by $\mathbf{f}_{t}$. By construction, the signal and noise terms are
orthogonal in that 
\begin{eqnarray*}
&&E\left[ \mathbf{w}_{\psi }^{\ast \prime }\left( \mathbf{f}_{t}-\mathcal{P}%
\left[ \mathbf{u}_{t}|\mathbf{f}_{t}\right] \right) \left( \mathbf{u}_{t}-%
\mathcal{P}\left[ \mathbf{u}_{t}|\mathbf{f}_{t}\right] \right) ^{\prime }%
\mathbf{w}_{\psi }^{\ast }\right] \\
&=&\mathbf{w}_{\psi }^{\ast \prime }E\left[ \left( \mathbf{f}_{t}-\mathbf{%
\Omega }_{u}\left( \mathbf{\Psi }+\mathbf{\Omega }_{u}\right) ^{-1}\mathbf{f}%
_{t}\right) \left( \mathbf{u}_{t}-\mathbf{\Omega }_{u}\left( \mathbf{\Psi }+%
\mathbf{\Omega }_{u}\right) ^{-1}\mathbf{f}_{t}\right) ^{\prime }\right] 
\mathbf{w}_{\psi }^{\ast }=0.
\end{eqnarray*}%
Simple calculation shows that the variance of the noise component is 
\begin{equation*}
\mathrm{var}\left[ \mathbf{w}_{\psi }^{\ast \prime }\left( \mathbf{u}_{t}-%
\mathcal{P}\left[ \mathbf{u}_{t}|\mathbf{f}_{t}\right] \right) +u_{y,t+1}%
\right] =\mathbf{w}_{\psi }^{\ast \prime }\left[ \mathbf{\Omega }_{u}-%
\mathbf{\Omega }_{u}\left( \mathbf{\Psi }+\mathbf{\Omega }_{u}\right) ^{-1}%
\mathbf{\Omega }_{u}\right] \mathbf{w}_{\psi }^{\ast }+\sigma _{y}^{2}\text{.%
}
\end{equation*}%
This, along with the fact that $\mathrm{var}[y_{t+1}]=\mathbf{w}_{\psi
}^{\ast \prime }\mathbf{\Psi }\mathbf{w}_{\psi }^{\ast }+\sigma _{y}^{2},$
implies that SNR here can be defined as 
\begin{eqnarray*}
\mathrm{SNR} &=&\frac{\mathrm{var}[y_{t+1}]-\mathrm{var}\left[ \mathbf{w}%
_{\psi }^{\ast \prime }\left( \mathbf{u}_{t}-\mathcal{P}\left[ \mathbf{u}%
_{t}|\mathbf{f}_{t}\right] \right) +u_{y,t+1}\right] }{\mathrm{var}\left[ 
\mathbf{w}_{\psi }^{\ast \prime }\left( \mathbf{u}_{t}-\mathcal{P}\left[ 
\mathbf{u}_{t}|\mathbf{f}_{t}\right] \right) +u_{y,t+1}\right] } \\
&=&\frac{\mathbf{w}_{\psi }^{\ast \prime }\left[ \mathbf{\Psi -\Omega }_{u}+%
\mathbf{\Omega }_{u}\left( \mathbf{\Psi }+\mathbf{\Omega }_{u}\right) ^{-1}%
\mathbf{\Omega }_{u}\right] \mathbf{w}_{\psi }^{\ast }}{\mathbf{w}_{\psi
}^{\ast \prime }\left[ \mathbf{\Omega }_{u}-\mathbf{\Omega }_{u}\left( 
\mathbf{\Psi }+\mathbf{\Omega }_{u}\right) ^{-1}\mathbf{\Omega }_{u}\right] 
\mathbf{w}_{\psi }^{\ast }+\sigma _{y}^{2}}.
\end{eqnarray*}

\subsection{Simulation Results for DGP 2 by Using the 5-fold CV}

\label{con5fold}

Table \ref{TableA1} reports the MSFEs for DGP 2 by using the conventional
5-fold CV to choose $\tau $ as in Box 2. Note that only Lasso, Ridge, and $%
\ell _{2}$-relaxation depend on tuning parameters. The results of other
estimators with no data-driven tuning parameters are the same as those in
Panel C and D of Table \ref{table1}. We find that MSFEs from the 5-fold CV
generally provides outcomes slightly better than the time series CV as in
Box 3. In theory, both the randomly formed CV folds and the time series CV
folds allow consistent estimation of VC from the training data as $%
T\rightarrow \infty $. In practice, however, the former uses $4/5$ of the
sample for training in each fold, while the latter utilizes $1/5,2/5,3/5,4/5$
respectively for each fold. The smaller practical training data sample sizes
in the time series scheme tend to yield more noisy VC estimates, and thereby
larger MSFEs in Table \ref{table1}.

{\ 
\begin{table}[th]
\caption{The MSFE for DGP 2 by Using the 5-fold CV }
\label{TableA1}\centering 
{\ {\scriptsize 
\begin{tabular}{cccccccccccccccc}
\hline
& $T$ & $N$ & $K$ & Oracle & SA & Lasso & Ridge & \multicolumn{3}{c}{PC} & 
& \multicolumn{3}{c}{$\ell_2$-relax} &  \\ \cline{9-11}\cline{13-15}
&  &  &  &  &  &  &  & $q = 5$ & $q=10$ & $q=20$ &  & $\widehat{\boldsymbol{%
\Sigma}}_s$ & $\widehat{\boldsymbol{\Sigma}}_1$ & $\widehat{\boldsymbol{%
\Sigma}}_2$ &  \\ \hline
\multicolumn{10}{l}{\emph{Panel A: Low SNR}} &  &  &  &  &  &  \\ 
& 50 & 100 & 2 & 0.292 & 1.032 & 0.404 & 1.131 & 0.787 & 0.800 & 0.763 &  & 
0.386 & 0.390 & 0.364 &  \\ 
& 100 & 200 & 4 & 0.133 & 3.052 & 0.232 & 1.090 & 1.134 & 1.275 & 1.531 &  & 
0.219 & 0.232 & 0.265 &  \\ 
& 200 & 300 & 6 & 0.066 & 3.699 & 0.141 & 0.398 & 1.109 & 1.050 & 1.173 &  & 
0.114 & 0.124 & 0.082 &  \\ 
&  &  &  &  &  &  &  &  &  &  &  &  &  &  &  \\ 
\multicolumn{10}{l}{\emph{Panel B: High SNR}} &  &  &  &  &  &  \\ 
& 50 & 100 & 2 & 0.262 & 0.993 & 0.394 & 1.387 & 0.702 & 0.747 & 0.751 &  & 
0.282 & 0.305 & 0.267 &  \\ 
& 100 & 200 & 4 & 0.146 & 3.210 & 0.370 & 1.376 & 1.030 & 1.257 & 1.428 &  & 
0.292 & 0.281 & 0.315 &  \\ 
& 200 & 300 & 6 & 0.106 & 4.524 & 0.162 & 0.430 & 1.343 & 1.601 & 1.639 &  & 
0.177 & 0.193 & 0.167 &  \\ \hline
\end{tabular}
} }
\end{table}
}

\subsection{MAFE for the Simulations}

\label{mafe}

In this section we report the mean absolute forecast error (MAFE) in the
simulations. The MAFE is defined as 
\begin{equation*}
\text{MAFE}=E\left[ \left\vert y_{T+1}-\widehat{\mathbf{w}}^{\prime }\mathbf{%
f}_{T+1}\right\vert \right] -\sigma _{y}\sqrt{2/\pi },
\end{equation*}%
where the unpredictable component, $\sigma _{y}\int_{-\infty }^{\infty }|x|%
\frac{1}{\sqrt{2\pi }}\exp (-x^{2}/2) \mathrm{d} x=\sigma _{y}\sqrt{2/\pi }$%
, is subtracted. In simulations we know $\sigma_y$ albeit it is unknown in
empirical applications.

{\ The results are collected in Table \ref{overall}. Results by the Oracle,
Lasso, Ridge, and the $\ell _{2}$-relaxation tend to decrease with }$\left( {%
T,N}\right) ${, yet results by SA and PC may diverge as }$\left( {T,N}%
\right) ${\ increases. Similar to the MSFE results in the main text, the $%
\ell _{2}$-relaxation is always the best feasible estimator in all cases. }

{\ 
\begin{table}[tph]
\caption{Results of Prediction Accuracy by MAFE}
\label{overall}\centering 
{\ {\scriptsize 
\begin{tabular}{ccccccccccccccc}
\hline
& $T$ & $N$ & $K$ & Oracle & SA & Lasso & Ridge & \multicolumn{3}{c}{PC} & 
\multicolumn{3}{c}{$\ell_2$-relax} &  \\ 
&  &  &  &  &  &  &  & $q = 5$ & $q=10$ & $q=20$ & $\hat\Sigma_0$ & $%
\hat\Sigma_1$ & $\hat\Sigma_2$ &  \\ \hline
\multicolumn{10}{l}{\emph{Panel A: DGP1 with Low SNR}} &  &  &  &  &  \\ 
& 50 & 100 & 2 & 0.125 & 0.318 & 0.160 & 0.413 & 0.242 & 0.248 & 0.232 & 
0.154 & 0.143 & 0.135 &  \\ 
& 100 & 200 & 4 & 0.063 & 0.949 & 0.098 & 0.323 & 0.370 & 0.440 & 0.436 & 
0.089 & 0.090 & 0.087 &  \\ 
& 200 & 300 & 6 & 0.034 & 1.063 & 0.055 & 0.142 & 0.386 & 0.439 & 0.423 & 
0.054 & 0.050 & 0.043 &  \\ 
&  &  &  &  &  &  &  &  &  &  &  &  &  &  \\ 
\multicolumn{10}{l}{\emph{Panel B: DGP1 with High SNR}} &  &  &  &  &  \\ 
& 50 & 100 & 2 & 0.337 & 0.696 & 0.381 & 0.497 & 0.586 & 0.596 & 0.586 & 
0.351 & 0.349 & 0.344 &  \\ 
& 100 & 200 & 4 & 0.221 & 1.267 & 0.255 & 0.339 & 0.695 & 0.765 & 0.801 & 
0.234 & 0.234 & 0.229 &  \\ 
& 200 & 300 & 6 & 0.182 & 1.510 & 0.210 & 0.222 & 0.790 & 0.802 & 0.805 & 
0.205 & 0.206 & 0.197 &  \\ 
&  &  &  &  &  &  &  &  &  &  &  &  &  &  \\ 
\multicolumn{10}{l}{\emph{Panel C: DGP2 with Low SNR}} &  &  &  &  &  \\ 
& 50 & 100 & 2 & 0.104 & 0.318 & 0.137 & 0.407 & 0.250 & 0.252 & 0.253 & 
0.140 & 0.139 & 0.131 &  \\ 
& 100 & 200 & 4 & 0.063 & 0.820 & 0.097 & 0.320 & 0.366 & 0.389 & 0.467 & 
0.091 & 0.102 & 0.117 &  \\ 
& 200 & 300 & 6 & 0.020 & 0.917 & 0.048 & 0.138 & 0.348 & 0.329 & 0.366 & 
0.061 & 0.050 & 0.048 &  \\ 
&  &  &  &  &  &  &  &  &  &  &  &  &  &  \\ 
\multicolumn{10}{l}{\emph{Panel D: DGP2 with High SNR}} &  &  &  &  &  \\ 
& 50 & 100 & 2 & 0.336 & 0.718 & 0.386 & 0.474 & 0.590 & 0.615 & 0.611 & 
0.349 & 0.351 & 0.344 &  \\ 
& 100 & 200 & 4 & 0.231 & 1.361 & 0.272 & 0.348 & 0.715 & 0.782 & 0.845 & 
0.248 & 0.254 & 0.241 &  \\ 
& 200 & 300 & 6 & 0.192 & 1.589 & 0.214 & 0.231 & 0.826 & 0.906 & 0.907 & 
0.209 & 0.208 & 0.199 &  \\ 
&  &  &  &  &  &  &  &  &  &  &  &  &  &  \\ 
\multicolumn{10}{l}{\emph{Panel E: DGP3 with Low SNR}} &  &  &  &  &  \\ 
& 50 & 100 & 2 & 0.160 & 0.350 & 0.345 & 0.477 & 0.291 & 0.324 & 0.305 & 
0.191 & 0.179 & 0.162 &  \\ 
& 100 & 200 & 4 & 0.102 & 0.949 & 0.238 & 0.443 & 0.517 & 0.520 & 0.581 & 
0.122 & 0.135 & 0.116 &  \\ 
& 200 & 300 & 6 & 0.098 & 1.105 & 0.181 & 0.278 & 0.553 & 0.619 & 0.662 & 
0.101 & 0.117 & 0.107 &  \\ 
&  &  &  &  &  &  &  &  &  &  &  &  &  &  \\ 
\multicolumn{10}{l}{\emph{Panel F: DGP3 with High SNR}} &  &  &  &  &  \\ 
& 50 & 100 & 2 & 0.420 & 0.775 & 0.620 & 0.626 & 0.683 & 0.706 & 0.717 & 
0.442 & 0.440 & 0.429 &  \\ 
& 100 & 200 & 4 & 0.348 & 1.373 & 0.436 & 0.483 & 0.896 & 0.939 & 0.978 & 
0.367 & 0.352 & 0.353 &  \\ 
& 200 & 300 & 6 & 0.338 & 1.635 & 0.380 & 0.405 & 1.046 & 1.064 & 1.101 & 
0.359 & 0.368 & 0.347 &  \\ \hline
\end{tabular}
} }
\end{table}
}

\subsection{MAFE for the Empirical Applications}

\label{mafe2}

Tables \ref{emp_a1} and \ref{emp_a2} report the relative (to the benchmark)
MAFEs in the movie and HICP forecasting applications, respectively. The
results and patterns are robust in comparison with those based on MSFE. {\ 
\begin{table}[tph]
\caption{Relative MAFE of Movie Forecasting}
\label{emp_a1}\centering
{\ {\ \ 
\begin{tabular}{cccccccccc}
\hline
$n_{\mathrm{ev}}$ & PMA & CSR$_{10}$ & CSR$_{15}$ & peLasso & Lasso & Ridge
& \multicolumn{3}{c}{$\ell_2$-relax} \\ 
&  &  &  &  &  &  & $\widehat{\boldsymbol{\Sigma}}_s$ & $\widehat{%
\boldsymbol{\Sigma}}_1$ & $\widehat{\boldsymbol{\Sigma}}_2$ \\ \hline
10 & 1.000 & 1.144 & 1.164 & 2.707 & 1.013 & 1.018 & 0.984 & 0.985 & 0.983
\\ 
20 & 1.000 & 1.139 & 1.189 & 2.610 & 1.024 & 1.030 & 0.985 & 0.988 & 0.985
\\ 
30 & 1.000 & 1.068 & 1.138 & 2.432 & 1.019 & 1.039 & 0.984 & 0.982 & 0.980
\\ 
40 & 1.000 & 1.013 & 1.098 & 2.269 & 1.008 & 1.039 & 0.995 & 0.990 & 0.970
\\ \hline
\multicolumn{5}{l}{\scriptsize Note: The MAFE of PMA is normalized as 1.} & 
&  &  &  & 
\end{tabular}%
} }
\end{table}
}

{\ 
\begin{table}[tph]
\caption{Relative MAFE of HICP Forecasting}
\label{emp_a2}\centering
{\ {\ 
\begin{tabular}{lcccccc}
\hline
Horizon & SA & Lasso & Ridge & \multicolumn{3}{c}{$\ell_2$-relax} \\ 
&  &  &  & $\widehat{\boldsymbol{\Sigma}}_s$ & $\widehat{\boldsymbol{\Sigma}}%
_1$ & $\widehat{\boldsymbol{\Sigma}}_2$ \\ \hline
One-year-ahead & 1.000 & 0.909 & 0.922 & 0.933 & 0.847 & 0.885 \\ 
Two-year-ahead & 1.000 & 0.828 & 0.924 & 0.817 & 0.850 & 0.756 \\ \hline
\multicolumn{5}{l}{\scriptsize Note: The MAFE of SA is normalized as 1.} & 
& 
\end{tabular}%
} }
\end{table}
}

\newpage %\subsection{More Robustness Checks}

\subsection{Converging Path of Weights}

\begin{figure}[h!]
\centering
\includegraphics[scale = 0.5]{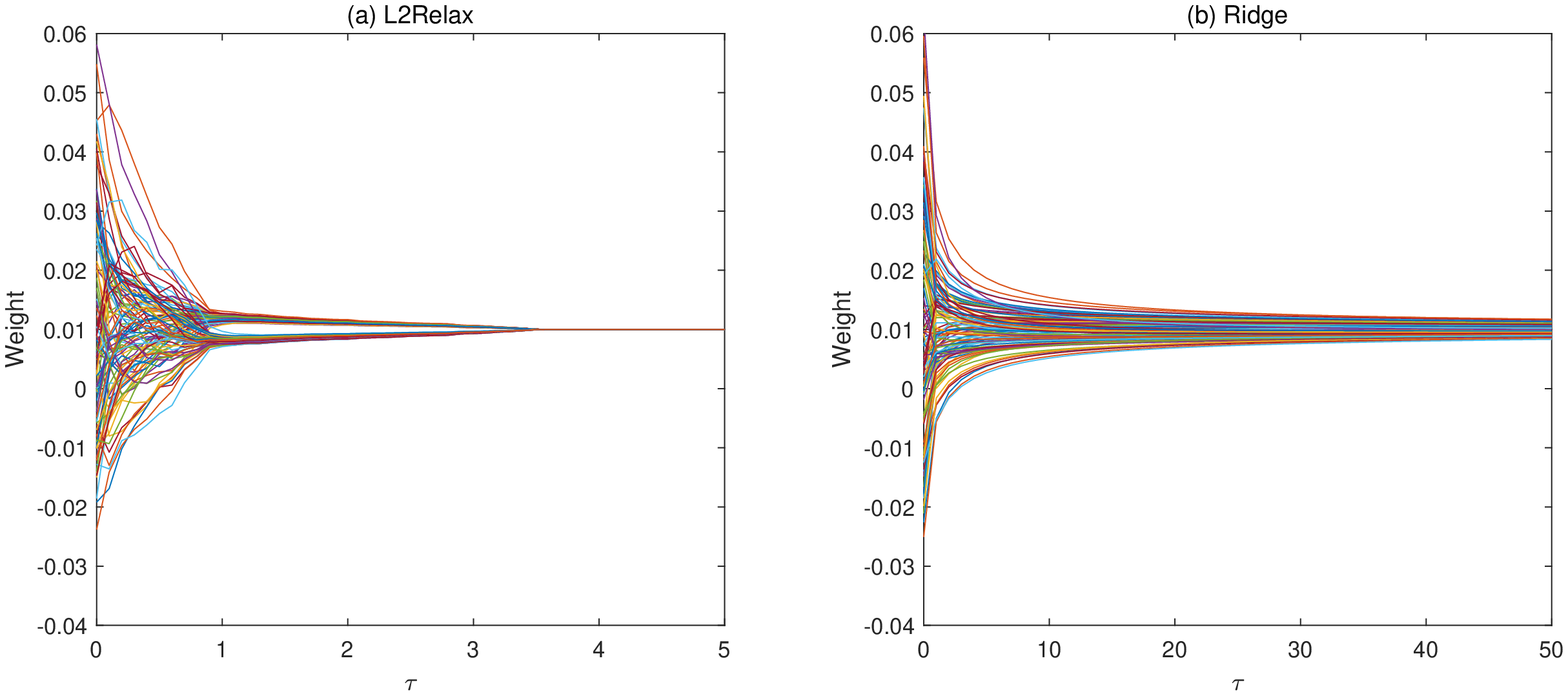}
\caption{Converging Path of Weights over Different Values of $\protect\tau$}
\label{path}
\end{figure}

In this section, we estimate the weights of the $\ell _{2}$-relaxation
estimator over different values of $\tau $ and compare the converging path
with the Ridge weights. The data is generated from DGP 1 in the simulation
with 50 observations and 100 forecasters that are categorized into two
latent groups. The two subplots of Figure \ref{path} illustrate the
converging paths of the $\ell _{2}$-relaxation and Ridge weights over
different values of $\tau $, respectively. The horizontal axis represents
the values of $\tau $ and the vertical axis shows the values of weights for
the 100 forecasters.

When $\tau =0$, all the weights are scattered. As $\tau $ increases, the $%
\ell _{2}$-relaxation weights quickly converge to two group centers, which
can be clearly observed for $\tau \in \lbrack 1,3.5]$. When $\tau >3.5$, the
tuning parameter is so large that the weights converge to the simple average
(SA) weights $1/N$. In contrast, the Ridge weights do not converge to two
group clusters even though they are centered around the SA weights.

\end{document}